\documentclass[journal,10pt]{IEEEtran}
\usepackage{fontspec}
\usepackage{amsmath,mathtools,booktabs,array,tabularx,longtable,graphicx}
\usepackage{unicode-math}
\usepackage{microtype,xurl}
\usepackage[hidelinks,unicode]{hyperref}
\usepackage{enumitem,algorithm,algpseudocode,stfloats,placeins}
\usepackage{xcolor}
\setlist{nosep,leftmargin=*}
\newcolumntype{C}{>{\centering\arraybackslash}X}
\newcolumntype{Y}{>{\raggedright\arraybackslash}X}
\newcolumntype{L}[1]{>{\raggedright\arraybackslash}p{#1}}

\newcommand{\E}{\mathbb E}\newcommand{\R}{\mathcal R}

\newcommand{\norm}[1]{\left\lVert#1\right\rVert}
\newcommand{\rev}[1]{#1}\newenvironment{revisionblock}{}{}
\ifdefined\showchanges
  \definecolor{revisionblue}{RGB}{0,70,190}
  \renewcommand{\rev}[1]{\textcolor{revisionblue}{#1}}
  
  \expandafter\def\csname revisedbib@hvarfner2024\endcsname{}
  \expandafter\def\csname revisedbib@gao2023ace\endcsname{}
  \expandafter\def\csname revisedbib@vetter2025\endcsname{}
  \expandafter\def\csname revisedbib@schmitt2024sc\endcsname{}
  \expandafter\def\csname revisedbib@mishra2026sc\endcsname{}
  \expandafter\def\csname revisedbib@elfwing2017silu\endcsname{}
  \let\originalbibitem\bibitem
  \renewcommand{\bibitem}[1]{\color{black}\ifcsname revisedbib@#1\endcsname\color{revisionblue}\fi\originalbibitem{#1}}
\fi

\ifdefined\showedits
  \definecolor{editblue}{RGB}{0,70,190}
  
  \expandafter\def\csname editedbib@amaran2016simulation\endcsname{}
  \expandafter\def\csname editedbib@beaumont2019abc\endcsname{}
  \let\editOriginalBibitem\bibitem
  \renewcommand{\bibitem}[1]{\color{black}\ifcsname editedbib@#1\endcsname\color{editblue}\fi\editOriginalBibitem{#1}}
\fi
\algrenewcommand\algorithmicrequire{\textbf{Input}}
\algrenewcommand\algorithmicensure{\textbf{Output}}
\algrenewcommand\alglinenumber[1]{\footnotesize #1.}
\makeatletter
\def\@IEEEsectpunct{.\ }
\makeatother

\newcommand{\roundedit}[1]{#1}
\ifdefined\RoundBlue
  \definecolor{roundblue}{RGB}{0,70,190}
  \renewcommand{\roundedit}[1]{\textcolor{roundblue}{#1}}
\fi
\begin{document}
\title{Amortized Optimization for Calibrating Stochastic Social Simulators}
\author{Muyao~Zhong and Peng~Yang,~\IEEEmembership{Senior Member,~IEEE}%
\thanks{This work was supported by National Natural Science Foundation of China (Grant 62272210). Corresponding author: Dr. Peng Yang (email: yangp@sustech.edu.cn).}%
\thanks{M. Zhong is with Harbin Institute of Technology, Harbin 150001, China.}%
\thanks{M. Zhong and P. Yang are with the Guangdong Key Laboratory of Brain-Inspired Intelligent Computation, Department of Computer Science and Engineering, Southern University of Science and Technology, Shenzhen 518055, China.}%
\thanks{P. Yang is also with the Department of Statistics and Data Science, Southern University of Science and Technology, Shenzhen 518055, China.}}

\maketitle
\begin{abstract}
Stochastic simulators need suitable parameter values to reproduce observed behaviour in complex systems. Searching for these values through repeated simulations makes frequent recalibration computationally expensive. We develop an amortized optimizer that uses simulations collected in advance to learn how to choose parameters that make simulated outcomes agree with observations. Once fitted, it maps new observations to feasible parameters without further simulator calls. The method retains alternative parameter settings that can explain similar observations and compares their responses, avoiding the conflicting labels that arise when training a single parameter regressor. Historical observations also provide supervision by showing which simulated responses agree with the data, even when the underlying parameters are unknown. \roundedit{Experiments across four simulators compare calibration accuracy and query time.} \roundedit{Dedicated cost analyses quantify traffic preparation and the recorded financial reuse costs.} \roundedit{On new traffic targets, local correction more than halves query time relative to empirical and surrogate selection; a learned inverse is faster still, with similar mean error.} \roundedit{In a retrospective comparison on real financial data, our method returns parameters about $1.07\times10^{7}$ times faster than a 10-minute Gaussian-process Bayesian optimization search, with mean calibration error 3.06\% higher.} Reusing simulation work across observations supports repeated parameter updates when the available time is too short for a new search.
\end{abstract}
\begin{IEEEkeywords}Optimization algorithms, amortized optimization, stochastic simulators, simulation memory, amortized surrogate.\end{IEEEkeywords}
\section{Introduction}\label{sec:intro}

Simulation provides a controlled way to study complex social systems whose behaviour is difficult to investigate through direct experiments alone. It helps examine how individual interactions give rise to collective behaviour. Simulators of traffic flow~\cite{treiber2000congested}, neural circuits~\cite{goncalves2020} and financial markets~\cite{cont2010stochastic}, for example, generate system responses from specified parameters and initial conditions. Applying a simulator to an observed system requires parameter values that reproduce its measured behaviour. Calibration estimates these values by comparing simulated and observed outcomes~\cite{kennedy2001bayesian,thiele2014calibration}. As new observations become available, calibration may need to be repeated, making the cost of each parameter search a practical concern.

For a stochastic simulator, this search must account for variation between repeated runs. Let $z = S(\theta,c,\xi)$ denote a simulator response, where $\theta$ belongs to the feasible parameter set $\Theta$. Here, $c$ gives the known conditions and $\xi$ represents simulation randomness. Let $o$ denote the information available at calibration time. For an observed target $x$, we have $o = (x,c)$, and a discrepancy $D(z,x)$ defines the expected calibration error

\begin{equation}\label{eq:introcalibration}
\R(\theta\mid o)=\E_{\xi}\!\left[D(S(\theta,c,\xi),x)\right].
\end{equation}

Calibration seeks $\theta^*(o)\in\arg\min_{\theta\in\Theta}\R(\theta\mid o)$. The expectation averages the discrepancy over simulator runs at the same parameter values.

\begin{figure*}[!t]
\setlength{\abovecaptionskip}{1.5pt}
\centering\includegraphics[width=\textwidth]{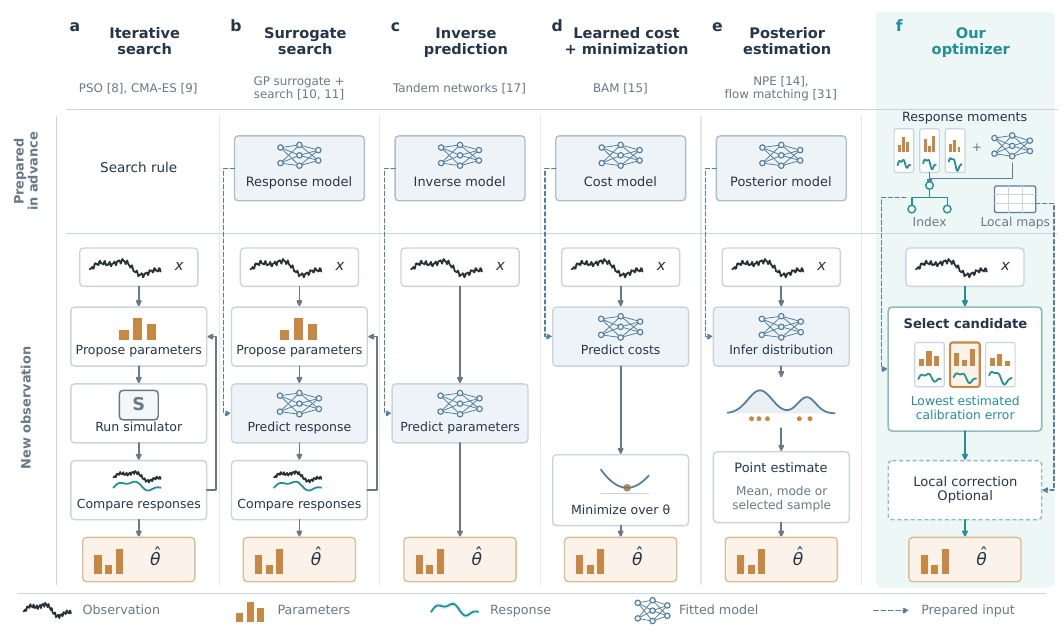}
\par\nointerlineskip
\caption{\roundedit{Representative workflows for obtaining calibration parameters. Columns share observation, parameter, response and fitted-model symbols. Iterative and surrogate search repeat evaluations; inverse prediction returns parameters directly. Learned costs require minimization, and posterior estimates require a point readout. Column (f) shows the observed-target route in Fig.~\ref{fig:method} for fixed conditions. Stored and predicted response moments form its candidate index, which selects a parameter by estimated calibration error before optional local correction. Dashed arrows denote reused quantities. Symbols are schematic.}}\label{fig:comparison}
\end{figure*}

Finding these parameters is an instance of simulation-based optimization (SBO)~\cite{amaran2016simulation}, which minimizes an expected objective estimated through stochastic simulation. Iterative methods propose parameters $\theta$, simulate their responses and use the resulting estimates of $\R$ to guide further proposals~\cite{amaran2016simulation,shahriari2016taking}. Population methods such as particle swarm optimization~\cite{kennedy1995pso} and covariance matrix adaptation evolution strategy~\cite{hansen2001cmaes} perform this search through successive candidate evaluations. Surrogate-assisted methods~\cite{shahriari2016taking,jin2011surrogate} reduce this evaluation burden by learning responses or objective values and using those predictions to guide the search. These estimates must also account for simulation noise. Repeated runs at the same $\theta$ help distinguish changes in expected error from random variation~\cite{ankenman2010stochastic}. When the available observations $o$ change, solving each objective $\R(\theta\mid o)$ separately repeats the search effort~\cite{amos2023amortized}.

These repeated searches concern the same simulator, so earlier simulations contain information that can be reused when $o$ changes. Amortization exploits this shared structure by preparing models for many later observations~\cite{amos2023amortized}. Bayesian simulation-based inference (SBI)~\cite{cranmer2020frontier,papamakarios2016fast}, for example, learns a conditional parameter distribution from simulated pairs and reuses it for later observations. Using this distribution for calibration requires choosing a parameter according to the expected error $\R$. For a nonlinear simulator, neither the posterior mean nor a density mode necessarily minimizes this error~\cite{gorecki2024bam}.

Amortized optimization~\cite{amos2023amortized} addresses parameter choice directly by learning a map from problem inputs to approximate solutions, here $\theta^{\ast}(o)$. Learning this map from parameter labels becomes more difficult when the simulator inverse is not unique. Different parameters $\theta_{1} \neq \theta_{2}$ can produce similar responses, giving an inverse regressor conflicting labels for similar inputs. Minimizing squared parameter error can then average distinct solutions into a parameter whose simulated response fits none of them~\cite{bishop1994mixture,liu2018tandem}. Training the map to minimize $\R$ can avoid this label conflict, but requires estimates of simulated response error. Obtaining these estimates through repeated simulator calls can be costly. The practical challenge is therefore to reuse simulations both to assess parameter choices and to return a parameter for each new observation. Figure~\ref{fig:comparison} compares how these approaches turn an observation into a parameter estimate and which computations they repeat for each observation.

We develop an amortized optimizer $\widehat\theta = \mathcal{A}_{\phi}(o)$ that reuses simulated responses to select and refine parameters for new observations (Fig.~\ref{fig:comparison}f). We retain feasible candidates $\left\{ \theta_{j} \right\}_{j = 1}^{M}$ and their repeated responses in a simulation memory. These simulations also train an amortized surrogate to estimate response means and variances. Together, stored responses and predicted moments let the optimizer compare candidates by estimated calibration error while preserving distinct parameter alternatives. For fixed conditions, we prepare candidate retrieval and local inverse maps to make this comparison and subsequent refinement efficient. At query time, the optimizer selects a candidate and can apply the prepared local correction when $x$ is known. Stored quantities and fitted models form $\phi$, so each query returns a feasible parameter without new simulator evaluations.

Because candidates are compared by response agreement, real observations can also provide supervision. Their generating parameters are usually unknown, but measured outcomes show which candidate responses agree with the system. We use these comparisons to train a score predictor. When the target outcome follows calibration, the predictor takes $o=(h,c)$, where $h$ is the available history, and estimates candidate scores from prior records. The history-based route in Fig.~\ref{fig:method} then selects a parameter before the new outcome is known.

This work makes two contributions to repeated stochastic calibration.

\begin{itemize}

\item \roundedit{\textbf{Indexed selection with precomputed local correction.}} \roundedit{For fixed conditions, we combine an exact candidate index with bounded affine maps prepared from repeated simulator responses.} \roundedit{A query retrieves a feasible anchor and applies one stored map to return a continuous parameter estimate.} \roundedit{This combines a finite candidate memory with continuous refinement without fitting a separate inverse predictor.}

\item \roundedit{\textbf{Candidate comparison for non-unique inverses and incomplete responses.}} \roundedit{For observed targets, we combine empirical repeats and learned moments in the criterion that selects among feasible alternatives, including availability probabilities when features are missing.} \roundedit{Keeping these alternatives separates response matching from fitting one inverse output to competing parameter labels.}

\end{itemize}

We evaluate the optimizer across predator--prey, traffic, neural-circuit and financial tasks. \roundedit{The prepared controls show when wider coverage, learned response availability and local refinement change the accuracy and cost of parameter selection.} \roundedit{Separate confirmation targets assess fixed rules, while the financial comparisons examine retrospective quality and the workload needed to recover preparation.}

The remainder of the paper is organized as follows. Section~\ref{sec:related} reviews related work, and Section~\ref{sec:method} develops the method. Sections~\ref{sec:setup} and~\ref{sec:results} present the experimental setup and results. Sections~\ref{sec:discussion} and~\ref{sec:conclusion} discuss the findings and conclude the paper.

\section{Related work}\label{sec:related}

\subsection{Simulation-based optimization methods}

Calibration connects simulator parameters to observable behaviour through optimization or probabilistic estimation~\cite{thiele2014calibration,grazzini2017bayesian}. PSO~\cite{kennedy1995pso} and CMA-ES~\cite{hansen2001cmaes} use simulation feedback to update a population or search distribution. Bayesian optimization~\cite{shahriari2016taking} instead learns an objective model and chooses evaluations through an acquisition criterion. TuRBO~\cite{eriksson2019scalable} restricts this search to adaptive local regions. For expensive simulators, the cost of evaluating each candidate limits how far these searches can proceed within a fixed budget.

Surrogate-assisted optimization~\cite{jin2011surrogate} reduces this evaluation cost by using predicted responses or objective values to guide the search. For stochastic simulators, stochastic kriging~\cite{ankenman2010stochastic} separates variation between repeated runs from uncertainty about the mean response in a Gaussian-process model. A fitted response model can be reused for a new target, but an iterative optimizer must still search it for suitable parameters. Our optimizer also prepares the quantities needed for candidate comparison and local parameter updates. We therefore compare its calibration error with that of iterative searches under different time budgets.

\subsection{Amortized optimization methods}

Amortized optimization~\cite{amos2023amortized} shares the solution process across related objectives. One approach is inverse regression from observations to parameters, but a non-unique inverse creates incompatible training labels. Tandem networks~\cite{liu2018tandem} address this difficulty by training an inverse model through a separately fitted response model. Other approaches learn the costs used to compare possible solutions. Bayesian amortized decision making (BAM)~\cite{gorecki2024bam} learns expected action costs and selects actions by minimizing them. Amortized cost estimation (ACE)~\cite{gao2023ace} learns expected simulator-to-observation distances and can include real observations as training targets for generalized Bayesian inference. These methods establish ways to learn from response agreement or decision costs when parameter error is an unsuitable training objective.

The simulations used for learning can also be retained as parameter--response pairs in reference tables. Approximate Bayesian computation (ABC)~\cite{beaumont2019abc} uses such pairs to approximate parameter posteriors, with extensions including regression adjustment~\cite{beaumont2002abc} and random-forest parameter inference~\cite{raynal2019abcrf}. The range of simulated parameters affects what reference tables and fitted models can represent. Related work on offline model-based optimization addresses extrapolation beyond the available data through conservative objective models~\cite{trabucco2021coms}. For financial simulators, multiple response criteria can distinguish similar parameter settings~\cite{wang2025nonidentifiability}, while learned inverse models support repeated inference in agent-based systems~\cite{vargas2025confidence}.

For point calibration, posterior estimates require a choice under $\R$, while learned cost functions require minimization over parameters. Our optimizer combines stored responses with learned response moments to compare feasible candidates by expected calibration error. Under fixed conditions, a prepared index retrieves the candidate with the lowest estimated error, and a local map based on damped least squares~\cite{marquardt1963} can refine it. The experiments examine how candidate coverage and local refinement affect this procedure.

\subsection{Posterior estimation methods}

Bayesian SBI~\cite{cranmer2020frontier,lueckmann2021benchmarking} learns parameter distributions when likelihood evaluation is difficult. Neural posterior estimation uses mixture density networks~\cite{papamakarios2016fast} and normalizing flows~\cite{papamakarios2017masked,durkan2019neural}. Flow matching~\cite{wildberger2023flow}, score models~\cite{geffner2023compositional} and consistency models~\cite{schmitt2024consistency,lu2025consistency} provide further conditional samplers. To use these distributions for point calibration, a readout must return a parameter. This may be a mean, a mode or a sample selected by response error. Our experiments measure response quality and complete query time for the specified readout, with additional mode and response-ranked controls.

Posterior estimators can also learn from real observations without parameter labels. Self-consistency losses~\cite{schmitt2024sc,mishra2026sc} use relationships among posterior, likelihood and prior models to train with such observations. We use measured outcomes to obtain differences between candidate mean-response errors. A score predictor learns these differences from the preceding history, allowing parameter selection before the next outcome is observed. To assess this choice of prediction target, we use a matched conditional surrogate control with the same observations and response model. It predicts the outcome before selecting parameters, whereas the proposed predictor estimates candidate scores directly.

\section{Method}\label{sec:method}

\subsection{Optimization objective and method overview}\label{sec:problem}\label{sec:overview}\label{sec:parameterreturn}

Let $z = S(\theta,c,\xi) \in {\mathbb{R}}^{d}$ be a stochastic simulator response, with feasible parameters $\theta \in \Theta \subset {\mathbb{R}}^{p}$, known conditions $c$ and randomness $\xi$. The response may be a trajectory or a vector of summary features. Calibration uses the information $o$ available at query time, including $c$, to choose parameters that reproduce a response $Y$. For independent simulator draws $Z_{\theta},Z_{\theta}^{\prime}$ and a response loss $\ell$, we define the expected calibration error as

\begin{equation}\label{eq:risk}
\R(\theta\mid o)=\E\!\left[\ell(Z_\theta,Z'_\theta,Y)\mid o\right].
\end{equation}

We seek $\theta^{\ast}(o) \in \arg\min_{\theta \in \Theta}\R(\theta \mid o)$ through an approximate solution map $\widehat\theta = \mathcal{A}_{\phi}(o)$. Its prepared quantities $\phi$ include separately fitted models and stored simulations. The available information determines how the objective is evaluated. For an observed target, $o = (x,c)$ and $Y = x$. Choosing $\ell(Z,Z^{\prime},Y) = D(Z,Y)$ gives the expected discrepancy from that target. When the outcome follows the query, $o = (h,c)$ contains history and conditions, and the expectation also averages over $Y \mid h,c$.

\begin{figure*}[!t]
\centering\includegraphics[width=0.92\textwidth]{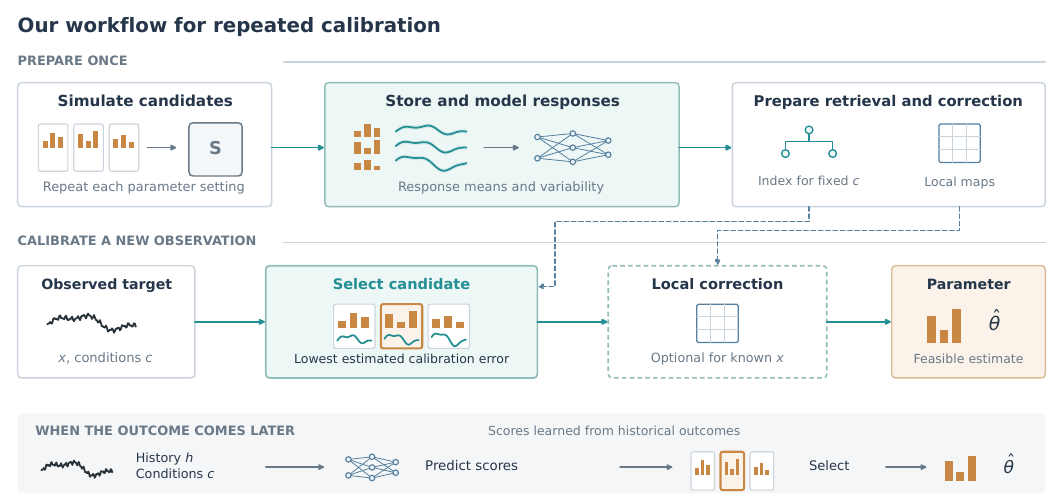}
\caption{\roundedit{Preparation and repeated use of the proposed optimizer. Repeated simulations and a fitted response model supply candidate means and variability. For fixed conditions, their combined moments form the index in Fig.~\ref{fig:comparison}f, while local simulations prepare correction maps. The observed-target route selects a candidate and optionally refines it. When the outcome follows calibration, historical outcomes train a mean-response proxy score predictor whose query inputs are history and conditions. Solid arrows show operations; dashed arrows show reused quantities. Prepared quantities are reused until their conditions, coverage or fitted models require updating.}}\label{fig:method}
\end{figure*}

Figure~\ref{fig:method} separates preparation from calibration. During preparation, we retain feasible candidates and repeated responses in simulation memory (Section~\ref{sec:candidatechoice}), fit the response surrogate (Section~\ref{sec:modeltraining}), and prepare candidate selection and local correction (Sections~\ref{sec:responselearning} and~\ref{sec:proposals}). \roundedit{For an observed target, stored responses and predicted moments estimate expected squared response error.} \roundedit{The financial historical predictor learns a squared-mean-response proxy ordering (Section~\ref{sec:realsupervision}).} \roundedit{Its returned parameters are evaluated separately with the energy score.} Both routes assign each candidate a score $q_j(o)$, with larger values indicating better estimated response agreement. Selection then returns

\begin{equation}\label{eq:selection}
a(o)=\arg\max_{1\leq j\leq M}q_j(o),\qquad
\widehat\theta=\theta_{a(o)}.
\end{equation}

An optional bounded correction can replace the selected candidate with a nearby feasible parameter. Algorithms~\ref{alg:prepare} and~\ref{alg:query} summarize preparation and the complete query. Preparation and independent evaluation call the simulator. The query itself uses only fitted models and stored quantities.

\begin{algorithm}[t]
\caption{\rev{Preparing the amortized optimizer}}\label{alg:prepare}
\begin{algorithmic}[1]
\Require \rev{Grouped simulations $\mathcal M$, historical outcomes when available, development observations $\mathcal D_{\rm dev}$}
\Ensure \rev{Candidates, input transforms, score representation, optional local maps}
\State \rev{Store candidates and fit normalization on training groups (Section~\ref{sec:candidatechoice})}
\State \rev{Fit the amortized surrogate using $L_F$ or $L_N$ (Section~\ref{sec:modeltraining})}
\State \roundedit{Construct mixed-moment expected-error scores or historical mean-response proxy scores (Sections~\ref{sec:responselearning} and~\ref{sec:realsupervision})}
\State \rev{Fix scoring settings on development data and store the moment index or score predictor and decoder}
\If{local correction is configured}
  \State \rev{Prepare $A_a,b_a,\mu_a,r_a$ at each anchor and select $\eta$ on $\mathcal D_{\rm dev}$ (Section~\ref{sec:proposals})}
\EndIf
\State \rev{Store the physical parameter transform}
\end{algorithmic}
\end{algorithm}

\begin{algorithm}[t]
\caption{\rev{Calibrating a new observation}}\label{alg:query}
\begin{algorithmic}[1]
\Require \rev{Prepared quantities and available observations $o$}
\Ensure Physical parameter $\widehat\theta\in\Theta$
\State Apply the stored input transforms to $o$
\State Select $a=\arg\max_j q_j(o)$ by indexed retrieval or prediction and decoding (Section~\ref{sec:responselearning})
\State Set $\widehat u\gets u_a$
\If{target $x$ and a configured local map are available}
  \State $\delta\gets A_a(x-\mu_a)-b_a$
  \State \rev{Apply the radius cap and projection in Eq.~\eqref{eq:cap}}
\EndIf
\State \Return the physical transform of $\widehat u$
\end{algorithmic}
\end{algorithm}

\subsection{Simulation memory}\label{sec:candidatechoice}

Preparation begins with a memory of $M$ physical candidates $\theta_{j}$, their unit-box coordinates $u_{j} \in \lbrack 0,1\rbrack^{p}$, conditions $c_{j}$ and repeated responses $z_{jr}$. Retaining candidates keeps distinct parameter settings available even when their responses are similar. Their placement determines which regions the optimizer can select. We therefore assess candidate coverage separately from scoring and local correction.

Within each parameter/condition group, repeated responses provide estimates of the mean and variance. Training groups determine feature centres and scales, which are then reused for development and evaluation. The traffic and predator--prey experiments separate these groups before fitting. With this normalization fixed, the response moments provide the empirical estimates used in candidate comparison.

\subsection{Amortized surrogate}\label{sec:modeltraining}

The surrogate complements these empirical estimates by sharing information across candidate responses. For complete vector outputs, a network maps normalized parameters to means $\mu_\phi$ and log variances $g_\phi$. Repeated simulations provide the sample mean $\overline z$ and unbiased sample variance $\widehat v$. We form the log-variance target $\widehat g=\log\max(\widehat v,\epsilon_v)$ and train the two output heads to minimize

\begin{equation}\label{eq:momenttraining}
L_F=\operatorname{MSE}(\mu_\phi,\overline z)
+\lambda_{\rm var}\operatorname{MSE}(g_\phi,\widehat g).
\end{equation}

The variance floor $\epsilon_{v}$ keeps the logarithmic targets finite. Predicted log variances are clipped before exponentiation. When conditions change, a multi-output model instead predicts candidate means from the current condition. These predictions support both direct comparison with a known target and supervision from historical outcomes.

\paragraph{Incomplete responses}\label{sec:validitymethod}

The neural-circuit simulator can omit rhythm features. Its value--validity network therefore predicts a mean $\mu_{k}$, log variance $g_{k}$ and validity logit $\gamma_{k}$ for each of the $d$ features. Given standardized value $v_{k}$ and availability indicator $a_{k} \in \{ 0,1\}$, we fit numeric values only where they are present, using

\begin{equation}\label{eq:neuralnumeric}
L_{\mathrm{num}}=\frac1d\sum_{k=1}^{d}
\frac{a_k}{2}\left[g_k+(v_k-\mu_k)^2e^{-g_k}\right].
\end{equation}

The availability indicators are fitted for every feature using binary cross-entropy (BCE),

\begin{equation}\label{eq:neuralvalidity}
L_{\mathrm{valid}}=\frac1d\sum_{k=1}^{d}
\left[\log(1+e^{\gamma_k})-a_k\gamma_k\right],
\end{equation}

and the two losses receive equal weight in the complete training objective,

\begin{equation}\label{eq:neuraltraining}
L_N=L_{\mathrm{num}}+L_{\mathrm{valid}}.
\end{equation}

The resulting probabilities $p_{k} = (1 + e^{- \gamma_{k}})^{- 1}$ and conditional numeric moments predict the response encoding $(a \odot v,5a)$. The second block assigns a discrepancy to missing features, so candidate comparison accounts for availability as well as numeric values. Although the encoding has $2d$ coordinates, its discrepancy is normalized by the original feature count $d$. Supplement~K gives the corresponding expected error, and Section~\ref{sec:selectionprotocol} specifies fitting settings for both surrogate types.

\subsection{Candidate scoring and selection}\label{sec:responselearning}

The stored responses and surrogate predictions now provide two estimates of candidate error. For an available target $x$, we use squared error in standardized response coordinates, $D(z,x)=\norm{z-x}^2/\nu$, with $\nu=d$ in the reported tasks. A candidate can be scored from its $n_j$ stored repeats or from predicted moments $(\mu_j,\sigma_j^2)$ as

\begin{align}
\widehat r_j^{\,\mathrm{mem}}(x)
 &=\frac{1}{\nu n_j}\sum_{r=1}^{n_j}\norm{z_{jr}-x}^2,\label{eq:empirical}\\
\widehat r_j^{\,\mathrm{sur}}(x)
 &=\frac{\norm{\mu_j(c)-x}^2+\sum_k\sigma^2_{jk}(c)}{\nu}.\label{eq:forward}
\end{align}

The surrogate estimate follows from the mean--variance decomposition of squared error. We combine the two estimates, giving the empirical one more weight as the number of repeats increases,

\begin{equation}\label{eq:mix}
\widehat r_j^{\,\mathrm{mix}}=(1-w_j)\widehat r_j^{\,\mathrm{mem}}
+w_j\widehat r_j^{\,\mathrm{sur}},\qquad w_j=\frac{\tau}{n_j+\tau}.
\end{equation}

Development response error selects the effective repeat count $\tau$ from $\left\{ 0,4,16,64,\infty \right\}$. The endpoints recover empirical-only and surrogate-only selection. Negating the mixed error, $q_{j}(o) = - {\widehat{r}}_{j}^{\rm mix}(x)$, gives the score used in Eq.~\eqref{eq:selection}.

Under fixed conditions, the same comparison can be performed through exact nearest-neighbour retrieval. We mix the first moments $m_j$ and scalar second moments $s_j$ with the same weights. Removing the common $\norm{x}^2$ term leaves $s_j-2m_j^\top x$. With $v_j=s_j-\norm{m_j}^2$ and $b_0=\min(0,\min_jv_j)$, the squared distance from $[x,0]$ to $[m_j,\sqrt{v_j-b_0}]$ is $\nu\widehat r_j^{\,\mathrm{mix}}(x)-b_0$. Nearest-neighbour retrieval therefore minimizes the same estimated calibration error. We store this representation in a median-split spatial tree with axis-aligned bounds and 16 candidates per leaf. Branch-and-bound prunes comparisons according to the score geometry, with linear worst-case work in $M$.

For scores learned from historical records, selection instead evaluates the predictor and fixed decoder described in Section~\ref{sec:realsupervision}. It then takes the largest of the $M$ scores. This route implements the same candidate-selection step when the target is not yet observed.

\subsection{Optional local correction}\label{sec:proposals}

Selection returns one of the stored candidates. When the target is known, a prepared local inverse map can move that candidate toward it without a new optimization run. We construct the map in coordinates $v = \kappa(u - 0.5)$ using central parameter perturbations of size 0.01. Random streams are shared across the centre and coordinate directions, and clipped bounds use the actual perturbation separation. Each of $R_{\rm loc} = 4$ streams gives a centre response $z_{r}$ and a transposed finite-difference matrix $B_{r} \in {\mathbb{R}}^{p \times d}$. Their moments are

\begin{align}\label{eq:localupdate}
\overline B&=R_{\rm loc}^{-1}\sum_rB_r,\qquad
\overline z=R_{\rm loc}^{-1}\sum_rz_r,\\
C_B&=R_{\rm loc}^{-1}\sum_r(B_r-\overline B)(B_r-\overline B)^\top,\nonumber\\
c_B&=R_{\rm loc}^{-1}\sum_r(B_r-\overline B)(z_r-\overline z).\nonumber
\end{align}

Let $\mu_{a}$ denote the mixed response mean at the selected anchor. The local approximation $\widetilde z_r(\delta)=\mu_a+(z_r-\overline z)+B_r^\top\delta$ retains variation across streams. We choose the displacement by minimizing its average target error with a quadratic penalty,

\begin{equation}\label{eq:localsolve}
\frac{1}{R_{\rm loc}}\sum_r\norm{x-\widetilde z_r(\delta)}^2
+\lambda_J\norm{\delta}^2.
\end{equation}

The normal equations can be solved during preparation, giving

\begin{align}\label{eq:primarymap}
H_a&=\overline B\overline B^\top+C_B+\lambda_JI,\\
A_a&=H_a^{-1}\overline B,\qquad b_a=H_a^{-1}c_B,\nonumber\\
\delta(x)&=A_a(x-\mu_a)-b_a.\nonumber
\end{align}

We use $\lambda_J=\max(10^{-12},10^{-3}\operatorname{tr}(\overline B\overline B^\top)/p)$ to regularize weak response directions. Once these quantities are stored, a new target requires only evaluation of the affine map.

To limit the correction, we cap its displacement relative to the distance $r_{a}$ to the 64th nearest anchor in local coordinates. Projection into the feasible box then gives

\begin{equation}\label{eq:cap}
\widehat u=\Pi_{[0,1]^p}\!\left[u_a+\frac{\delta}{\kappa}
\min\left\{1,\frac{\eta r_a}{\norm{\delta}}\right\}\right].
\end{equation}

Zero displacement remains zero, and the physical transform maps $\widehat{u}$ into $\Theta$. The coordinate scale $\kappa$ is $\sqrt{12}$ for traffic and 10 for predator--prey. Development data select $\eta \in \{ 0,0.1,0.25,0.5\}$, with zero retaining the selected candidate without correction. Supplement~A.3 describes the additional correction variants.

\subsection{Supervision from historical outcomes}\label{sec:realsupervision}

When the target occurs after calibration, it cannot be used directly to score candidates at query time. Historical triples $\left( h_{i},c_{i},Y_{i} \right)$ provide the training information by pairing available observations with later outcomes. The response surrogate predicts each candidate mean $\mu_{j}(c_{i})$. Using common training-fitted baselines and scales, we compare its squared error with that of a fixed reference $j_{0}$,

\begin{equation}\label{eq:costdiff}
\Delta_{ij}=\frac{\norm{\mu_j(c_i)-Y_i}^2-
\norm{\mu_{j_0}(c_i)-Y_i}^2}{\nu},\qquad q_{ij}=-\Delta_{ij}.
\end{equation}

Subtracting the reference cancels the common $\left\| Y_{i} \right\|^{2}$ term without changing candidate ordering. We normalize the score columns using weighted training means $\overline{q}$ and standard deviations $d_{q}$. A weighted singular-value decomposition supplies a basis $V_{k} \in {\mathbb{R}}^{M \times k}$, whose coordinates are learned by a regressor $f(h,c)$. At query time, the decoder reconstructs candidate scores as

\begin{equation}\label{eq:scoredecode}
\widehat q=\overline q+d_q\odot[V_kf(h,c)],\qquad q_j(o)=\widehat q_j.
\end{equation}

Equation~\eqref{eq:selection} uses these scores to return a candidate. The subsequent outcome $Y$ supplies training targets, but the query requires only the preceding history and known conditions.

To relate score prediction to outcome prediction, assume finite second moments and let $m_Y=\E[Y\mid h,c]$ and $v_Y=\operatorname{tr}\operatorname{Cov}(Y\mid h,c)$. Then

\begin{equation}\label{eq:proxymean}
\E[\norm{\mu_j(c)-Y}^2\mid h,c]
=\norm{\mu_j(c)-m_Y}^2+v_Y.
\end{equation}

Because $v_{Y}$ is common to all candidates, predicting scores and predicting $m_{Y}$ before comparing candidate means target the same population ordering. The latter procedure forms the conditional surrogate control used in the experiments.

Financial evaluation uses the energy score to assess expected calibration error,

\begin{equation}\label{eq:predictiverisk}
\begin{aligned}
\R_{\rm pred}(\theta\mid h,c)
=\E\bigl[2\norm{Z_\theta-Y}-\norm{Z_\theta-Z'_\theta}\mid h,c\bigr].
\end{aligned}
\end{equation}

This evaluation criterion differs from the squared mean-response criterion approximated by the learned scores. It also accounts for dispersion and can therefore distinguish candidates with identical means.

\section{Experimental settings}\label{sec:setup}

The four tasks span different response dimensions, parameter spaces and observation sources (Table~\ref{tab:protocol}). All methods return feasible parameters, evaluated with independent simulator responses for calibration error and complete query time.

\roundedit{The four applications were selected during exploration; an earlier susceptible--infectious--recovered task was excluded after empirical memories matched or exceeded the proposed rule (Supplement~B.3).}

\begin{table*}[t]
\centering\small
\caption{\rev{Four-task calibration protocol.} \rev{The first three tasks use independently simulated observations.} \rev{Finance uses real market records.} \rev{Feature dimensions and losses are task-specific.}}\label{tab:protocol}
\begin{tabularx}{\textwidth}{@{}L{24mm}L{40mm}L{40mm}Y@{}}
\toprule
\rev{Task} & \rev{Simulator and parameters} & \rev{Observation / available input} & \rev{Training and evaluation}\\\midrule
\rev{Predator--prey} & \rev{Lotka--Volterra growth, predation, death and reproduction rates (4 parameters)} & \rev{18 noisy population features} & \roundedit{2 response banks; 512 targets in each of A/B for development and components; 512 new set-C targets after fixing the original rules}\\\midrule
\rev{Traffic} & \rev{Intelligent-driver desired speed, headway and spacing (3 parameters)} & \rev{36 gap/speed features from 24 vehicles} & \roundedit{2 original banks, 512 targets each in A/B for development and components; 2 new 64-target cohorts with fixed rules and one small bank}\\\midrule
\rev{Neural circuit} & \rev{Pyloric membrane and synaptic conductances (31 parameters)} & \rev{18 rhythm features and availability indicators} & \rev{Enriched simulation pool, 177 complete-rhythm targets, separate 154-target budget study}\\\midrule
\rev{Financial} & \rev{Order-driven market quote scales, buy probabilities, arrival/cancellation rates and intensity (7 parameters)} & \rev{Past 5-minute summaries and current book (48 inputs), 8 subsequent-response features} & \rev{1,524 historical training windows, 475 evaluation windows, 75-window search subset}\\
\bottomrule
\end{tabularx}
\end{table*}

\subsection{Simulated observations}

Predator--prey calibration estimates four Lotka--Volterra rates from 18 noisy population features~\cite{lueckmann2021benchmarking}. Traffic calibration estimates three intelligent-driver parameters from 36 gap and speed summaries~\cite{treiber2000congested}. Supplement~M.1 gives the dynamics, physical ranges and observation construction.

For both tasks, the main comparison uses two independent response banks and 512 set-A targets, with 64 evaluation responses per returned parameter. Each traffic bank has 10,035 uniformly sampled anchors. Sets A and B supported component analysis and selection of the primary local correction. To assess the chosen configuration, predator--prey set C was generated after the correction and wider candidate allocation were fixed. Supplements~B.1 and~B.8 give parameter ranges, bank designs and cohort roles.

The pyloric neural-circuit model extends the evaluation to 24 membrane and seven synaptic conductances~\cite{goncalves2020}. Calibration uses 18 rhythm features and indicators of whether each feature is available. The main comparison evaluates 177 complete-rhythm targets generated after fixing the original models and scoring settings. Each returned parameter is evaluated with 16 new simulated responses. Earlier budget studies use a separate 154-target cohort, while an eight-fit follow-up examines initialization variability. We keep these cohorts distinct in the analyses.

\subsection{Real financial observations}\label{sec:financialdata}

The order-driven financial simulator~\cite{belcak2022fast} has seven calibrated parameters. Eight summaries of the preceding five minutes and forty current-book coordinates determine parameters for the next 300 s. Evaluation compares price changes, variability, activity, spread, depth and imbalance. Supplements~B.7 and~M.1 give the simulator and feature definitions.

Training combines 1,159 SZ000001 windows from 2019 and 365 windows from AAPL, AMZN, GOOG, INTC and MSFT on 8 December 2025. The two sources receive equal total weight. Nasdaq TotalView-ITCH records provide the target-market books. Evaluation uses 19 non-overlapping windows per stock/date on 9--12 December 2025 and 12 June 2026, giving 475 windows. Within this cohort, three fixed starts per stock/date define the 75-window search subset.

The candidate dictionary contains 96 inherited candidates and 20 additions, one per target-training context. To obtain each addition, CMA-ES searches against the training outcome using squared mean-response error. We retain the first distinct proposal in score order with valid responses on all 84 training contexts. Each query then selects one of the resulting 116 parameter vectors. Evaluation dates were excluded from model fitting but informed dictionary and readout development, so the financial comparison is retrospective. The dictionary, rank-nine score representation and selection settings remain fixed throughout the reported analyses (Supplements~B.8 and~I).

\subsection{Model fitting and selection}\label{sec:selectionprotocol}

Traffic and predator--prey use two-layer SiLU surrogates~\cite{elfwing2017silu} fitted with AdamW~\cite{loshchilov2019adamw}. The original retrieval surrogate is fitted on 10,035 four-response anchor groups per bank, with 1,843 separate groups for inherited feature scales and 205 held-out groups for development. Retrieval weights and local radii were fixed before sets A and B. The financial forests use 100 trees, with the rank-nine score representation retained during retrospective development. Supplement~M.2 gives the architectures, optimization schedules and the distinction between fixed and selected settings.

\subsection{Comparison methods and controls}\label{sec:comparators}

\paragraph{Iterative optimization}

PSO, CMA-ES, GP-BO and TuRBO-1 have a 10-minute limit and four responses per candidate, with 512 evaluations in finance and 128 in other tasks. Searches start at the unit-box centre, counting that evaluation toward the budget.

Financial searches minimize standardized squared error against the same future-response prediction from 48 inputs; subsequent outcomes are reserved for evaluation. Incumbent histories give returns at one-minute checkpoints. Additional financial comparators are budget-adapted NL-SHADE-LBC~\cite{stanovov2022nlshade}, SNO-PSO and SNO-DE~\cite{tsai2026sno}. Supplements~C.2 and~M.4 specify the search streams and budgets.

\paragraph{Posterior methods}

The posterior comparison includes neural posterior estimation with MDN~\cite{papamakarios2016fast}, MAF~\cite{papamakarios2017masked} and NSF~\cite{durkan2019neural}. The remaining four methods are FMPE~\cite{wildberger2023flow}, NPSE~\cite{geffner2023compositional}, CMPE~\cite{schmitt2024consistency} and conditional sCM~\cite{lu2025consistency}. To obtain a parameter for calibration, each method returns the arithmetic mean of 32 actual samples in physical coordinates. Query timing therefore includes sampling, transformation and averaging. Supplementary neural comparisons also examine MAP and response-ranked readouts.

For finance, posterior fitting conditions on the current book and simulated response residuals. At query time, the fixed history model predicts the unknown future residuals. These models use 67,392 simulated responses over 16,848 parameter/condition pairs. Our response surrogate uses 38,976 responses over 9,744 pairs from its dictionary. Thus, the workflows share query information and evaluation scales but use distinct simulation designs, as documented in Supplement~G.

\paragraph{Fixed comparator settings}

GP-BO uses noisy log expected improvement~\cite{ament2023logei}. Comparator configurations were fixed without selection on evaluation error. Posterior fitting uses a one-hour cap and internal validation or predetermined final checkpoints. Supplement~M.3 reports the search settings, fitting schedules and sampling steps.

\paragraph{Prepared controls}

Prepared controls compare empirical and predicted responses, wider coverage, local refinement and feature availability. Financial controls share the observations and response surrogate, or reuse global and stock-specific parameters selected on training data (Supplement~M.4).

\subsection{Calibration error, timing and uncertainty}\label{sec:metrics}

For target $i$, independent simulator responses estimate the calibration error $\ell_i$; tables report $N^{-1}\sum_i\ell_i$. Lower error means better response agreement, distinct from the surrogate-fitting objectives $L_F$ and $L_N$. Traffic and predator--prey use 64 responses and training-scaled squared errors over 36 and 18 features, respectively. Neural evaluation uses 16 responses and the $(a\odot v,5a)$ encoding, divided by $d=18$, with missing-feature penalty $25/d$. Numeric error and response completeness are also reported.

Finance uses 16 responses transformed with the same training scales and current-book baseline as the observed outcome. Its error is
\begin{equation}\label{eq:energy}
\ell_i=\widehat{\mathrm{ES}}_i=2D_{{\rm obs},i}-D_{{\rm sim},i},
\end{equation}
where $D_{{\rm obs},i}$ is the mean Euclidean simulator--observation distance and $D_{{\rm sim},i}$ the mean Euclidean distance between distinct simulator draws. This is twice the conventional energy score, a proper multivariate scoring rule~\cite{gneiting2007strictly}. Supplement~M.5 gives the finite-sample formulas.

Query timing uses one AMD EPYC 9755 core at batch size one. It includes normalization, inference or selection, physical conversion and financial record summarization after warm-up; preparation, loading and disk access precede timing. Tables report mean ± sample SD across complete-cohort run means. Prepared timings first use each target's median. Paired financial intervals use 20,000 crossed date/stock bootstrap draws after averaging fitted rules within windows, preserving stock/date dependence in this five-stock, five-date cohort. Supplement~M.6 specifies repetitions, fit counts and aggregation.

To compare time at matched calibration quality, we record the earliest saved search return reaching the paired direct result,

\begin{equation}\label{eq:matchtime}
T_{\rm match}=\min\{t\in\mathcal T\mid\ell_i(\theta_t^{\rm search})\leq\ell_i(\widehat\theta^{\rm direct})\},
\end{equation}

Both returns are evaluated independently at recorded incumbent times $\mathcal T$. Matching fractions include non-matches; median times and ratios use successful matches. Cohort comparisons use the first one-minute checkpoint below the direct mean (Supplement~M.7).

\subsection{Repeated-use cost and feature diagnostics}\label{sec:diagnosticsprotocol}

For $Q$ queries, the cost of repeated calibration includes preparation as well as query time,

\begin{equation}\label{eq:totalcost}
\roundedit{C_m(Q)=P_m+Qt_m.}
\end{equation}

\roundedit{Here, $P_m$ includes simulation, normalization, model fitting, construction of maps and indices, serialization, and one process start and load for the query.}
\roundedit{The small traffic experiment records each operation and reports wall time and processor time separately.}
\roundedit{An additional calculation starts with the same response groups already in memory and removes only their generation time.}

For selectors sharing a response surrogate, let $T_{\rm add}$ denote the extra time to fit and save the direct score model. Its query savings offset that additional preparation after

\begin{equation}\label{eq:headbreakeven}
Q^*=\frac{T_{\rm add}}{t_{\rm selector}-t_{\rm direct}}.
\end{equation}

This is an incremental comparison that assigns zero preparation time to the selector's future-response model. It excludes shared simulation and surrogate fitting, compilation, loading and refresh.

Finally, we examine whether aggregate financial error hides differences between response features. These diagnostics use all recorded returns on 475 windows across three banks and five fits. Replacing the vector norm in Eq.~(A26) with scalar absolute distance gives a marginal energy score for each feature. The paired date/stock bootstrap follows the same within-window averaging, with no adjustment for multiple feature comparisons.

\section{Experimental results}\label{sec:results}

\subsection{Calibration quality across four tasks}\label{sec:neuralnpe}

\roundedit{Supplementary Table~A42 describes the four-task benchmark, with development targets for predator--prey and traffic, independent neural targets and retrospective financial windows.} \roundedit{Across these cohorts, the amortized optimizer returns parameters in 0.00995--0.0689 ms.} Its mean calibration error is lower than that of all seven posterior methods with the specified physical-mean readout. Among the four iterative optimizers and our method, it also gives the lowest mean error on predator--prey, traffic and neural-circuit calibration. For these three tasks, respectively, error falls from 0.5361 to 0.0426, from 0.9563 to 0.9518, and from 3.1006 to 0.8907 relative to the best listed iterative optimizer.

\roundedit{The retrospective financial comparison shows a different trade-off.} \roundedit{Direct scoring with the full response teacher takes 0.0562 ms and attains error 35.6001 ± 0.1335 on its 75-window search subset.} GP-BO attains 34.5429 ± 0.0556 after 10 minutes of search. Direct calibration therefore has 3.06\% higher mean error, with a 95\% date/stock bootstrap interval of 0.50--6.27\%. Among the posterior-mean workflows, FMPE gives the lowest financial error, 36.9147. Direct calibration has 3.56\% lower mean error than this workflow.

These posterior comparisons depend on how samples are converted into a parameter for the simulator. \roundedit{For the original box-coordinate neural MAF, selecting a sample by response error lowers mean error from 6.2871 to 4.5964 (Supplementary Table~A29).} \roundedit{This comparison keeps that model fixed; Supplementary Table~A42 uses a separate bounded-coordinate fit.} Our optimizer prepares the information needed for this comparison in advance, so it can use response-based selection whenever new observations require a parameter update.

\subsection{Search time and matched quality}\label{sec:attainmentresults}

\begin{figure*}[t]
\centering\includegraphics[width=0.8\textwidth]{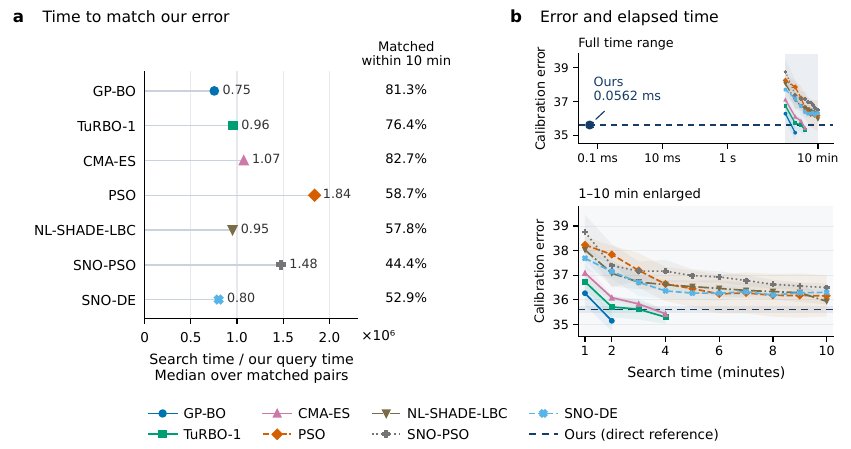}
\caption{\roundedit{Search and query times in the retrospective financial comparison on 75 windows.} (a) Dots show each optimizer's time to match our calibration error divided by our query time, summarized by the median over matched window/run pairs. Percentages show the fraction matched within 10 minutes out of 225 pairs (75 windows across three runs). (b) The upper view places our measured query time of 0.0562 ms and the search checkpoints on the same logarithmic time axis. The lower view enlarges the shaded 1--10 minute interval. Each search curve ends at its first checkpoint below our mean error, or at 10 minutes if no crossing occurs. The dashed line marks our mean error of 35.6001. Error bars and shaded bands show SD across three complete-cohort runs.}\label{fig:attainment}
\end{figure*}

We next compare the search time needed to match individual direct returns and to improve on their mean calibration error (Fig.~\ref{fig:attainment}). GP-BO matches the paired direct result in 183 of 225 window/run pairs, with a median first-match time of 42,930 ms among successful pairs. TuRBO-1 matches in 172 pairs, with a median of 54,120 ms. Their median paired search/direct time ratios are $7.54 \times 10^{5}$ and $9.58 \times 10^{5}$. Across the seven optimizers, 44.4--82.7\% of pairs match within 10 minutes. Each fraction in Fig.~\ref{fig:attainment}a retains all 225 pairs in its denominator.

At the cohort level, GP-BO first attains a lower mean error than the direct method at the 2-minute checkpoint, reaching 35.1430. TuRBO-1 and CMA-ES first cross the direct mean at 4 minutes, with errors of 35.2925 and 35.4250 (Fig.~\ref{fig:attainment}b). The other four optimizers do not reach a lower cohort mean within 10 minutes.

\roundedit{Attainment times are determined retrospectively from independently evaluated returns.} \roundedit{The evaluation outcome does not enter the search trajectory, and its error need not decrease at every saved return.} Search chooses candidates through a finite-response proxy, whereas evaluation uses fresh simulator draws and the subsequent market outcome. The gap in computation time matters when parameters must be updated before the next observation arrives. Reusing simulation work permits calibration at update frequencies that leave too little time for repeated search.

\subsection{Prepared selection and learned components}\label{sec:coverage}

\roundedit{The prepared controls compare how simulation effort is allocated and how the resulting responses are used to return parameters (Table~\ref{tab:preparedcontrols}).}
\roundedit{On the independently generated predator--prey set C, local correction lowers the error of a concentrated memory from 0.07547 to 0.04894, while wider empirical coverage gives 0.03905.}
\roundedit{A further comparison matches the primary rule's minimum budget of 401,400 responses and gives errors of 0.04192 for wider empirical selection and 0.04894 for local correction.}
\roundedit{\mbox{The latter is} 16.75\% higher, with a paired 95\% interval of 2.72--30.32\%.}
\roundedit{Thus, the local map improves returns from an existing concentrated memory, whereas allocating simulations to broader coverage is more effective in this task.}

\begin{table}[t]
\centering
\ifdefined\RoundBlue\color{roundblue}\fi
\caption{\roundedit{Matched-cohort comparisons of prepared methods. Values are mean $\pm$ sample SD across recorded bank, fit or profiling means. $\dagger$ denotes one shared empirical rule with no between-fit loss SD; its timing varies across profiling runs. A dash denotes unrecorded timing.}}
\label{tab:preparedcontrols}
\setlength{\tabcolsep}{3.5pt}
\footnotesize
\begin{tabular}{@{}lrr@{}}
\toprule
Prepared method & Calibration error & Time (ms)\\
\midrule
\multicolumn{3}{@{}l}{\textit{Predator--prey, set C, 401,400 responses}}\\
Wider empirical memory & $\mathbf{0.04192\pm0.00045}$ & $\mathbf{0.01350}$\\
Local correction & $0.04894\pm0.00789$ & $0.01397$\\
\addlinespace
\multicolumn{3}{@{}l}{\textit{Traffic, set A}}\\
Empirical selection & $0.9768\pm0.0001$ & $0.716\pm0.024$\\
Surrogate selection & $\mathbf{0.9515\pm0.0007}$ & $0.1377\pm0.0037$\\
Local correction & $0.9518\pm0.0019$ & $\mathbf{0.0689\pm0.0071}$\\
\addlinespace
\multicolumn{3}{@{}l}{\textit{Neural circuit, 177 targets, two fits}}\\
Empirical selection & $1.0497^{\dagger}$ & $0.01589\pm0.00021$\\
Validity-only mixture & $\mathbf{0.8714\pm0.0489}$ & ---\\
Full mixture & $0.8907\pm0.0569$ & $\mathbf{0.00995\pm0.00085}$\\
\addlinespace
\multicolumn{3}{@{}l}{\textit{Finance, 475 windows, shared compressed teacher}}\\
Conditional selection & $\mathbf{34.0526\pm0.0550}$ & $0.06041\pm0.00038$\\
Direct scores & $34.0602\pm0.0580$ & $\mathbf{0.05543\pm0.00046}$\\
\bottomrule
\end{tabular}
\end{table}

\roundedit{Traffic shows a different trade-off.}
\roundedit{On set A, local correction takes 0.0689 ms compared with 0.1377 ms for surrogate selection, with mean errors of 0.95184 and 0.95154.}
\roundedit{The query is approximately twice as fast while the mean error changes by 0.032\%.}
\roundedit{The prepared local maps allow a smaller anchor set to return continuous parameters, so the observed speed reflects retrieval and correction together.}

\roundedit{The joint-response mean-field controls isolate the covariance terms within the earlier multi-candidate variant (Supplementary Table~A19).}
\roundedit{Relative to these controls, the full variant has 0.40\% higher error on predator--prey, with a paired 95\% interval of $[-0.44\%,1.57\%]$, and 0.02\% lower error on traffic, with an interval of $[-0.04\%,0.01\%]$ for the signed difference.}
\roundedit{Both intervals include zero, and these exploratory controls leave the covariance terms' separate benefit in the primary single-return rule undetermined.}
\roundedit{The prepared retrieval and continuous correction are therefore evaluated as a combined procedure.}

\roundedit{The neural controls identify the contribution of estimating whether simulated features will be available.}
\roundedit{On 177 targets generated after fixing the original models and scoring settings, the full mixture lowers mean error from 1.0497 to 0.8907, a 15.15\% reduction with a paired 95\% interval of 3.18--37.48\%.}
\roundedit{Its mean and 90th-percentile errors fall while its median error rises.}
\roundedit{In an eight-fit follow-up on this cohort, validity-only mixing and full mixing give errors of 0.87946 and 0.87958, respectively, and increase complete responses from 97.07\% to 98.92\% and 99.04\%.}
\roundedit{Most of this gain is retained when only the availability estimates are learned.}
\roundedit{Full mixing further reduces missing-feature penalties but raises numeric error, making the required response completeness relevant to the choice of scoring rule.}

\label{sec:neuralcomponents}

\subsection{Confirmation on new traffic targets}\label{sec:trafficconfirmation}

\roundedit{A separate small-budget experiment fixes all method settings and fitted models before generating 64 new traffic targets.}
\roundedit{The three methods share 16,384 bank responses, with simple controls allowed to select all 3,584 anchor and perturbation locations while the local rule uses 512 anchors.}
\roundedit{This confirmation uses the same exact cKDTree index for all three rules, replacing the original benchmark index.}
\roundedit{Its paired target intervals, and those of the following inverse comparison, condition on the shared bank and fixed fitted rules.}
\roundedit{Local correction gives mean error 0.96926 compared with 0.99096 for empirical selection, a 2.19\% reduction with a paired 95\% interval of 0.68--3.57\%.}
\roundedit{Relative to surrogate selection, the estimated reduction is 1.35\%, with an interval of $[-0.06\%,2.64\%]$.}
\roundedit{Queries take 0.02631 ms, approximately 2.35 and 2.14 times faster than the empirical and surrogate controls, respectively.}
\roundedit{These fixed-bank comparisons support local refinement with limited preparation, with protocols and all paired results in Supplement~L.2.}

\roundedit{A further comparison trains a tandem-style inverse~\cite{liu2018tandem} through the frozen expanded response surrogate, using expected squared response error without parameter-label supervision of the inverse.}
\roundedit{Two predetermined fits and the three existing rules are evaluated on a second set of 64 targets generated after all five rules are fixed.}
\roundedit{Table~\ref{tab:closesttraffic} places the five rules on the same accuracy, query-time and preparation-cost comparison.}
\roundedit{The local rule's estimated reduction relative to the two-fit mean error is 0.55\%, with a paired 95\% interval of $[-0.30\%,1.43\%]$.}
\roundedit{The inverse queries take approximately 0.01832 ms each, compared with 0.02676 ms for local correction.}
\roundedit{Thus, direct inverse prediction gives the shortest queries at similar mean error in this comparison, with additional fitting whose cost is examined in Section~\ref{sec:offlinecost}.}
\roundedit{Supplement~L.3 reports both fits, all paired comparisons and the common response budget.}

\begin{table}[t]
\centering
\ifdefined\RoundBlue\color{roundblue}\fi
\caption{\roundedit{Prepared workflows on the second independent traffic cohort of 64 targets, using one shared bank. $P$ sums required preparation and one fresh process start with loading. Query times include complete parameter returns. Both inverse fits are retained.}}
\label{tab:closesttraffic}
\setlength{\tabcolsep}{3pt}
\footnotesize
\begin{tabular}{@{}lrrr@{}}
\toprule Method & Mean error & Query (ms) & $P$ (s)\\\midrule
Local correction & 0.99295 & 0.02676 & 15.065\\
Expanded empirical & 1.02282 & 0.06346 & 11.935\\
Expanded surrogate & 0.99944 & 0.05938 & 14.089\\
Tandem inverse, seed 7 & 0.99752 & 0.01832 & 22.374\\
Tandem inverse, seed 19 & 0.99929 & 0.01831 & 23.192\\
\bottomrule
\end{tabular}
\end{table}

\subsection{Learning from historical observations}\label{sec:realtraining}

For calibration from historical observations, we first compare query-specific estimates with fixed parameters. In the single-run 75-window control, error is 39.8001 for one training-selected global parameter, 36.9663 for stock-specific parameters and 35.4728 for the direct selector. A random prior draw gives 45.4885. The direct result is 10.9\% and 4.0\% lower than the global and stock-specific errors, respectively. These gains support adapting parameter estimates to the observations available for each query.

The benefit of adding training windows is less clear. Increasing target-day training from 95 to 365 windows gives an estimated 0.56\% reduction in mean error, with a bootstrap interval spanning improvement and degradation. Repeating the original windows or using 185 non-overlapping windows also gives no consistent gain. These controls therefore support observation-dependent selection more clearly than they support increasing the number of training windows.

To isolate the role of direct scoring, we compare it with selecting candidates through a predicted future response. Across 475 financial windows, direct scoring and conditional surrogate selection give mean errors of 34.0602 and 34.0526 when they share the same compressed response teacher. The matched direct model takes 0.05543 ms per query, compared with 0.06041 ms for conditional surrogate selection, an 8.25\% reduction. \roundedit{For the squared-mean-response proxy, both target the same population ordering in Eq.~\eqref{eq:proxymean}.} Their close errors and different query times identify the practical role of predicting candidate scores directly, which reduces the computation repeated for each observation.

\subsection{Response features and changing dates}\label{sec:behaviour}

The similar aggregate errors of these selectors can conceal differences in the behaviours they reproduce (Supplementary Fig.~A1a). With the same response surrogate, direct scoring raises ask-side log-depth error from 2.4802 to 2.6988 (+8.81\%) and lowers bid-side log-depth error from 2.5868 to 2.4321 (−5.98\%). Exploratory 95\% intervals are {[}0.34\%, 16.77\%{]} and {[}−11.92\%, −0.15\%{]}, respectively. Mean-spread error falls by 4.12\%, with an interval spanning both directions. Studies of liquidity should therefore check both sides of the order book alongside aggregate agreement.

Differences between evaluation dates are larger than the aggregate difference between the selectors (Supplementary Fig.~A1b). For the matched direct model, joint error rises from 18.9006 on 9 December 2025 to 46.3797 on 12 June 2026. Relative differences from conditional selection range from −0.55\% to +0.45\% across dates. Both selectors encounter the same difficult dates, indicating a shared challenge in reproducing those market conditions. Evaluating later dates is therefore central to deciding whether a prepared calibrator remains suitable for reuse.

\subsection{When reuse offsets preparation}\label{sec:offlinecost}

\roundedit{The traffic cost estimates sum once-measured preparation stages, one fresh process start and the complete query latency measured after warm-up, at each rule's observed error.}
\roundedit{Table~\ref{tab:closesttraffic} reports preparation and query costs in the common deployment runtime.}
\roundedit{Each workflow is charged the same 24,576 preparation responses, including normalization and development data.}
\roundedit{Simulation uses eight workers, while fitting and queries use one core.}
\roundedit{At 100,000 queries, cumulative elapsed costs are 17.74 s for local correction and 24.21 and 25.02 s for the inverses.}
\roundedit{Local correction offsets its extra preparation after approximately 85,300 queries against empirical selection and 29,900 against surrogate selection.}
\roundedit{The inverses' shorter queries offset their additional preparation after approximately 0.87 and 0.96 million queries, respectively.}
\roundedit{The inverse therefore pays off after sufficiently long reuse at the observed, similar errors.}
\roundedit{These crossing points depend on the measured runtime and the number of queries before preparation is repeated.}
\roundedit{Stage-level wall and processor times, the simple controls and reuse calculations are reported in Supplement~L.5.}

\roundedit{The financial records provide a partial simulation count and a separate measurement of the extra direct score model.}
\roundedit{The bank and candidate-generation records account for 47,056 responses before inherited-dictionary construction and rejected-candidate screening.}
\roundedit{For the matched compressed-teacher comparison, preparing and saving the direct score model takes 2.896 s and saves 0.004983 ms per query, offsetting that extra preparation after approximately 581,276 queries.}
\roundedit{With 95 queries between score-model refreshes, head preparation plus direct queries takes 2.902 s compared with 0.00574 s for conditional-selector queries.}
\roundedit{At one million queries, these incremental totals are 58.32 and 60.41 s.}
\roundedit{Shared simulation and response-model preparation are additional costs for both rules (Supplement~L.6).}
\roundedit{These financial estimates combine preparation and query measurements from separate sessions on the same processor family.}
\roundedit{The expected workload before refresh therefore determines whether a faster query also reduces cumulative computation.}

\section{Discussion}\label{sec:discussion}

\subsection{Reusing the calibration solution}

\roundedit{The prepared index and local maps turn a set of simulated responses into a reusable calibration rule.}
\roundedit{The index selects a feasible anchor, and the local map solves a regularised response-matching problem around it.}
\roundedit{This provides continuous parameter corrections without fitting an inverse regressor to parameter labels (Supplement~F.3).}

\roundedit{For fixed conditions, deployed candidates, the index and local maps occupy $O(Mp+Md+Mpd)$ storage, with worst-case query work $O(Md+pd)$.}
\roundedit{The historical scorer stores and applies an $O(Mk)$ decoder in addition to its predictor.}
\roundedit{Conditional surrogate selection adds $O(Md)$ candidate comparison to prediction.}
\roundedit{These costs make the candidate count and response dimension explicit considerations when preparing a calibrator.}

\subsection{Planning repeated calibration}\label{sec:limitations}

\roundedit{The prepared index and local maps serve successive targets $x$ under the same conditions $c$.}
\roundedit{When $c$ changes, conditional response and score models incorporate it as an input, while the index-based route updates its response estimates, index and local maps.}
\roundedit{Preparation can therefore be organised around the conditions and target ranges expected during use.}

\roundedit{The choice of calibration procedure also depends on the workload before refresh.}
\roundedit{Query latency determines whether a request meets its deadline, while preparation and subsequent queries together determine cumulative computation.}
\roundedit{Frequent refreshes favour modest preparation, whereas sustained reuse can justify additional fitting to shorten each query.}

\subsection{Research materials}

\roundedit{We will release the implementation and experiment scripts in a GitHub repository and provide the necessary data on Hugging Face.}

\section{Conclusion}\label{sec:conclusion}

We developed an amortized optimizer for repeated calibration of stochastic simulators. \roundedit{For observed targets, it combines a simulation memory with a separately fitted surrogate to select candidates by expected squared response error and optionally apply a prepared local correction.} Retaining distinct candidates avoids averaging conflicting inverse labels. \roundedit{For financial calibration, historical outcomes supervise a mean-response proxy, and independent simulator responses evaluate the returned parameters with the energy score.}

The framework extends the reuse of simulations from predicting responses to returning calibrated parameters. \roundedit{Within the conditions represented during preparation, the same stored quantities and fitted models can serve successive observed targets.} \roundedit{The component comparisons identify how coverage, response availability and the workload before refresh guide the choice of a prepared calibrator.}

\section*{Acknowledgment}
OpenAI ChatGPT/Codex and Google Gemini assisted with language polishing. OpenAI Codex also assisted with parts of the analysis and plotting code.

\nocite{elfwing2017silu,loshchilov2019adamw,lu2025consistency,schmitt2024consistency,stanovov2022nlshade,tsai2026sno}

\bibliographystyle{IEEEtran}
\bibliography{references}

@article{amaran2016simulation,
  author = {Amaran, Satyajith and Sahinidis, Nikolaos V. and Sharda, Bikram and Bury, Scott J.},
  title = {Simulation Optimization: A Review of Algorithms and Applications},
  journal = {Annals of Operations Research},
  volume = {240}, pages = {351--380}, year = {2016},
  doi = {10.1007/s10479-015-2019-x}
}

@inproceedings{ament2023logei,
  author={Ament, Sebastian and Daulton, Samuel and Eriksson, David and Balandat, Maximilian and Bakshy, Eytan},
  title={Unexpected Improvements to Expected Improvement for Bayesian Optimization},
  booktitle = {Advances in Neural Information Processing Systems},
  year={2023}, volume={36}, pages={20577--20612}
}

@article{amos2023amortized,
 author = {Brandon Amos},
 title = {Tutorial on Amortized Optimization},
 journal = {Foundations and Trends in Machine Learning},
 volume = {16}, number = {5}, pages = {592--732}, year = {2023}
}

@article{ankenman2010stochastic,
  title = {Stochastic Kriging for Simulation Metamodeling},
  author = {Ankenman, Bruce and Nelson, Barry L. and Staum, Jeremy},
  journal = {Operations Research},
  volume = {58},
  number = {2},
  pages = {371--382},
  year = {2010}
}

@article{beaumont2002abc,
 author={Beaumont, Mark A. and Zhang, Wenyang and Balding, David J.},
 title={Approximate Bayesian Computation in Population Genetics},
 journal={Genetics}, year={2002}, volume={162}, number={4}, pages={2025--2035}
}

@incollection{belcak2022fast,
  title = {Fast Agent-Based Simulation Framework with Applications to Reinforcement Learning and the Study of Trading Latency Effects},
  author = {Belcak, Peter and Calliess, Jan-Peter and Zohren, Stefan},
  booktitle = {Multi-Agent-Based Simulation {XXII}},
  series = {Lecture Notes in Computer Science},
  volume = {13128},
  pages = {42--56},
  publisher = {Springer},
  year = {2022}
}

@techreport{bishop1994mixture,
  author={Bishop, Christopher M.}, title={Mixture Density Networks},
  institution={Aston University}, number={NCRG/94/004}, year={1994}
}

@article{cont2010stochastic,
  title = {A Stochastic Model for Order Book Dynamics},
  author = {Cont, Rama and Stoikov, Sasha and Talreja, Rishi},
  journal = {Operations Research},
  volume = {58},
  number = {3},
  pages = {549--563},
  year = {2010}
}

@article{cranmer2020frontier,
  title = {The Frontier of Simulation-Based Inference},
  author = {Cranmer, Kyle and Brehmer, Johann and Louppe, Gilles},
  journal = {Proceedings of the National Academy of Sciences},
  volume = {117},
  number = {48},
  pages = {30055--30062},
  year = {2020}
}

@inproceedings{durkan2019neural,
  title = {Neural Spline Flows},
  author = {Durkan, Conor and Bekasov, Artur and Murray, Iain and Papamakarios, George},
  booktitle = {Advances in Neural Information Processing Systems},
  volume = {32},
  pages = {7509--7520},
  year = {2019}
}

@misc{elfwing2017silu,
  author={Elfwing, Stefan and Uchibe, Eiji and Doya, Kenji},
  title={Sigmoid-Weighted Linear Units for Neural Network Function Approximation in Reinforcement Learning},
  year={2017}, eprint={1702.03118}, archivePrefix={arXiv}, note={arXiv:1702.03118}
}

@inproceedings{eriksson2019scalable,
  title = {Scalable Global Optimization via Local Bayesian Optimization},
  author = {Eriksson, David and Pearce, Michael and Gardner, Jacob and Turner, Ryan D. and Poloczek, Matthias},
  booktitle = {Advances in Neural Information Processing Systems},
  volume = {32},
  pages = {5497--5508},
  year = {2019}
}

@inproceedings{geffner2023compositional,
  title = {Compositional Score Modeling for Simulation-Based Inference},
  author = {Geffner, Tomas and Papamakarios, George and Mnih, Andriy},
  booktitle = {Proc. 40th International Conference on Machine Learning (ICML)},
  series = {Proceedings of Machine Learning Research},
  volume = {202},
  pages = {11098--11116},
  year = {2023},
  publisher = {PMLR}
}

@article{gneiting2007strictly,
  title = {Strictly Proper Scoring Rules, Prediction, and Estimation},
  author = {Gneiting, Tilmann and Raftery, Adrian E.},
  journal = {Journal of the American Statistical Association},
  volume = {102},
  number = {477},
  pages = {359--378},
  year = {2007}
}

@article{goncalves2020,
 title={Training deep neural density estimators to identify mechanistic models of neural dynamics},
 author={Gonçalves, Pedro J. and Lueckmann, Jan-Matthis and Deistler, Michael and Nonnenmacher, Marcel and Öcal, Kaan and Bassetto, Giacomo and Chintaluri, Chaitanya and Podlaski, William F. and Haddad, Sara A. and Vogels, Tim P. and Greenberg, David S. and Macke, Jakob H.},
 journal={eLife}, volume={9}, pages={e56261}, year={2020}
}

@article{gorecki2024bam,
 title={Amortized Bayesian Decision Making for simulation-based models},
 author={Gorecki, Mila and Macke, Jakob H. and Deistler, Michael},
 journal={Transactions on Machine Learning Research}, year={2024}
}

@article{hansen2001cmaes,
  title = {Completely Derandomized Self-Adaptation in Evolution Strategies},
  author = {Hansen, Nikolaus and Ostermeier, Andreas},
  journal = {Evolutionary Computation},
  volume = {9},
  number = {2},
  pages = {159--195},
  year = {2001}
}

@article{jin2011surrogate,
  title = {Surrogate-Assisted Evolutionary Computation: Recent Advances and Future Challenges},
  author = {Jin, Yaochu},
  journal = {Swarm and Evolutionary Computation},
  volume = {1},
  number = {2},
  pages = {61--70},
  year = {2011}
}

@inproceedings{kennedy1995pso,
  title = {Particle Swarm Optimization},
  author = {Kennedy, James and Eberhart, Russell},
  booktitle = {Proc. ICNN'95 - International Conference on Neural Networks},
  volume = {4},
  pages = {1942--1948},
  year = {1995},
  organization = {IEEE}
}

@article{kennedy2001bayesian,
  title = {Bayesian Calibration of Computer Models},
  author = {Kennedy, Marc C. and O'Hagan, Anthony},
  journal = {Journal of the Royal Statistical Society: Series B (Statistical Methodology)},
  volume = {63},
  number = {3},
  pages = {425--464},
  year = {2001}
}

@article{liu2018tandem,
  title = {Training Deep Neural Networks for the Inverse Design of Nanophotonic Structures},
  author = {Liu, Dianjing and Tan, Yixuan and Khoram, Erfan and Yu, Zongfu},
  journal = {ACS Photonics},
  volume = {5},
  number = {4},
  pages = {1365--1369},
  year = {2018}
}

@inproceedings{loshchilov2019adamw,
  author={Loshchilov, Ilya and Hutter, Frank}, title={Decoupled Weight Decay Regularization},
  booktitle = {International Conference on Learning Representations}, year={2019}
}

@inproceedings{lu2025consistency,
  title = {Simplifying, Stabilizing and Scaling Continuous-Time Consistency Models},
  author = {Lu, Cheng and Song, Yang},
  booktitle = {International Conference on Learning Representations},
  year = {2025}
}

@inproceedings{lueckmann2021benchmarking,
  title = {Benchmarking Simulation-Based Inference},
  author = {Lueckmann, Jan-Matthis and Boelts, Jan and Greenberg, David S. and Goncalves, Pedro J. and Macke, Jakob H.},
  booktitle = {Proc. 24th International Conference on Artificial Intelligence and Statistics (AISTATS)},
  series = {Proceedings of Machine Learning Research},
  volume = {130},
  pages = {343--351},
  year = {2021},
  publisher = {PMLR}
}

@article{marquardt1963,
 author={Marquardt, Donald W.}, title={An Algorithm for Least-Squares Estimation of Nonlinear Parameters},
 journal={Journal of the Society for Industrial and Applied Mathematics}, year={1963}, volume={11}, number={2}, pages={431--441}
}

@inproceedings{mishra2026sc,
 title={Robust Amortized Bayesian Inference with Self-Consistency Losses on Unlabeled Data},
 author={Mishra, Aayush and Habermann, Daniel and Schmitt, Marvin and Radev, Stefan T. and Bürkner, Paul-Christian},
 booktitle = {International Conference on Learning Representations}, year={2026}
}

@inproceedings{papamakarios2016fast,
  title = {Fast $\epsilon$-free Inference of Simulation Models with Bayesian Conditional Density Estimation},
  author = {Papamakarios, George and Murray, Iain},
  booktitle = {Advances in Neural Information Processing Systems},
  volume = {29},
  year = {2016}
}

@inproceedings{papamakarios2017masked,
  title = {Masked Autoregressive Flow for Density Estimation},
  author = {Papamakarios, George and Pavlakou, Theo and Murray, Iain},
  booktitle = {Advances in Neural Information Processing Systems},
  volume = {30},
  year = {2017}
}

@article{raynal2019abcrf,
  author = {Raynal, Louis and Marin, Jean-Michel and Pudlo, Pierre and Ribatet, Mathieu and Robert, Christian P. and Estoup, Arnaud},
  title = {{ABC random forests for Bayesian parameter inference}},
  journal = {Bioinformatics},
  year = {2019},
  volume = {35},
  number = {10},
  pages = {1720--1728}
}

@inproceedings{schmitt2024consistency,
  title = {Consistency Models for Scalable and Fast Simulation-Based Inference},
  author = {Schmitt, Marvin and Pratz, Valentin and K{\"o}the, Ullrich and B{\"u}rkner, Paul-Christian and Radev, Stefan T.},
  booktitle = {Advances in Neural Information Processing Systems},
  volume = {37},
  year = {2024}
}

@inproceedings{schmitt2024sc,
 title={Leveraging Self-Consistency for Data-Efficient Amortized Bayesian Inference},
 author={Schmitt, Marvin and Ivanova, Desi R. and Habermann, Daniel and Koethe, Ullrich and Bürkner, Paul-Christian and Radev, Stefan T.},
 booktitle = {Proc. 41st International Conference on Machine Learning (ICML)},
 series={Proceedings of Machine Learning Research}, volume={235}, pages={43723--43741}, year={2024}
}

@article{shahriari2016taking,
  title = {Taking the Human Out of the Loop: A Review of Bayesian Optimization},
  author = {Shahriari, Bobak and Swersky, Kevin and Wang, Ziyu and Adams, Ryan P. and de Freitas, Nando},
  journal = {Proceedings of the IEEE},
  volume = {104},
  number = {1},
  pages = {148--175},
  year = {2016}
}

@inproceedings{stanovov2022nlshade,
  author={Stanovov, Vladimir and Akhmedova, Shakhnaz and Semenkin, Eugene},
  title={{NL-SHADE-LBC} algorithm with linear parameter adaptation bias change for {CEC 2022} Numerical Optimization},
  booktitle = {2022 IEEE Congress on Evolutionary Computation (CEC)},
  year={2022}, pages={1--8}
}

@inproceedings{trabucco2021coms,
  title = {Conservative Objective Models for Effective Offline Model-Based Optimization},
  author = {Trabucco, Brandon and Kumar, Aviral and Geng, Xinyang and Levine, Sergey},
  booktitle = {Proc. 38th International Conference on Machine Learning (ICML)},
  series = {Proceedings of Machine Learning Research},
  volume = {139},
  pages = {10358--10368},
  year = {2021},
  publisher = {PMLR}
}

@article{treiber2000congested,
  title = {Congested Traffic States in Empirical Observations and Microscopic Simulations},
  author = {Treiber, Martin and Hennecke, Ansgar and Helbing, Dirk},
  journal = {Physical Review E},
  volume = {62},
  number = {2},
  pages = {1805--1824},
  year = {2000}
}

@article{tsai2026sno,
  author={Tsai, Chun-Wei and Yang, Yi-Cheng and Tang, Tzu-Chieh and Hsu, Che-Wei},
  title={Space Net Optimization for Single Objective Bound Constrained Problems},
  journal={IEEE Transactions on Evolutionary Computation},
  year={2026}, volume={30}, number={5}, pages={1833--1847}
}

@inproceedings{wildberger2023flow,
  title = {Flow Matching for Scalable Simulation-Based Inference},
  author = {Wildberger, Jonas and Dax, Maximilian and Buchholz, Simon and Green, Stephen R. and Macke, Jakob H. and Sch{\"o}lkopf, Bernhard},
  booktitle = {Advances in Neural Information Processing Systems},
  volume = {36},
  pages = {16837--16864},
  year = {2023}
}

@article{wang2025nonidentifiability,
 author={Wang, Chenkai and Ren, Junji and Yang, Peng},
 title={Alleviating Nonidentifiability: A High-Fidelity Calibration Objective for Financial Market Simulation With Multivariate Time Series Data},
 journal={IEEE Transactions on Computational Social Systems}, year={2025},
 volume={12}, number={6}, pages={4910--4922}
}

@article{vargas2025confidence,
 author={Vargas-P\'erez, V\'ictor A. and Gir\'aldez-Cru, Jes\'us and Mesejo, Pablo and Cord\'on, Oscar},
 title={Unveiling Agents' Confidence in Opinion Dynamics Models via Graph Neural Networks},
 journal={IEEE Transactions on Computational Social Systems}, year={2025},
 volume={12}, number={2}, pages={725--737}
}

@article{thiele2014calibration,
 author = {Thiele, Jan C. and Kurth, Winfried and Grimm, Volker},
 title = {Facilitating Parameter Estimation and Sensitivity Analysis of Agent-Based Models: A Cookbook Using {NetLogo} and {R}},
 journal = {Journal of Artificial Societies and Social Simulation},
 volume = {17}, number = {3}, pages = {11}, year = {2014}
}

@article{grazzini2017bayesian,
 author = {Grazzini, Jakob and Richiardi, Matteo and Tsionas, Mike},
 title = {{Bayesian} Estimation of Agent-Based Models},
 journal = {Journal of Economic Dynamics and Control},
 volume = {77}, pages = {26--47}, year = {2017}
}

@inproceedings{gao2023ace,
 author = {Gao, Richard and Deistler, Michael and Macke, Jakob H},
 booktitle = {Advances in Neural Information Processing Systems},
 pages = {80191--80219},
 publisher = {Curran Associates, Inc.},
 title = {Generalized Bayesian Inference for Scientific Simulators via Amortized Cost Estimation},
 volume = {36},
 year = {2023}
}

@article{beaumont2019abc,
  author = {Beaumont, Mark A.},
  title = {Approximate {Bayesian} Computation},
  journal = {Annual Review of Statistics and Its Application},
  volume = {6},
  pages = {379--403},
  year = {2019}
}

\end{document}


\appendix
\renewcommand{\thesection}{\Alph{section}}
\renewcommand{\theequation}{A\arabic{equation}}
\renewcommand{\thepage}{A\arabic{page}}
\renewcommand{\thefigure}{A\arabic{figure}}
\renewcommand{\thetable}{A\arabic{table}}
\begin{center}{\Large\bfseries Supplementary Material}\\[6pt]
{\large Amortized Optimization for Calibrating Stochastic Social Simulators}\\[6pt] Muyao Zhong and Peng Yang\end{center}
This supplement develops the response-error estimates and local corrections used by the amortized optimizer, documents all four calibration tasks, and reports the complete experiments and protocols. References cited in this supplement are listed at its end.
\section{Response errors and calibration objectives}\label{app:proof}
\subsection{Parameter selection and its approximation}
The amortized optimizer returns $\widehat\theta=\mathcal A_\phi(o)$ from available observations using separately fitted models and stored quantities. The ideal learning objective averages calibration risk across queries,
\begin{equation}\label{eq:inverselearning}
\min_\phi\,\E_o\bigl[\R(\mathcal A_\phi(o)\mid o)\bigr].
\end{equation}
Response models, candidate scores and optional local maps approximate this objective. The basic selection step is
\begin{equation}\label{eq:return}
a(o)=\arg\max_j q_j(o),\qquad\widehat\theta=\theta_{a(o)}.
\end{equation}
A configured local correction can refine the candidate using a known target. Algorithms~1 and~2 specify the separately prepared modules and complete query.

In the financial task, $o=(h,c)$ precedes the response $Y$ that the selected parameters should reproduce. Let $Z_\theta$ and $Z'_\theta$ be independent simulator responses at $(\theta,c)$, with all three response vectors transformed by the same baseline and scales. The financial energy objective is
\begin{equation}\label{eq:financialconditionalrisk}
\mathcal R_{\rm fin}(\theta\mid h,c)
=\E\!\left[2\norm{Z_\theta-Y}-\norm{Z_\theta-Z'_\theta}\mid h,c\right].
\end{equation}
The expectation averages over $Y\mid h,c$ and simulation randomness, independently sampled conditional on the query. Training compares candidates through the mean-response proxy $\E[\norm{\mu_j(c)-Y}^2/8\mid h,c]$, while independent evaluation uses 16 responses to estimate $\mathcal R_{\rm fin}$ at the returned parameter. The energy objective also accounts for response dispersion, so its candidate ordering can differ from that of the training proxy.

\subsection{Expected squared error from response moments}
For a fixed observed response $x$, let $Z_j=S(\theta_j,c,\xi)$ have finite second moments. With $m_j=\E Z_j$ and $s_j=\E\norm{Z_j}^2$,

\begin{equation}
\E\norm{Z_j-x}^2=s_j-2m_j^\top x+\norm{x}^2
=\norm{m_j-x}^2+\operatorname{tr}\operatorname{Cov}(Z_j).
\end{equation}

Replacing $m_j,s_j$ with sample or predicted moments yields the empirical and surrogate errors. Because both have affine dependence on $x$ after the common quadratic term is removed, their weighted combination can be evaluated from moments mixed during preparation. The exact index retrieves the same minimum as an exhaustive comparison of these errors, with pruning determined by their geometry and linear worst-case work in the candidate count.

\begin{revisionblock}
For $d$ partially observed response features, the augmented vector is $(a\odot v,\omega a)$, with validity weight $\omega$ as in Section~\mbox  {III-C}0a. The surrogate model predicts validity probability $p_\ell$, mean $\mu_\ell$ conditional on validity, and corresponding variance $\sigma_\ell^2$. Its moments are
\end{revisionblock}
\begin{revisionblock}
\begin{align}
\E[a_\ell v_\ell]&=p_\ell\mu_\ell, &
\E[(a_\ell v_\ell)^2]&=p_\ell(\mu_\ell^2+\sigma_\ell^2),\\
\E[\omega a_\ell]&=\omega p_\ell, & \E[(\omega a_\ell)^2]&=\omega^2p_\ell.
\end{align}
\end{revisionblock}
\begin{revisionblock}
These marginal identities hold for any response distribution with finite second moments. The continuous response loss uses only valid features, while the validity head uses all observed validity indicators. Candidate scoring includes both contributions.
\end{revisionblock}

\subsection{Local fields and the full local variant}\label{app:localfields}
The two derivative estimates use the same perturbation coordinates but different random-stream coupling. For the mean-field control $J_a$, direction $k$ uses 4 plus/minus response pairs on a direction-specific stream. Averaging those differences gives row $k$ of $J_a$. The streams for different directions are independent. Let $\lambda_M=\max(10^{-12},10^{-3}\operatorname{tr}(J_aJ_a^\top)/p)$. Its mean-only update is

\begin{equation}\label{eq:meanupdate}
\delta_M=(J_aJ_a^\top+\lambda_M I)^{-1}J_a(x-\mu_a).
\end{equation}

For the second field, 4 streams are shared across the center and every derivative direction. The direction-0 pairs are reused. The center and other directions receive new runs on those streams. Let $B_r$ be the resulting joint derivative matrix, $\overline B$ its repeat mean and $\overline z$ the center-response mean. Define

\begin{align}\label{eq:localmoments}
C_B&=\E_r[(B_r-\overline B)(B_r-\overline B)^\top],\\
c_B&=\E_r[(B_r-\overline B)(z_r-\overline z)].\nonumber
\end{align}

With $H_J=\overline B\overline B^\top$ and $\lambda_J=\max(10^{-12},10^{-3}\operatorname{tr}(H_J)/p)$, the evaluated second field is

\begin{equation}\label{eq:stochasticupdate}
\delta_R(x)=(H_J+C_B+\lambda_J I)^{-1}[\overline B(x-\mu_a)-c_B].
\end{equation}

It minimizes the average squared residual of the recentered local model $\widetilde Z_r(\delta)=\mu_a+(z_r-\overline z)+B_r^\top\delta$, plus $\lambda_J\norm{\delta}^2$. Here $\mu_a$ is the retrieval mean, not necessarily $\overline z$. Recentring deliberately shifts the local response intercept by $\mu_a-\overline z$. It preserves empirical fluctuations while using the mean estimate used in the anchor score. Even with empirical retrieval, the original anchor repeats and new joint-center repeats differ. With mixed retrieval, surrogate bias also enters this intercept. The normal equation solves this recentered approximation exactly. Native response evaluation measures the resulting correction quality.

The joint-response mean-field control drops $C_B$ and $c_B$ from Eq.~\eqref{eq:stochasticupdate}, keeping $\overline B$, $\lambda_J$, $\mu_a$ and the radius multiplier of the stochastic field. It therefore does not reproduce $\delta_M$ because $J_a$ and $\overline B$ differ in their response streams, and their ridges are calculated separately. On set B, the capped joint-mean and original mean proposals coincided on none of the 2,048 returns per task, even under a maximum unit-coordinate difference tolerance of $10^{-10}$. Median standardized displacement differences were 0.0411 and 0.0224 in the 2 predator--prey banks. The matched mechanism results support the additional candidate, while covariance-specific effects remain unresolved.

The retrieval mean and joint-center mean use separate repeats. Their sampling difference, possible surrogate bias and local linearization all contribute to approximation error. The matched controls below assess the combined return.

Let $v=\kappa(u-0.5)$, where $\kappa=10$ for predator--prey and $\kappa=\sqrt{12}$ for traffic, and let $r_a$ be the Euclidean distance in $v$ space to the 64th nearest anchor. A proposed standardized displacement $\delta$ returns

\begin{equation}
\widehat u=\operatorname{clip}_{[0,1]^p}\left[u_a+\kappa^{-1}\min\{1,\eta r_a/\lVert\delta\rVert\}\delta\right],
\end{equation}

with zero displacement unchanged. Thus the radius acts before conversion to unit coordinates and feasibility clipping. The conservative candidate shrinks the clipped mean proposal. Development observations select the retrieval weight and radius. The displacement calculation in Supplement~\ref{app:implementation} fixes shrinkage. Retrieval, radius and acceptance coefficients were fixed before sets A and B. The subsequent A/B component analyses motivated selecting $u_R$ alone as the primary local return. Set C was generated after that choice, as recorded in Supplement~\ref{app:evaluationcohorts}.

\subsection{Auxiliary acceptance correction}\label{app:acceptance}
For surrogate scores $\widehat r_0,\widehat r_M,\widehat r_R$, define
\begin{equation}\label{eq:accept}
\rho_j=\frac{\widehat r_j-\widehat r_0}{\widehat r_j+\widehat r_0+10^{-8}},\qquad
s_j=\operatorname{atanh}(\operatorname{clip}(\rho_j))+a\rho_j+b,\qquad j\in\{M,R\}.
\end{equation}
Clipping is to $[-0.9999,0.9999]$. The selected candidate minimizes $(0,s_M,s_R)$.

The acceptance model is fitted on paired native discrepancies $R_0,R_j$. Write $t_j=(R_j-R_0)/(R_j+R_0+10^{-8})$ and let $\rho_j$ be the analogous surrogate-score difference. Ridge regression with an intercept fits $\operatorname{atanh}(\operatorname{clip}(t_j))-\operatorname{atanh}(\operatorname{clip}(\rho_j))$ from a standardized $\rho_j$. The slope penalty is $0.001$ times the number of training pairs. Restoring the coordinate scale gives $a,b$ in Eq.~\eqref{eq:accept}. Candidate ties follow the order conservative, mean and stochastic risk. All fitted slopes satisfy $a>-1$. On the observed unclipped range, $s'(\rho)=1/(1-\rho^2)+a>0$. The correction therefore changes acceptance relative to the conservative candidate while preserving the ranking of the two alternatives.

\begin{longtable}{@{}llrrr@{}}
\caption{How often acceptance changes the returned parameter. Equality refers to the actual parameter vector, not a branch label. In traffic bank 1 the mean and shrinkage candidates coincide because the mean-step radius is zero, returning that parameter need not indicate an active rejection by the gate. Changed targets count each observation once across its 4 fits.}\label{tab:gatefrequency}\\
\toprule
Task & Set & Equal to $u_0$ & Different from always-select & Changed targets\\\midrule
\endfirsthead
\toprule
Task & Set & Equal to $u_0$ & Different from always-select & Changed targets\\\midrule
\endhead
Predator--prey & A & 2/2048 & 2/2048 & 2/512\\
Predator--prey & B & 4/2048 & 4/2048 & 2/512\\
Traffic & A & 383/2048 & 2/2048 & 2/512\\
Traffic & B & 392/2048 & 5/2048 & 5/512\\
\bottomrule
\end{longtable}

\subsection{Candidate coverage and score accuracy}
For any query-specific candidate set $\mathcal C$ and true response risk $R$, the error of a selected parameter decomposes as

\begin{align}
R(\widehat\theta)-\inf_{\theta\in\Theta}R(\theta)
={}&\left[R(\widehat\theta)-\min_{\theta\in\mathcal C}R(\theta)\right]\\
&+\left[\min_{\theta\in\mathcal C}R(\theta)-\inf_{\theta\in\Theta}R(\theta)\right].\nonumber
\end{align}

The first term is selection error. The second is candidate coverage error. If the estimated score differs from $R$ by at most $\varepsilon$ on every candidate, its minimum has selection error at most $2\varepsilon$. Local proposals affect coverage error, while their score accuracy controls selection error. If a preliminary empirical screen retains a subset, its lost best-candidate quality forms an additional screening term. The original neural component table includes full-pool empirical, surrogate and mixed scoring. Its top-8 control tests this additional screening loss.

For financial labels, subtracting the fixed reference cancels $\norm{Y}^2$ and gives

\begin{equation}
\E[\Delta_j\mid h,c]=\frac{\norm{\mu_j(c)}^2-\norm{\mu_0(c)}^2
-2\E[Y\mid h,c]^\top[\mu_j(c)-\mu_0(c)]}{\nu}.
\end{equation}

The reference is fixed at stored candidate index 8. Replacing it adds a common within-example offset to all exact candidate scores, so pairwise score differences and their ordering are invariant. Column scaling and rank truncation subsequently define the targets seen by the fitted forest. Invariance of the exact comparison does not establish invariance of that fitted approximation. The current direct model fits labels computed from realized historical subsequent responses. Their conditional expectation has the expression above. The surrogate selection separately estimates the future-response mean. The bounds apply to the squared mean-response training objective. Native energy-score evaluation additionally measures response-distribution shape.

\section{Task definitions and preparation}\label{app:settings}
\subsection{Predator--prey and traffic models}\label{app:additional}
The predator--prey model uses initial populations $(X,Y)=(30,1)$ and equations $\dot X=\alpha X-\beta XY$, $\dot Y=-\gamma Y+\delta XY$. Independent log rates have means $(-0.125,-3,-0.125,-3)$ and standard deviation 0.5. Populations are observed at $0,2.1,\ldots,18.9$ with multiplicative lognormal noise of log standard deviation 0.1. The initial observations are removed. The deterministic trajectory is shared across independent observation-noise draws. The DOP853 solver uses relative tolerance $10^{-8}$ and absolute tolerance $10^{-10}$. Log populations retain positivity, and outputs are numerically bounded to $[10^{-10},10^4]$ by the task adapter. Optimizer and returned unit coordinates map to log rates within 5 prior standard deviations. Target generation uses the untruncated original prior, which permits parameters outside the return box. The realized targets in sets A, B and C lie inside it.

Traffic uses 24 vehicles on a 260 m ring with a periodic bottleneck at the first vehicle, 180 steps and time step 0.2 seconds. Desired speed, desired time headway and standstill spacing have uniform priors on $[12,32]$, $[0.4,2.4]$ and $[1,9]$. Vehicle length is 4.5, maximum acceleration 1.4, comfortable braking 2.0 and bottleneck strength 0.8 in the simulator units. Process acceleration noise has standard deviation 0.03. Measurement noise has standard deviation 0.015 after gap and speed are divided by 260 and 35. 8 evenly spaced probes supply means over successive 10-step blocks, giving 18 gap and 18 speed summaries.

For vehicle $i$ with cyclic leader $i+1$, position $q_i$ and speed $v_i$, the update uses

\begin{align}
g_i&=\max\{(q_{i+1}-q_i-4.5)\bmod260,0.5\},\\
g_i^*&=\max\{s_0+v_iT+v_i(v_i-v_{i+1})/(2\sqrt{1.4\times2}),s_0\},\\
a_i&=1.4[1-(v_i/v_0)^4-(g_i^*/g_i)^2]\nonumber\\
&\quad-\mathbf1_{i=0}\,0.8[0.6+0.4\sin(2\pi k/45)]+\epsilon_{ik},\\
v_i^+&=\operatorname{clip}(v_i+0.2a_i,0,v_0),\qquad q_i^+=(q_i+0.2v_i^+)\bmod260,
\end{align}

where $k=0,\ldots,179$ and independent $\epsilon_{ik}\sim N(0,0.03^2)$. The code field \path{reaction_time} represents desired time headway $T$. Initially, $q_i=260i/24+N(0,0.15^2)$ and $v_i=\operatorname{clip}(0.58v_0+N(0,0.05^2),0,v_0)$. Probe identifiers are integer-truncated points evenly spaced from 0 to 23. The observed pair is the pre-update gap $g_i/260$ and updated speed $v_i^+/35$, with independent $N(0,0.015^2)$ noise per coordinate and clipping to $[0,1]$. At each step, probe observations are averaged, and consecutive 10-step blocks produce the 36 summaries. The position update uses the newly updated speed.

\begin{revisionblock}
Each low-dimensional response bank contains 10,035 anchors. Paired finite differences use step 0.01 in the standardized parameter coordinate and common random streams across each plus/minus pair. The original mean field uses direction-specific paired streams. The stochastic field adds jointly coupled responses. Expanded-memory controls receive both sets, with reused direction-0 records counted once. The surrogate architecture has 2 256-unit SiLU layers, outputting response means and log variances. The retrieval network for each primary bank is fitted from scratch on its 10,035 4-response anchor groups. The earlier 1,843-group training split supplies the inherited feature scales, and the same 205 held-out groups supply development responses. Its moment loss combines mean squared error and log-variance squared error with weight $\lambda_{\rm var}=0.01$. Log variances are clipped to $[-12,8]$ for scoring. AdamW uses learning rate 0.001, weight decay $10^{-4}$, batch size 256 and gradient clipping at 5. The expanded surrogate memory and the ranking network of the full variant are fitted from scratch by uniform sampling of the pooled augmented parameter-group moments. They use a maximum of 4,000 updates with patience 8 checks at 100-update intervals and a 600-second fit cap. A separate additional-decision-data control updates this network by mixing old and new groups. Only that control retains the existing checkpoint when no development improvement occurs.
\end{revisionblock}

The acceptance-budget experiment crosses 512 or 4,096 training queries with 4 or 16 native responses per proposal. Each configuration fits the same 2-coefficient rule. A common 512-query development set with 64 responses per proposal is separate from the new target sets. In the smallest budget, predator--prey has 6,144 additional response records per fitted configuration. Traffic has 4,096--6,144 because identical proposals within a query reuse a response group. The same added records are supplied to the information-matched memory and refitted-surrogate selections. Comparing the four budgets tests incremental decision supervision, while preserving the shared response and development investment.

\subsection{Local settings and query procedure}\label{app:implementation}
The anchor is one stored parameter minimizing the affine form of Eq.~(10) on standardized response features. The retrieval grid is $\tau\in\{0,4,16,64,\infty\}$ and the radius grid is $\eta\in\{0,0.1,0.25,0.5\}$, selected on 205 development inputs, each using the first recorded response from each of the original 205 held-out simulator groups, separate from sets A and B. These groups come from an initial 2,048-parameter design split by permutation seed 92301 into 1,843 training and 205 development groups, with 4 responses each. Predator--prey parameters are a fixed subset of prior draws (subset seed 89501), with independent lognormal observation noise. Traffic parameters are uniform unit-cube draws (seed 923030000), with the stochastic responses specified above. Feature scales use only the 1,843 training groups. The 4 development responses supply surrogate-model early stopping, and the first supplies each calibration-development input. The exact encoded inputs and group identifiers are supplied in the design archive. Each new bank draws 10,035 anchors from clipped $0.5+0.1N(0,I)$ for predator--prey and uniform unit-cube draws for traffic. NumPy generator seeds are $3000000000+10000b+t$, with bank $b\in\{0,1\}$ and fixed task codes $t=1,2$. Coordinates are $v=10(u-0.5)$ for predator--prey and $v=\sqrt{12}(u-0.5)$ for traffic. Finite differences perturb each standardized coordinate by $\pm0.01$, with clipping at parameter boundaries. The derivative denominator is the actual plus/minus separation. The separate derivative field uses $\lambda_M=\max(10^{-12},10^{-3}\operatorname{tr}(J_aJ_a^\top)/p)$. The joint field uses $\lambda_J$ defined above. The response moments in the stochastic derivative terms average 4 paired repeats.

Shrinkage matches a development displacement scale. If $d_i$ is the capped mean update and $\alpha_i$ its prepared directional-damping multiplier, the fitted constant is $\gamma=[\sum_i\alpha_i^2\lVert d_i\rVert^2/\sum_i\lVert d_i\rVert^2]^{1/2}$, or zero when the denominator is zero. The multiplier is fitted from proposal displacements. Let $r_i=x_i-\mu_{a_i}$, $p_i=B_{a_i}^{\top}d_i$, and let $\mathcal N_i$ contain the 64 nearest anchors in standardized parameter distance. Define $L_i^2=\max_{k\in\mathcal N_i}\lVert(B_k-B_{a_i})^\top d_i\rVert^2/\lVert v_k-v_{a_i}\rVert^2$ and $t_i=L_i\lVert d_i\rVert$. The multiplier is zero if $r_i^\top p_i\leq0$ or the displacement is zero, and otherwise $\alpha_i=\min\{1,(r_i^\top p_i)/(\lVert p_i\rVert^2+t_i(\lVert r_i\rVert+\lVert p_i\rVert)+t_i^2/4)\}$. This displacement construction sets the fixed multiplier for the auxiliary conservative candidate. Table~\ref{tab:localsettings} gives every resulting setting and acceptance coefficient. The acceptance fit standardizes its scalar feature, uses a ridge penalty of $0.001$ times the pair count on the slope, and leaves the intercept unpenalized.

\begin{longtable}{@{}llrrrrrr@{}}
\caption{Fitted local-proposal settings. Bank and seed identify the 4 fitted rules. Mixing and radius settings were selected on development inputs, none is selected from sets A or B. The last two columns are learned acceptance coefficients.}\label{tab:localsettings}\\
\toprule
Task & Bank / seed & $\tau$ & $\eta_M$ & $\eta_R$ & $\gamma$ & $a$ & $b$\\\midrule
\endfirsthead
\toprule
Task & Bank / seed & $\tau$ & $\eta_M$ & $\eta_R$ & $\gamma$ & $a$ & $b$\\\midrule
\endhead
Predator--prey & 0 / 7 & 0 & 0.50 & 0.50 & 0.2040 & -0.6566 & -0.15062\\
Predator--prey & 0 / 17 & 0 & 0.50 & 0.50 & 0.2040 & -0.6874 & -0.14961\\
Predator--prey & 1 / 7 & 0 & 0.25 & 0.25 & 0.4120 & -0.7572 & -0.09005\\
Predator--prey & 1 / 17 & 0 & 0.25 & 0.25 & 0.4120 & -0.8594 & -0.09224\\
Traffic & 0 / 7 & 64 & 0.25 & 0.50 & 0.2599 & -0.6840 & -0.00075\\
Traffic & 0 / 17 & 64 & 0.25 & 0.50 & 0.2671 & -0.6824 & -0.00088\\
Traffic & 1 / 7 & $\infty$ & 0.00 & 0.25 & 0.0000 & -0.4729 & -0.00037\\
Traffic & 1 / 17 & $\infty$ & 0.00 & 0.50 & 0.0000 & -0.4321 & -0.00074\\
\bottomrule
\end{longtable}

For the primary query, (1) standardize the observation with the training response moments, (2) retrieve the minimum-score anchor, (3) apply its stored joint-stream affine map $A_a(x-\mu_a)-b_a$, (4) cap the displacement in standardized coordinates, map to the unit box and clip, and (5) convert the result to physical parameters. The query uses only these prepared arrays and the exact index. The expanded full-rule control instead applies both prepared maps, forms the additional fixed-shrinkage proposal, evaluates the surrogate model on the three candidates and applies the learned acceptance scores. Its no-acceptance deletion selects the better mean/risk alternative using the unchanged surrogate scores.

\subsection{Application selection}\label{app:selectionhistory}
The four reported applications were selected during exploration. An earlier susceptible--infectious--recovered (SIR) experiment was excluded after strong empirical memories matched or exceeded the proposed rule. The reported suite therefore represents an adaptively chosen application set. Its stronger predator--prey spread memory and other competitive component controls remain in the comparisons.

\subsection{Configuration and hyperparameter selection ledger}\label{app:configledger}
Tables~\ref{tab:modelstructures} and~\ref{tab:hyperparameters} give the model dimensions, training configurations and development-selection rules. The neural models are task-specific multilayer perceptrons built from fully connected layers with biases and SiLU activations. The SiLU and AdamW citations identify the activation and optimization algorithms~\cite{elfwing2017silu,loshchilov2019adamw}. The layer sizes, learning rates, weight decay and numerical bounds below are the fixed configurations used in this study.

\begin{table*}[t]
\centering\normalsize
\caption{Fitted models used by the four task instances. Layer widths include input and output dimensions. All hidden neural layers use SiLU and include biases. Parameter counts include biases. The financial models use 100 trees each. Tree size is determined by fitting. Surrogate networks prepare stored response moments. They are not evaluated by the primary local or neural query.}\label{tab:modelstructures}
\begin{tabularx}{180mm}{@{}L{29mm}L{43mm}L{31mm}Y@{}}\toprule
Task / component & Architecture & Output & Fitting objective / role\\\midrule
Predator--prey surrogate & $4\to256\to256\to36$, 76,324 parameters & 18 means, 18 log variances & Moment MSE, Eq.~(4), offline response scores\\
Traffic surrogate & $3\to256\to256\to72$, 85,320 parameters & 36 means, 36 log variances & Same moment objective, offline retrieval moments\\
Neural circuit surrogate & $31\to256\to256\to256\to54$, 153,654 parameters & 18 means, log variances and validity logits & Gaussian loss plus validity BCE, Eq.~(7)\\
Financial response teacher & Random forest: $40\to928$ & 8 summaries for each of 116 candidates & Simulated conditional mean responses\\
Financial direct calibrator & Random forest: $48\to9$, linear decoder $9\to116$ & Negative relative-error scores & Weighted real-window candidate-error labels\\
Financial surrogate selection & Random forest: $48\to8$, response representation $40\to9$ & Future mean and compressed candidate responses & Same real inputs/windows, conditional response comparison\\\bottomrule
\end{tabularx}
\end{table*}

Neural initialization uses seeded PyTorch linear-layer defaults. Minibatches are sampled with replacement. Both optimizers use AdamW with its default first/second-moment coefficients $(0.9,0.999)$ and $\epsilon=10^{-8}$. The explicitly supplied learning rates and weight decay are listed below. Gradient-norm clipping at 5 limits the update contribution from large gradients. The fixed log-variance bounds limit the range of exponentiation and inverse-variance weighting, while the low-dimensional sample-variance floor makes logarithmic targets finite. These fixed bounds provide numerical safeguards.

\begin{revisionblock}
\begin{longtable}{@{}p{30mm}p{44mm}p{80mm}@{}}
\caption{\rev{Component parameters, fitting or selection rules, and data boundaries. Validation/selection uses the specified development data. Fixed recipe choices are distinguished from selected values.}}\label{tab:hyperparameters}\\
\toprule Component & Parameter / candidate values & Fitting or selection procedure and retained setting\\\midrule
\endfirsthead\toprule Component & Parameter / candidate values & Fitting or selection procedure and retained setting\\\midrule\endhead
Low-dimensional network & 2 hidden layers, width 256, AdamW $10^{-3}$, weight decay $10^{-4}$, batch 256 & Fixed architecture/optimizer recipe, minimize mean MSE plus 0.01 log-variance MSE. At most 4,000 updates, check every 100, patience 8, restore minimum development loss. Primary fit: 10,035 4-response anchor groups per bank. Inherited scale/development split: 1,843/205 groups, split seed 92301. Expanded control: uniform pooled-group sampling, from-scratch fit.\\
Low-dimensional normalization & Training response mean and SD, network input $10(u-0.5)$ & Estimate response scaling only from the 1,843 training groups and reuse it for candidates, targets and all comparisons. Physical maps and local coordinate scales are given in Supplement~\ref{app:additional}.\\
Low-dimensional variance & Sample-variance floor $10^{-5}$, prediction log variance $[-12,8]$, loss weight $\lambda_{\rm var}=0.01$ & Fixed numerical bounds and moment-loss balance. Floor the unbiased sample variance before taking its log. Clip the predicted log variance only when converting it into a positive variance.\\
Shared neural optimization & AdamW $\beta=(0.9,0.999)$, $\epsilon=10^{-8}$, gradient-norm cap 5 & Moment coefficients and epsilon are implementation defaults, learning rate/weight decay are explicit study settings. Sample minibatches with replacement. Restore the checkpoint with the lowest development loss.\\
Retrieval & $\tau\in\{0,4,16,64,\infty\}$ & Minimize development native discrepancy on 205 held-out query responses. Predator--prey: 0 in both banks, traffic: 64 in bank 0 and infinity in bank 1.\\
Local field & 4 joint repeats, step 0.01, ridge $\max(10^{-12},10^{-3}\operatorname{tr}(\overline B\overline B^\top)/p)$ & Repeat count, step and ridge multiplier fixed. Solve Eq.~(13) offline, share each stream across the center and all plus/minus directions. Boundary derivatives use actual clipped separation.\\
Displacement cap & 64th-neighbour radius, $\eta\in\{0,0.1,0.25,0.5\}$ & Neighbour count fixed, radius factor minimizes development native error. Exact bank/seed values in Table~\ref{tab:localsettings}, primary rule inherits the stochastic-field radius without retuning on sets A/B/C.\\
Full-rule acceptance control & 2 coefficients, ridge $0.001\times$ pair count on slope & Fit relative native-error labels, unpenalized intercept. Primary reported variant uses 512 queries and 4 repeats per proposal. Compare 512/4,096 query and 4/16 repeat budgets, acceptance is absent from the primary query.\\
Spread-memory allocation & Repeats $\{4,8,16,32\}$ at a fixed retained-response budget & Select repeat allocation by native error on the 205 development inputs. Empirical selection chooses 4 in both banks, surrogate selection chooses 32 in bank 0 and 8 in bank 1. Freeze before set C.\\
Neural network & 3 hidden layers, width 256, AdamW $5\times10^{-4}$, weight decay $10^{-4}$, batch 512 & Fixed recipe, loss in Eq.~(7), maximum 12,000 updates, checks every 200, patience 10, best checkpoint, 1,200 s fit cap. Early stopping uses 307 4-response groups plus 3,277 single-response groups.\\
Neural input and variance & Input $\sqrt{12}(u-0.5)$, log variance $[-6,6]$ & Fixed unit-box scaling and log-variance bounds, used during loss evaluation and moment prediction. Missing standardized numeric values are stored as 0 and excluded by the validity mask in Eq.~(5).\\
Neural mixture & $\tau\in\{0,4,16,64,\infty\}$ & Minimize mean independent native error on 155 complete queries from the 308 remaining development groups, ties to smaller $\tau$. Both original fitting seeds choose 4, inherited by later fits.\\
Neural loss preference & $\omega\in\{1,2.5,5,10,20\}$ & Primary 5 is a declared completeness preference. Other weights are sensitivity experiments, each with new selections and native evaluation, none is selected for best test performance.\\
Financial forests & 100 trees, bootstrap, all features, unlimited depth, leaf minimum 1, split minimum 2 & Fixed settings for teacher, direct and future-response regression. Weighted squared-error splits, fitting seed is the forest random state. Query uses 1 thread. 3-run seeds 7/17/27, 5-fit bank seeds 7/17/27/37/47.\\
Financial labels & Source/target total weights $1/2$ each, weighted SD floor 0.001 & Source 1,159 real windows plus 365 target-day windows. Training future observations generate relative-error labels, subsequent evaluation outcomes never enter a query.\\
Financial compression & 9 or 18 coordinates, full outputs, response 99.5\% variance control & Weighted SVD, rank 9 retained during retrospective development. Report matched rank/teacher controls rather than an independent held-out rank-selection claim.\\
Financial dictionary & 96 inherited plus 20 training-context candidates, CMA-ES $\sigma_0=0.3$, population 10, 10 generations & 4 responses for each of 101 proposals per context, select first distinct proposal passing all 84 permitted-context feasibility checks. Final dictionary has 116 vectors. Construction seeds are $2026937+k$.\\
SBI comparisons & MDN/MAF/NSF/FMPE/NPSE package defaults, sample counts 8/32/256 & No additional architecture search. Main table fixes 32 samples, physical sample mean or shared-surrogate ranking. Training/readout caps and complete versus failed fits are retained in Supplement~\ref{app:sbi}.\\
\bottomrule\end{longtable}
\end{revisionblock}

The local inverse coefficients are solved from saved finite-difference moments. Their array dimensions per anchor are $A_a:p\times d$, $b_a:p$ and $\mu_a:d$. The radius adds one scalar. Store the physical parameter transform, feature center/scale, anchor coordinates, score moments and exact-index tree with these arrays. A neural instance additionally stores the 31-conductance candidate vectors and prepared 36-coordinate moments. A financial instance stores the fixed 116-candidate dictionary, input/response normalization, 9-coordinate forest and score decoder. These are the complete quantities needed for the primary query once preparation finishes.

Retrieval weights and local radii are selected by development native error over $\tau\in\{0,4,16,64,\infty\}$ and $\eta\in\{0,0.1,0.25,0.5\}$. Predator--prey selects empirical retrieval in both banks. Traffic selects $\tau=64$ in bank 0 and surrogate-only retrieval in bank 1. This difference follows the same selection rule on separate response banks. The stochastic radius is 0.50 in predator--prey bank 0 and 0.25 in bank 1. Traffic values depend on bank/fit as listed in Supplementary Table~\ref{tab:localsettings}. Neural mixing uses $\tau=4$ selected on 155 complete development observations. The table separates these development-selected coefficients from the later choice of the primary return described in Supplement~\ref{app:evaluationcohorts}.

\paragraph{Fitting and optimizer seeds}\label{app:seedledger}
We use fitting seeds 7 and 17 for the low-dimensional and original neural comparisons. The neural follow-up adds 27, 37, 47, 57, 67 and 77. Financial analyses use seeds 7, 17 and 27 for the 3-run cohort, and 7, 17, 27, 37 and 47 for each of the 3 475-window response banks. The common iterative comparisons use optimizer seeds 7, 17 and 27. The low-dimensional posterior fits use seeds 7 and 17 in each of 2 banks. Neural and financial posterior fits also use seeds 7 and 17. For finance, both posterior fits condition on the seed-7 history predictor. The financial optimizer comparison jointly varies the history-predictor, direct-model and search seeds over 7, 17 and 27. Supplement~\ref{app:materials} defines the aggregation and resampling units for these fits.

\subsection{Exact indexing of prepared scores}\label{app:index}
All full-pool empirical, surrogate and mixed memories use the same exact index implementation. Write a prepared score as $\lVert x-\mu_j\rVert^2+v_j$, where $v_j$ is its scalar second-moment offset. Let $h=\min(0,\min_j v_j)$. The lifted candidate $[\mu_j,\sqrt{v_j-h}]$ and query $[x,0]$ have squared distance equal to the score minus $h$. A binary spatial tree splits the widest coordinate at its median, stores axis-aligned bounds, and stops at 16 candidates per leaf. Branch-and-bound prunes only nodes whose lower bound cannot improve the incumbent. Leaf comparisons use the original affine score and deterministic candidate-index ties. This preserves the exhaustive argmin. Worst-case work remains linear in pool size.

The different score geometries lead to different pruning costs even with this common implementation. On the current neural full pool, empirical, surrogate and mixed selection evaluated an average of 215.1, 52.2 and 85.0 leaf candidates per query, respectively. Their corresponding times were 0.0199, 0.0100 and 0.0126 ms. The observed microsecond differences reflect geometry-dependent pruning on the same index and hardware, including the larger traffic empirical workload.

\subsection{Neural circuit model and candidate design}\label{app:pyloric}
The neural task uses the \texttt{pyloric} simulator released by the original authors for AB/PD, LP and PY units. Its 31 parameters comprise 8 membrane conductances per unit and 7 synaptic conductances. Unit coordinates map to the prior specified by the original authors, with membrane conductances uniform in linear space and synaptic conductances uniform in logarithmic space. Simulations run for 11,000 ms with a 0.025 ms integration step at 283 K and produce 18 standard rhythm features. A simulation takes approximately 0.65--0.85 seconds locally. The candidate design begins with 32,768 prior locations and adds 4,096 local perturbations around complete training responses and 2,048 perturbations of older inverse-model returns on training observations only. Perturbations act in logit coordinates with standard deviations selected from $\{0.12,0.35,0.9\}$ for local draws and $\{0.04,0.15\}$ for inverse returns (design seed 120001). Each added group receives 4 initial responses. Complete original anchors receive 8 additional responses. The existing prior split and 90\% of each added group type form training, with added-group split seed 120005. The resulting training pool has 35,020 candidates, including 5,832 parameter groups with repeated observations. The design archive supplies the exact candidate coordinates and group identifiers needed to instantiate this fixed pool. The full retained bank contains 69,367 responses. The lower budgets keep the first available repeats for the same parameter groups and preserve normalization and candidate-design decisions.

\begin{table}[ht]
\centering\normalsize
\caption{Neural membrane-conductance prior bounds in the conductance units used by the simulator. Each cell is a lower--upper range and is sampled uniformly. The 7 synaptic conductances are log-uniform over $[10^{-8},10^{-2}]$ for the first synapse in the ordering specified by the original authors and $[10^{-8},10^{-3}]$ for the other 6. These bounds define the default 31-dimensional prior from the original authors, with no additional $Q_{10}$ parameters.}\label{tab:neuralprior}
\begin{tabular}{@{}lrrr@{}}\toprule
Conductance & AB/PD & LP & PY\\\midrule
$g_{\mathrm{Na}}$ & 0--500 & 0--200 & 0--600\\
$g_{\mathrm{CaT}}$ & 0--7.5 & 0--2.5 & 0--12.5\\
$g_{\mathrm{CaS}}$ & 0--8 & 2--12 & 0--4\\
$g_{\mathrm{A}}$ & 0--60 & 10--60 & 30--60\\
$g_{\mathrm{KCa}}$ & 0--15 & 0--10 & 0--5\\
$g_{\mathrm{Kd}}$ & 25--150 & 0--125 & 50--150\\
$g_{\mathrm{H}}$ & 0--0.02 & 0--0.06 & 0--0.06\\
$g_{\mathrm{leak}}$ & 0--0.01 & 0.01--0.04 & 0--0.04\\
\bottomrule\end{tabular}
\end{table}

Training at each response budget uses the network, loss and optimization settings in Supplement~\ref{app:configledger}. Of 615 4-response development parameter groups, a fixed permutation assigns 307 to early stopping and 308 to a separate development diagnostic. 3,277 single-response development groups also contribute to early stopping. The final 154 targets are newly generated prior observations and enter neither fitting nor early stopping. Their fixed 16,384-draw batch uses design seed 17020261011 and response-seed base 1870000000.

The full-pool comparison fixes the mixing constant at 4. A previous development study used the 155 complete query responses among 308 held-out candidate groups. It compared $\tau\in\{0,4,16,64,\infty\}$ by mean independent native MSE of the full-pool returns, breaking exact ties toward smaller $\tau$. Both fitting seeds selected 4. That value was inherited without retuning on the new 154 targets. For a query, the 18 numeric features are standardized and concatenated with the weighted validity indicators. The prepared mixed affine scores are queried with the exact index, and the selected stored conductance vector is returned. Validity indicators are weighted by 5 before squared-error evaluation. Complete-response rate is the number of returned native responses with all 18 statistics divided by all required returned responses. Target eligibility is defined separately by complete observed rhythms. Incomplete returned responses enter the discrepancy. The fixed target batch and its native evaluation completed without execution failures.

\subsection{Financial records and parameter mapping}\label{app:finance}
Source training uses the existing 10-level SZ000001 records from 2019. Target-market records are reconstructed from Nasdaq TotalView-ITCH files for AAPL, AMZN, GOOG, INTC and MSFT. The permitted training day is 8 December 2025. Evaluation starts are every 20 minutes from 09:40 to 15:40, giving 19 non-overlapping 10-minute windows per stock/date. The search subset uses 3 of these starts. Evaluation days are 9, 10, 11 and 12 December 2025 and 12 June 2026. The ITCH timestamp is nanoseconds after local midnight in New York market time. Session times follow that clock, including its seasonal daylight-saving convention. Displayed orders are reconstructed in message order. Additions enter the price-level book, executions and cancellations reduce it, deletions remove orders, and replacements remove the old order before adding the new one. A 3-second grid retains the 10 best displayed price levels per side after all messages at or before that time. Windows require valid two-sided books and regular-session trading status. Input records are sampled at 3-second intervals. A 200-record window contains 101 prefix records through the 300-second decision time and 99 subsequent records. The reconstruction and window extraction are fixed before fitting the reported models. The measured cohort is conditional on complete valid windows.

Prices are shifted using the best bid at the beginning of the window so that it equals 99.99, and quantities are scaled using the initial mean of the 20 book levels so that this mean equals 50. Current-book prices are relative to the current mid-price in units of 0.01, and quantities use $\log(1+v/50)$. 8 summaries comprise final and mean mid-price change, standard deviation of successive changes, fraction of nonzero changes, mean spread, mean log ask depth, mean log bid depth and mean depth imbalance. Past summaries use records at 3--297 seconds. Future summaries use records at 303--597 seconds relative to the known current book. Training-response scales and a current-book response baseline are shared across methods.

For unit parameters $u\in[0,1]^7$, the physical mapping is

\begin{align}
\lambda_{\rm init}&=300^{u_1},&c_\lambda&=0.1\,500^{u_2},&\alpha&=0.1+0.8u_3,\\
\delta_s&=0.005+0.055u_4,&\mu&=0.1+0.8u_5,&\delta&=0.005+0.055u_6,\\
f&=5\,4^{u_7}.&&&&
\end{align}

The first two parameters set quote-offset scales. $\alpha$ and $\mu$ are buy probabilities for liquidity providers and demanders. $\delta_s$ is the demander arrival frequency. $\delta$ controls immediate cancellation, and $f$ controls trading intensity and quantity scaling. Candidate feasibility is evaluated on training contexts. The simulator uses its extended-6-parameter configuration with a seventh quantity/intensity factor. 100 setup agents populate the initial book. Observed levels set the first 10. Levels 11--100 extend prices by 1 tick per level and use the median depth of observed levels 6--10. Input quantities are multiplied by $f$, and simulated quantities are divided by $f$ before response summaries. There are 125 liquidity providers and 125 liquidity demanders. The exchange uses the Ashare book in the simulator. Incoming orders match the best crossing price first, then FIFO order within that level, and execute at the price of the resting order. Exhausted orders and empty levels are removed. Any limit-order remainder joins its price queue. This generic price/time book is used for both markets and does not reproduce all exchange-specific order types or rules. Duration is 300 seconds with 3-second observations. The fixed simulator configuration and book-initialization source are included in the design archive.

The fixed dictionary contains 116 feasible candidates from the first retained training-only construction. Candidate and readout alternatives were examined on the evaluation dates during subsequent retrospective development. Starting from an existing 96-candidate dictionary, construction selects 20 target-training contexts and runs CMA-ES from their stored solutions, with initial unit-coordinate scale 0.3, population 10 and 10 generations (seed $2026937+k$ for context $k$). Each context evaluates its initial solution and 100 proposals with 4 responses, giving $20\times101\times4=8,080$ native calls. The objective is squared error of the response mean against the training outcome. Proposals are traversed in score order. The first distinct one with valid responses on all 84 permitted contexts is retained for each context. The resulting 116 unit-coordinate vectors are supplied in the design archive, including the inherited 96-candidate subset. The final training bank uses 64 source contexts and 20 target training contexts, with 4 responses for each of 116 candidates. Rejected candidate screening is an additional construction cost, not part of the 38,976 final-bank count. The 2 further response banks vary permitted response records while holding the evaluation inputs fixed.

Financial response, future-response and direct-score models are 100-tree random forests with squared-error criterion and 1 query thread. A single response forest takes 40 condition coordinates and produces $116\times8=928$ outputs, an 8-dimensional mean for each candidate. The future model takes all 48 available input coordinates. The direct model takes the same 48 coordinates and predicts a 9-dimensional singular-value representation of standardized relative candidate scores. A fixed decoder restores their original scale. Candidate index 8 is the implementation reference. The full-teacher and compressed-teacher direct models share inputs, the mathematical definition of the labels, rank, fitting seeds and feasible candidates. Only the response teacher generating those labels changes. For score compression, each row of the label matrix is one historical window and each column a candidate. With normalized row weights $w_i$, each negative-relative-error column is centered by its weighted mean and divided by its weighted standard deviation, floored at $0.001$. SVD is applied to $\operatorname{diag}(\sqrt w)Z$. The first 9 right singular vectors encode $Z$ for forest training. Decoding reverses the column scaling and centering. Source and target rows each carry 0.5 of the total weight, uniform within each group. The future-response forest uses the same 1,159 plus 365 rows and weights. Each forest uses its stated fitting seed as \path{random_state}. The forests use bootstrap sampling, all input features at a split, no depth or leaf-count limit, minimum leaf size 1 and minimum split size 2. For a query, form the 48 input coordinates, evaluate the direct forest, decode 116 scores, and return the maximum negative-relative-error candidate. The fixed reference $j_0$ in the main text is stored candidate index 8. Subscript 0 in the supplementary derivation denotes this same reference. The score representation uses this fixed linear compression.

\begin{longtable}{@{}p{42mm}p{49mm}p{65mm}@{}}
\caption{Financial model names and output spaces. All use the same feasible dictionary. ``Full teacher'' concerns response-model outputs, score compression concerns the direct model. The same primary score-9 configuration is evaluated in both cohorts.}\label{tab:financialnames}\\
\toprule
Name & Response teacher & Query model / source identifier\\\midrule
\endfirsthead\toprule Name & Response teacher & Query model / source identifier\\\midrule\endhead
Ours (score-9, full teacher) & 928 candidate-response outputs & 9 score coordinates, ``Risk compression9 / Direct'', F116 on the single-run 75-window cohort\\
Score-9, compressed teacher & 9 response coordinates, decoded to 928 outputs & 9 score coordinates, ``Compressed teacher9 / Direct9'', F125\\
Surrogate selection & 9 response coordinates, decoded to 928 outputs & Predicted future and candidate means, ``Response compression9 / Surrogate'', F118\\
Uncompressed direct & 928 response outputs & 116 relative candidate scores, ``Dense365 / Direct'', F001\\
\bottomrule\end{longtable}

\subsection{Evaluation cohorts and response banks}\label{app:evaluationcohorts}
The low-dimensional banks each begin with 10,035 anchors and 2 fitting seeds. Paired perturbations give 642,240 retained predator--prey responses or 481,680 traffic responses per bank. Expanded-memory controls receive the same records. The full local variant additionally prepares acceptance labels from 512 decision-training inputs. Information-matched controls receive those records as well. Preparation accounting is given in Supplement~\ref{app:cost}.

Sets A and B contain 512 targets each and assess budget choices and component deletions. Their returned parameters are evaluated with 64 independent native responses. The earlier 100-target sets use 16 responses and support memory-initialized search. Predator--prey targets follow the untruncated benchmark prior. The realized targets in sets A, B and C fall within the return box. Traffic targets follow its uniform prior. Because the primary selection step was chosen after the A/B analyses, an additional predator--prey set C was generated after fixing the local rules and spread-memory allocation. Its 512 targets provide a subsequent coverage comparison. All targets are retained.

The supplied neural pool contains 35,020 candidate locations. Retaining at most 1, 2, 4 or all available responses gives 35,020, 40,852, 52,516 and 69,367 training rows. 2 fits per budget are evaluated on 154 complete observations from an independent 16,384-draw prior batch. After fixing the full-budget models, smoothers and weight grid, a second batch of the same size supplies 177 complete confirmation observations. Every returned parameter receives 16 new native responses, including incomplete responses in the loss. 6 additional fixed-configuration fits later test training variability on these same 177 targets. Laplace, global-Beta and neighbour smoothers retain empirical numeric moments. A separate 69,367-response prior memory tests uniform coverage at the same retained-response budget. The original enriched pool is supplied as a fixed design. This budget comparison excludes its earlier construction history (Supplement~\ref{app:neuralconfirmation}).

Training combines 1,159 windows from the existing 2019 SZ000001 data and 365 windows from AAPL, AMZN, GOOG, INTC and MSFT on 8 December 2025. The two sources each receive 0.5 of the total training weight. Target-training windows start every 5 minutes from 09:40 to 15:40 and overlap. Nasdaq ITCH records are reconstructed into 10-level books as described in Supplement~\ref{app:finance}. Evaluation covers the 5 stocks on 9--12 December 2025 and 12 June 2026, with 19 non-overlapping 10-minute windows per stock/date, giving 475 windows. 3 starts per stock/date form the 75-window search subset. All evaluation dates follow training, but candidate dictionaries and readouts were examined on them during development. These comparisons are retrospective.

The common dictionary has 116 7-parameter vectors. Its final bank contains 84 conditions and 4 repeats per candidate, or 38,976 responses. The primary direct model predicts 9 score coordinates using labels from the full 928-output response teacher. The surrogate selection compresses candidate responses to 9 coordinates. A second direct model uses this compressed teacher for a strictly matched comparison. Rank and dictionary were chosen during development. 3 response banks and 5 fitting seeds per bank evaluate the 475-window cohort. 3 joint model/search runs evaluate all seven optimizers on the search subset.

\section{Comparators and resource accounting}\label{app:baselines}
\subsection{Earlier default-training and posterior-draw grid}\label{app:sbi}
The earlier posterior grid uses the installed \texttt{sbi} defaults for MDN, MAF, NSF, FMPE and NPSE, with 2 response banks and fitting seeds 7 and 17 per task. Full-bank fits use all shared records. The R4 condition caps repeats per parameter at 4. Each fit has 3,600 seconds for initialization and training. The augmented bank supplies the training distribution, and the unit-box prior specifies parameter support. Evaluation compares native quality and query cost of the resulting point returns.

After fitting, the sample budgets 8, 32 and 256 are actual generated draws. Physical means and surrogate-ranked sample returns are evaluated separately. Timings include normalization, sampling, readout and conversion to physical parameters. Each completed model receives the same 100 retrospective and 512 independent set-A targets. Different readouts and cohorts reuse the same fitted model. All planned status rows remain in the accompanying CSVs, including resource limits and numerical failures. Full-cohort means are reported only for complete procedures. The predator--prey MDN failures report nonfinite mixture logits. Inputs already use training-fitted feature scales. The underlying numerical instability was not isolated. Those attempts did not retain a model after a numerical stop or training deadline. The checkpoint-retaining protocol in Supplement~\ref{app:comparisonprotocol} uses the same training defaults and budgets to recover a point-return model whenever a finite development checkpoint exists. The grid here preserves the original stopping policy and sample-budget comparisons. The records identify the default configuration and stopping policy associated with each outcome.

\begin{table*}[t]
\centering\normalsize
\caption{Posterior networks trained on the shared augmented response bank. The first independent 512-target set (set A) is evaluated with 64 native responses per return and 4 fits per reported family. This panel uses posterior training capped at 4 responses per shared parameter. Full local variant uses the full response bank plus decision labels. No full-bank predator--prey family completed all 4 fits. Complete full-bank traffic results and all partial fits are retained in the supplement. The panel evaluates these default, resource-limited point-return procedures, not a tuned prior-sampled SBI benchmark. Mean returns the physical mean of the stated number of samples. Rank selects among those samples using the shared surrogate model. Timings include actual sampling and return. The contemporaneously timed Full local variant column is used for each speed ratio. The spread-memory and primary rows supply same-cohort quality comparisons only. Their separate timing groups are not used to infer a posterior speed ratio. Families without 4 completed fits are reported individually in Supplement~\ref{app:sbi}. They are not averaged over surviving fits.}\label{tab:sbi}
\begin{tabular}{@{}lrrrr@{}}
\toprule
Method and readout & MSE $\downarrow$ & Time (ms) $\downarrow$ & Full local variant (ms) & Time/Full local variant\\\midrule
\multicolumn{5}{@{}l}{\emph{Predator--prey}}\\
NPE--MAF, 32 mean & 0.07256 & 1.794 & 0.088 & 20.43$\times$\\
NPE--MAF, 32 rank & 0.09573 & 1.925 & 0.088 & 21.92$\times$\\
FMPE, 32 mean & 0.10323 & 54.123 & 0.091 & 597.85$\times$\\
FMPE, 32 rank & 0.12997 & 54.287 & 0.091 & 599.65$\times$\\
Spread empirical memory (quality reference) & 0.03261 & --- & --- & ---\\
primary rule (quality reference) & 0.04263 & --- & --- & ---\\
Full local variant (quality reference) & 0.04193 & --- & --- & ---\\
\multicolumn{5}{@{}l}{\emph{Traffic}}\\
NPE--MDN, 8 mean & 0.96047 & 0.452 & 0.124 & 3.64$\times$\\
NPE--MDN, 32 mean & 0.96035 & 0.522 & 0.124 & 4.21$\times$\\
NPE--MDN, 32 rank & 0.95696 & 0.664 & 0.124 & 5.35$\times$\\
NPE--MAF, 32 mean & 0.95991 & 2.470 & 0.124 & 19.90$\times$\\
NPE--MAF, 32 rank & 0.95732 & 2.612 & 0.124 & 21.04$\times$\\
FMPE, 32 mean & 0.96052 & 76.395 & 0.129 & 590.29$\times$\\
FMPE, 32 rank & 0.95736 & 76.576 & 0.129 & 591.69$\times$\\
NPSE, 32 mean & 0.96018 & 222.971 & 0.129 & 1722.09$\times$\\
NPSE, 32 rank & 0.95743 & 223.164 & 0.129 & 1723.58$\times$\\
Full local variant (quality reference) & 0.95217 & --- & --- & ---\\
\bottomrule
\end{tabular}
\end{table*}

\begin{longtable}{@{}lllrrr@{}}
\caption{Completion of all planned posterior fits for the two low-dimensional tasks. Each row plans 2 banks and 2 fitting seeds. Resource-limited and numerical-error fits supply no complete quality result.}\label{tab:failures}\\
\toprule
Task & Bank & Family & Complete & Limit & Error\\\midrule
\endfirsthead
\toprule
Task & Bank & Family & Complete & Limit & Error\\\midrule
\endhead
Predator--prey & All & NPE-MDN & 0 & 0 & 4\\
Predator--prey & All & NPE-MAF & 3 & 1 & 0\\
Predator--prey & All & NPE-NSF & 0 & 4 & 0\\
Predator--prey & All & FMPE & 0 & 4 & 0\\
Predator--prey & All & NPSE & 0 & 4 & 0\\
Predator--prey & R4 & NPE-MDN & 1 & 0 & 3\\
Predator--prey & R4 & NPE-MAF & 4 & 0 & 0\\
Predator--prey & R4 & NPE-NSF & 0 & 4 & 0\\
Predator--prey & R4 & FMPE & 4 & 0 & 0\\
Predator--prey & R4 & NPSE & 3 & 1 & 0\\
Traffic & All & NPE-MDN & 4 & 0 & 0\\
Traffic & All & NPE-MAF & 4 & 0 & 0\\
Traffic & All & NPE-NSF & 0 & 4 & 0\\
Traffic & All & FMPE & 0 & 4 & 0\\
Traffic & All & NPSE & 0 & 4 & 0\\
Traffic & R4 & NPE-MDN & 4 & 0 & 0\\
Traffic & R4 & NPE-MAF & 4 & 0 & 0\\
Traffic & R4 & NPE-NSF & 2 & 2 & 0\\
Traffic & R4 & FMPE & 4 & 0 & 0\\
Traffic & R4 & NPSE & 4 & 0 & 0\\
\bottomrule
\end{longtable}

\begin{longtable}{@{}lllp{26mm}rr@{}}
\caption{Complete 4-fit posterior comparisons on the retrospective 100-target set (16 responses per return). All denotes the full permitted shared bank, R4 retains at most 4 responses per parameter. Incomplete families are documented in Table~\ref{tab:failures} and the accompanying per-fit CSV.}\label{tab:sbiA194}\\
\toprule
Task & Bank & Family & Samples / readout & MSE $\downarrow$ & Time (ms) $\downarrow$\\\midrule
\endfirsthead
\toprule
Task & Bank & Family & Samples / readout & MSE $\downarrow$ & Time (ms) $\downarrow$\\\midrule
\endhead
Predator--prey & R4 & NPE-MAF & 8 / mean & 0.02471 & 1.473\\
Predator--prey & R4 & NPE-MAF & 8 / rank & 0.02677 & 1.566\\
Predator--prey & R4 & NPE-MAF & 32 / mean & 0.02313 & 1.795\\
Predator--prey & R4 & NPE-MAF & 32 / rank & 0.02997 & 1.927\\
Predator--prey & R4 & NPE-MAF & 256 / mean & 0.02201 & 4.758\\
Predator--prey & R4 & NPE-MAF & 256 / rank & 0.03446 & 5.247\\
Predator--prey & R4 & FMPE & 8 / mean & 0.06929 & 42.179\\
Predator--prey & R4 & FMPE & 8 / rank & 0.06919 & 42.301\\
Predator--prey & R4 & FMPE & 32 / mean & 0.08406 & 54.378\\
Predator--prey & R4 & FMPE & 32 / rank & 0.05200 & 54.541\\
Predator--prey & R4 & FMPE & 256 / mean & 0.08992 & 158.879\\
Predator--prey & R4 & FMPE & 256 / rank & 0.07628 & 159.421\\
Traffic & All & NPE-MDN & 8 / mean & 0.95019 & 0.446\\
Traffic & All & NPE-MDN & 8 / rank & 0.94795 & 0.543\\
Traffic & All & NPE-MDN & 32 / mean & 0.95007 & 0.498\\
Traffic & All & NPE-MDN & 32 / rank & 0.94685 & 0.638\\
Traffic & All & NPE-MDN & 256 / mean & 0.95020 & 0.576\\
Traffic & All & NPE-MDN & 256 / rank & 0.94589 & 1.111\\
Traffic & All & NPE-MAF & 8 / mean & 0.94950 & 1.511\\
Traffic & All & NPE-MAF & 8 / rank & 0.94758 & 1.610\\
Traffic & All & NPE-MAF & 32 / mean & 0.94950 & 2.169\\
Traffic & All & NPE-MAF & 32 / rank & 0.94694 & 2.312\\
Traffic & All & NPE-MAF & 256 / mean & 0.94944 & 4.957\\
Traffic & All & NPE-MAF & 256 / rank & 0.94605 & 5.498\\
Traffic & R4 & NPE-MDN & 8 / mean & 0.94968 & 0.449\\
Traffic & R4 & NPE-MDN & 8 / rank & 0.94754 & 0.547\\
Traffic & R4 & NPE-MDN & 32 / mean & 0.94952 & 0.518\\
Traffic & R4 & NPE-MDN & 32 / rank & 0.94624 & 0.659\\
Traffic & R4 & NPE-MDN & 256 / mean & 0.94958 & 0.588\\
Traffic & R4 & NPE-MDN & 256 / rank & 0.94534 & 1.121\\
Traffic & R4 & NPE-MAF & 8 / mean & 0.94980 & 1.674\\
Traffic & R4 & NPE-MAF & 8 / rank & 0.94760 & 1.772\\
Traffic & R4 & NPE-MAF & 32 / mean & 0.94980 & 2.452\\
Traffic & R4 & NPE-MAF & 32 / rank & 0.94669 & 2.595\\
Traffic & R4 & NPE-MAF & 256 / mean & 0.94986 & 5.233\\
Traffic & R4 & NPE-MAF & 256 / rank & 0.94600 & 5.773\\
Traffic & R4 & FMPE & 8 / mean & 0.95005 & 50.465\\
Traffic & R4 & FMPE & 8 / rank & 0.94715 & 50.600\\
Traffic & R4 & FMPE & 32 / mean & 0.94989 & 73.842\\
Traffic & R4 & FMPE & 32 / rank & 0.94566 & 74.025\\
Traffic & R4 & FMPE & 256 / mean & 0.94999 & 199.580\\
Traffic & R4 & FMPE & 256 / rank & 0.94602 & 200.186\\
Traffic & R4 & NPSE & 8 / mean & 0.94980 & 159.196\\
Traffic & R4 & NPSE & 8 / rank & 0.94726 & 159.354\\
Traffic & R4 & NPSE & 32 / mean & 0.94964 & 225.219\\
Traffic & R4 & NPSE & 32 / rank & 0.94645 & 225.427\\
Traffic & R4 & NPSE & 256 / mean & 0.94965 & 561.830\\
Traffic & R4 & NPSE & 256 / rank & 0.94572 & 562.464\\
\bottomrule
\end{longtable}

\begin{longtable}{@{}lllp{26mm}rr@{}}
\caption{Complete 4-fit posterior comparisons on independent set A (512 targets, 64 responses per return). All denotes the full permitted shared bank, R4 retains at most 4 responses per parameter. Incomplete families are documented in Table~\ref{tab:failures} and the accompanying per-fit CSV.}\label{tab:sbiA195}\\
\toprule
Task & Bank & Family & Samples / readout & MSE $\downarrow$ & Time (ms) $\downarrow$\\\midrule
\endfirsthead
\toprule
Task & Bank & Family & Samples / readout & MSE $\downarrow$ & Time (ms) $\downarrow$\\\midrule
\endhead
Predator--prey & R4 & NPE-MAF & 8 / mean & 0.07506 & 1.477\\
Predator--prey & R4 & NPE-MAF & 8 / rank & 0.08365 & 1.569\\
Predator--prey & R4 & NPE-MAF & 32 / mean & 0.07256 & 1.794\\
Predator--prey & R4 & NPE-MAF & 32 / rank & 0.09573 & 1.925\\
Predator--prey & R4 & NPE-MAF & 256 / mean & 0.07158 & 4.762\\
Predator--prey & R4 & NPE-MAF & 256 / rank & 0.10135 & 5.251\\
Predator--prey & R4 & FMPE & 8 / mean & 0.11759 & 41.946\\
Predator--prey & R4 & FMPE & 8 / rank & 0.11860 & 42.069\\
Predator--prey & R4 & FMPE & 32 / mean & 0.10323 & 54.123\\
Predator--prey & R4 & FMPE & 32 / rank & 0.12997 & 54.287\\
Predator--prey & R4 & FMPE & 256 / mean & 0.10082 & 158.008\\
Predator--prey & R4 & FMPE & 256 / rank & 0.12895 & 158.548\\
Traffic & All & NPE-MDN & 8 / mean & 0.96016 & 0.457\\
Traffic & All & NPE-MDN & 8 / rank & 0.95826 & 0.555\\
Traffic & All & NPE-MDN & 32 / mean & 0.96009 & 0.515\\
Traffic & All & NPE-MDN & 32 / rank & 0.95721 & 0.656\\
Traffic & All & NPE-MDN & 256 / mean & 0.96006 & 0.585\\
Traffic & All & NPE-MDN & 256 / rank & 0.95601 & 1.118\\
Traffic & All & NPE-MAF & 8 / mean & 0.95990 & 1.547\\
Traffic & All & NPE-MAF & 8 / rank & 0.95829 & 1.645\\
Traffic & All & NPE-MAF & 32 / mean & 0.95977 & 2.225\\
Traffic & All & NPE-MAF & 32 / rank & 0.95750 & 2.367\\
Traffic & All & NPE-MAF & 256 / mean & 0.95977 & 5.009\\
Traffic & All & NPE-MAF & 256 / rank & 0.95656 & 5.547\\
Traffic & R4 & NPE-MDN & 8 / mean & 0.96047 & 0.452\\
Traffic & R4 & NPE-MDN & 8 / rank & 0.95829 & 0.549\\
Traffic & R4 & NPE-MDN & 32 / mean & 0.96035 & 0.522\\
Traffic & R4 & NPE-MDN & 32 / rank & 0.95696 & 0.664\\
Traffic & R4 & NPE-MDN & 256 / mean & 0.96031 & 0.594\\
Traffic & R4 & NPE-MDN & 256 / rank & 0.95547 & 1.130\\
Traffic & R4 & NPE-MAF & 8 / mean & 0.96005 & 1.764\\
Traffic & R4 & NPE-MAF & 8 / rank & 0.95816 & 1.862\\
Traffic & R4 & NPE-MAF & 32 / mean & 0.95991 & 2.470\\
Traffic & R4 & NPE-MAF & 32 / rank & 0.95732 & 2.612\\
Traffic & R4 & NPE-MAF & 256 / mean & 0.95991 & 5.281\\
Traffic & R4 & NPE-MAF & 256 / rank & 0.95625 & 5.821\\
Traffic & R4 & FMPE & 8 / mean & 0.96078 & 47.750\\
Traffic & R4 & FMPE & 8 / rank & 0.95868 & 47.882\\
Traffic & R4 & FMPE & 32 / mean & 0.96052 & 76.395\\
Traffic & R4 & FMPE & 32 / rank & 0.95736 & 76.576\\
Traffic & R4 & FMPE & 256 / mean & 0.96048 & 203.653\\
Traffic & R4 & FMPE & 256 / rank & 0.95638 & 204.255\\
Traffic & R4 & NPSE & 8 / mean & 0.96028 & 150.354\\
Traffic & R4 & NPSE & 8 / rank & 0.95820 & 150.499\\
Traffic & R4 & NPSE & 32 / mean & 0.96018 & 222.971\\
Traffic & R4 & NPSE & 32 / rank & 0.95743 & 223.164\\
Traffic & R4 & NPSE & 256 / mean & 0.96011 & 567.433\\
Traffic & R4 & NPSE & 256 / rank & 0.95617 & 568.039\\
\bottomrule
\end{longtable}

\subsection{Native optimizers and common information}\label{app:financialsearch}
Low-dimensional warm-start comparisons use CMA-ES, an equation-based CPU port of LSEO-S3, and a CPU port of LSEOA-MSS. Each is initialized either from an expanded-memory result or from the fixed-shrinkage proposal. Predator--prey receives 5 seconds and traffic 10 seconds, with a hard response cap. Their objective uses 4 responses per candidate. For the ODE task, 1 trajectory supports 4 observation-noise draws. The two CPU ports are exploratory adaptations of the original algorithms.

Financial PSO, GP-BO with log noisy expected improvement, CMA-ES, TuRBO-1, NL-SHADE-LBC, SNO-PSO and SNO-DE all receive the same training-fitted future predictor $\widehat y(h,c)$. It predicts the 8 standardized response residuals. For raw simulated summary vectors $s_r(\theta,c)$, the search objective is
\begin{equation}\label{eq:financialsearchobjective}
\widehat r_{\rm search}(\theta\mid h,c)=\frac{1}{8}\sum_{k=1}^{8}
\left[\frac{\tfrac14\sum_{r=1}^{4}s_{rk}(\theta,c)-b_k(c)}{\sigma_k}
-\widehat y_k(h,c)\right]^2.
\end{equation}
Here $b(c)$ is the current-book baseline and $\sigma_k$ the source-training scale given in Supplement~\ref{app:comparisonprotocol}. Responses are averaged before squaring. The search objective has no additional variance or within-simulation distance term. Its four query-specific streams are reused across candidates and optimizers. Dictionary construction uses the same squared-mean form against recorded training outcomes. Search uses a predicted outcome, and independent evaluation uses energy loss against the subsequent real outcome. Each run permits 600 seconds or 512 objective evaluations. The 60-second checkpoints through 600 seconds come from the same trajectory. Only evaluations completed within the active budget can replace an incumbent. The reported final completion time includes overrun by the last evaluation.

The repeated comparison uses model-fitting and optimizer seeds 7, 17 and 27 jointly. The future predictor for each seed is shared by all search methods. Query-specific objective streams and independent 16-response evaluation streams follow the original protocol. The additional 5-method runs reuse the corresponding fitted predictors but generate new search trajectories. Exact matches to saved input/parameter/independent-stream triples reuse their evaluation responses, while every search generates its own objective trajectory. Every planned window remains in the reported cohort. The accompanying summary tables include all 3 run means and the window-level checkpoint values.

The TuRBO implementation is based on the release from the original authors. NL-SHADE-LBC~\cite{stanovov2022nlshade} uses a population/budget adaptation, and SNO uses the simplified PSO/DE examples released by the original authors~\cite{tsai2026sno}, whose release describes them as explanatory implementations. These distinctions are retained in the complete financial table. Broader earlier neural comparisons include cold-start PSO, SACSO, CMA-ES, GP-EI, DSP-BO, TuRBO, MetaDE-small and posterior readouts. They concern the earlier 70-target development protocol and the earlier top-8 selector, so their complete tables are supplied in the separate historical-comparison data files, rather than transferred to the new 154-target mixture results. The earlier 0.6667 fixed-parameter result is also re-evaluated directly on the current cohort below.

\subsection{Online and offline cost}\label{app:cost}
Query timing starts with response features for fixed-response tasks and with the 101-record prefix for finance. It ends with the returned physical parameter. Loading, transport and offline preparation are outside the timer. Prepared methods use 5 warm-ups and 31 timed repetitions, with methods interleaved in randomized order on one AMD EPYC 9755 core. Per-target median times are averaged across targets and then across fits or banks. Current posterior procedures use 3 complete timed queries with actual sampling, and native searches are timed once per target and seed. Supplement~\ref{app:timinggroups} identifies the profiling groups.

Financial forests are fitted in scikit-learn and exported to contiguous tree arrays for a single-query C implementation. The query library is compiled with gcc using \texttt{-O3 -fno-fast-math -ffp-contract=off}. The timed Python call encodes the raw prefix, traverses the trees, decodes the 9-dimensional output into 116 candidate scores and returns the selected 7 physical parameters. Model loading, tree export and library compilation precede warm-up. Saved candidate choices were checked against the corresponding scikit-learn predictions and NumPy decoding. The primary 75-window time comes from the A120 interleaved profile. The 475-window teacher comparison uses A164.

If two methods have preparation costs $C_A,C_B$ and query times $t_A,t_B$, their total costs at $Q$ queries are $C_A+Qt_A$ and $C_B+Qt_B$. For $t_A<t_B$ and $C_A>C_B$, their formal crossover is $(C_A-C_B)/(t_B-t_A)$. It requires preparation costs measured under the same accounting boundary. Existing simulation records can be shared only when both methods actually use the same supplied records. The low-dimensional bank counts include both the additional decision labels and the common development evaluation in the conceptual total. These counts quantify stored training evidence. Elapsed preparation measurements are identified separately.

Neural response-budget counts hold candidate locations and normalization fixed, with earlier design costs treated as shared dependencies. Financial accounting separates the initial dictionary, candidate construction, screening, response bank, future model and direct head. One compressed-teacher direct fit took 2.825 seconds, including teacher loading, labels, singular-value decomposition, fitting and saving. Dividing this by the approximately five-microsecond query saving gives an illustrative scale of 565,000 queries. The preparation and query jobs used different execution conditions, so this ratio describes a scale for reuse rather than a measured crossover.

Time-to-quality uses independently evaluated saved incumbents. Because the optimizer selects using a noisy objective, first attainment and final attainment can differ. The low-dimensional analysis contains 100 targets with 4 fits, giving 400 paired runs. The financial search analysis contains 75 windows with 3 joint runs, giving 225 pairs. Supplement~\ref{app:materials} defines the corresponding statistical units.

\subsection{Preparation response accounting}\label{app:accounting}
Table~\ref{tab:preparation} distinguishes retained evidence from additional development and construction. Counts describe stochastic responses. In predator--prey simulation, observation-noise draws can share an ODE trajectory, so response counts alone are not proportional to wall time. The listed development counts are recorded components, not a reconstructed total for all exploratory alternatives. Shared-bank comparisons charge the same existing evidence to both methods. From-scratch use must additionally pay the shared investment.

\begin{longtable}{@{}p{29mm}p{42mm}p{84mm}@{}}
\caption{Known preparation costs and their accounting scope. Every proposed query uses zero new native responses. Counts cover the listed retained and development records.}\label{tab:preparation}\\
\toprule Task & Retained training evidence & Additional preparation / development\\\midrule
\endfirsthead\toprule Task & Retained training evidence & Additional preparation / development\\\midrule\endhead
Predator--prey & 642,240 responses per bank, including paired perturbation evidence & 6,144 decision-label responses per fit, nominal 98,304 acceptance-development responses per fit. Retrieval/radius development generated 404,416 and 397,632 responses in banks 0 and 1. Earlier design and other field-development work are additional.\\
Traffic & 481,680 responses per bank, including paired perturbation evidence & 4,096--6,144 decision-label responses per fit, nominal 98,304 acceptance-development responses per fit before duplicate reuse. Retrieval/radius development generated 633,664 and 475,328 responses. Earlier design and other field-development work are additional.\\
Neural circuit & 69,367 full-budget response rows at 35,020 locations & Candidate design, normalization and development responses are inherited. Retained rows alone imply about 12.5--16.4 serial simulation hours at the measured 0.65--0.85 seconds per run, before model fitting, this serial-time estimate is derived from the response count.\\
Financial & 38,976 final-bank responses from 116 candidates, 84 contexts and 4 repeats & 8,080 candidate-generation calls, plus rejected-candidate screening and the inherited 96-candidate design. The compressed-teacher direct head required 2.825 seconds in one recorded fitting job after its teacher existed.\\
\bottomrule\end{longtable}

For scale, financial construction plus the retained bank alone account for 47,056 native responses. A search query is capped at $512\times4=2,048$ responses, so at least 23 such maximum-budget queries are needed to equal even this partial preparation count. Actual calls can be lower under the time limit, and candidate-design, screening and fitting costs increase the total. This lower bound compares response counts. The preceding calculation separately defines a time-based crossover. For one low-dimensional bank and one fitted rule, the retained responses, recorded retrieval/radius development, 6,144 nominal labels and 98,304 nominal gate-development responses total approximately 1.14--1.15 million for predator--prey and 1.06--1.22 million for traffic. The warm-search cap is 20,000 native responses per query, corresponding to about 58 and 54--61 maximum-budget query counts. These counts include the gate variant and recorded development but omit earlier design costs. Actual searches often hit the time limit first. A complete preparation-time comparison would additionally require elapsed costs under matched execution conditions.

On the earlier predator--prey cohort, memory-initialized CMA-ES matched the full local variant in 238 of 400 runs within 5 seconds, including 54 matches at initialization. The matched-run median was 0.619 seconds, compared with 0.0868 ms for the prepared full local variant. Traffic had 208 of 400 immediate matches. These comparisons use the original concentrated bank. Table~\ref{tab:warm} reports endpoints and retained query times.

\begin{longtable}{@{}lp{79mm}rr@{}}
\caption{Warm-start optimization on the retrospective 100-target set. 4 fits use 2 independent response banks. CPU ports of LSEO-S3 and LSEOA-MSS are exploratory adaptations. Search times are full-budget endpoint times, not time to matched quality.}\label{tab:warm}\\
\toprule
Task & Procedure & MSE $\downarrow$ & Time (seconds)\\\midrule
\endfirsthead
\toprule
Task & Procedure & MSE $\downarrow$ & Time (seconds)\\\midrule
\endhead
Predator--prey & Full local variant & 0.02593 & 0.000087\\
Predator--prey & CMA-ES / memory start & 0.02898 & 5.001185\\
Predator--prey & CMA-ES / shrinkage start & 0.02535 & 4.994199\\
Predator--prey & LSEO-S3-port / memory start & 0.02755 & 5.001370\\
Predator--prey & LSEO-S3-port / shrinkage start & 0.02543 & 5.001393\\
Predator--prey & LSEOA-MSS-port / memory start & 0.04064 & 5.002990\\
Predator--prey & LSEOA-MSS-port / shrinkage start & 0.03717 & 5.003406\\
Traffic & Full local variant & 0.94367 & 0.000155\\
Traffic & CMA-ES / memory start & 0.95711 & 10.006900\\
Traffic & CMA-ES / shrinkage start & 0.95762 & 10.007077\\
Traffic & LSEO-S3-port / memory start & 0.95325 & 10.006893\\
Traffic & LSEO-S3-port / shrinkage start & 0.95526 & 10.006799\\
Traffic & LSEOA-MSS-port / memory start & 0.99810 & 10.007313\\
Traffic & LSEOA-MSS-port / shrinkage start & 0.99834 & 10.007092\\
\bottomrule
\end{longtable}

\section{Component evidence}\label{app:ablations}
The second independent 512-target set tests eight local proposal/scoring paths plus two expanded memories, using fixed models and the same native random streams. Removing the acceptance test preserves both alternatives and their surrogate scores. It saves the acceptance transformation but does not exploit a newly optimized two-score implementation. Its 1\% criterion was written before generating set B. Failure of its upper bound on predator--prey prevented uniform adoption. The first-set experiment changed two predator--prey returns and slightly improved their mean, motivating an independent set-B check of that simplification.

On set B, the two predator--prey targets that changed explain the uncertainty. The error on one target increased from 0.17880 to 0.53674 without acceptance, while the other fell from 0.28263 to 0.19287. The changes occurred in both fits of one response bank. The other bank returned identical parameters. The aggregate difference is therefore driven by two target-level effects in opposite directions. All per-target changes and the eight-path first-set results are retained in the data files.

\subsection{Neural budget and score-source controls}\label{app:neuralfull}
\begin{longtable}{@{}lp{68mm}rrrrr@{}}
\caption{Neural circuit budget and score-source controls. The numeric-value and missing-feature terms sum to MSE. Swapping either source changes part of the response-risk estimate, these are score-source controls rather than independently trained network-head ablations.}\label{tab:neuralfull}\\
\toprule
Budget & Score & Loss & Values & Missing & Complete (\%) & ms\\\midrule
\endfirsthead
\toprule
Budget & Score & Loss & Values & Missing & Complete (\%) & ms\\\midrule
\endhead
R1 & Empirical memory & 3.2401 & 0.5841 & 2.6560 & 78.90 & 0.0205\\
R1 & Surrogate memory & 1.3221 & 1.0290 & 0.2931 & 97.36 & 0.0097\\
R1 & Moment mixture & 1.3186 & 0.9398 & 0.3788 & 96.61 & 0.0102\\
R1 & Screen 8 / surrogate & 1.8428 & 0.5779 & 1.2649 & 89.41 & 0.0361\\
R1 & Empirical value / mixed validity & 1.6649 & 0.7777 & 0.8872 & 92.47 & 0.0251\\
R1 & Mixed value / empirical validity & 1.7781 & 0.6316 & 1.1465 & 90.44 & 0.0382\\
R2 & Empirical memory & 2.2441 & 0.5621 & 1.6820 & 85.63 & 0.0215\\
R2 & Surrogate memory & 1.5881 & 0.9680 & 0.6200 & 94.42 & 0.0087\\
R2 & Moment mixture & 1.0845 & 0.8500 & 0.2345 & 97.91 & 0.0107\\
R2 & Screen 8 / surrogate & 1.2149 & 0.5611 & 0.6539 & 94.54 & 0.0352\\
R2 & Empirical value / mixed validity & 0.9671 & 0.7202 & 0.2469 & 97.87 & 0.0261\\
R2 & Mixed value / empirical validity & 1.5874 & 0.5920 & 0.9954 & 91.90 & 0.0317\\
R4 & Empirical memory & 1.3817 & 0.5023 & 0.8793 & 92.65 & 0.0216\\
R4 & Surrogate memory & 1.0451 & 0.9482 & 0.0970 & 99.13 & 0.0098\\
R4 & Moment mixture & 0.8760 & 0.7633 & 0.1127 & 98.99 & 0.0127\\
R4 & Screen 8 / surrogate & 0.8459 & 0.5551 & 0.2909 & 97.50 & 0.0345\\
R4 & Empirical value / mixed validity & 0.8241 & 0.6764 & 0.1477 & 98.82 & 0.0236\\
R4 & Mixed value / empirical validity & 1.2066 & 0.5573 & 0.6494 & 95.05 & 0.0237\\
Full & Empirical memory & 0.7489 & 0.4817 & 0.2672 & 97.65 & 0.0199\\
Full & Surrogate memory & 1.0249 & 0.8851 & 0.1398 & 98.74 & 0.0100\\
Full & Moment mixture & 0.6367 & 0.5741 & 0.0626 & 99.45 & 0.0126\\
Full & Screen 8 / surrogate & 0.6783 & 0.5673 & 0.1110 & 99.05 & 0.0319\\
Full & Empirical value / mixed validity & 0.6511 & 0.5209 & 0.1302 & 99.01 & 0.0210\\
Full & Mixed value / empirical validity & 0.7375 & 0.5142 & 0.2232 & 98.07 & 0.0230\\
\bottomrule
\end{longtable}

For each budget, the six paths are empirical selection, surrogate selection, fixed moment mixing, empirical top-8 screening followed by surrogate ranking, empirical values with mixed validity, and mixed values with empirical validity. The independent native error, its numeric and missing-feature terms, complete-response rate and query latency are reported together. The fixed grid includes lower errors for mixed-validity controls at some budgets. The R4 mixture underperformed the full empirical memory, while full-budget mixing improved its own comparison. Both contrasts retain their original settings and outcomes.

\section{Complete financial comparisons}\label{app:fullreal}
Table~\ref{tab:finfull} retains the complete common-input financial comparison. The source CSV additionally reports every date and stock, the exact timing source and the direct method measured in the same timing group. Stored-density posterior readouts use a finite parameter grid and are distinguished from actual posterior sampling. Numerical failures and timeouts retain incomplete status in the full-cohort inventory.

The original 75-window single-run grid is retained below as a record of the configuration comparison. The current main-text search comparison extends all seven stochastic optimizers to seeds 7, 17 and 27 on those same windows. GP-BO and TuRBO already had 3 runs. PSO, CMA-ES, NL-SHADE-LBC, SNO-PSO and SNO-DE received the two missing runs. This adds 750 search queries without changing the method configurations. The 475-window response-bank experiment uses all 19 starts and assesses the full and compressed teachers. These repeated evaluations estimate response-bank and fitting variation on the same market windows. The 3 response banks and all 5 fitting seeds are kept in the supplied tables.

\begin{longtable}{@{}lp{92mm}rrr@{}}
\caption{Complete financial algorithm comparison on the common 75-window cohort, fitting seed 7. All paths use the same available information. A missing mean indicates at least one required return failed or exceeded its limit. Rows distinguish algorithm, training data, readout and checkpoint, they are not distinct algorithm counts. Times are seconds. The accompanying CSV retains exact configuration names and per-stock/date results.}\label{tab:finfull}\\
\toprule
ID & Procedure & ES $\downarrow$ & Time $\downarrow$ & Returned\\\midrule
\endfirsthead
\toprule
ID & Procedure & ES $\downarrow$ & Time $\downarrow$ & Returned\\\midrule
\endhead
F001 & Dense365 / Direct & 35.5254 & 0.000129 & 75/75\\
F002 & Dense365 / Surrogate & 35.6388 & 0.000151 & 75/75\\
F003 & PSO / 60 s & 37.9508 & 60.000000 & 75/75\\
F004 & GP-BO (qLogNEI) / 60 s & 36.0992 & 60.000000 & 75/75\\
F005 & CMA-ES / 60 s & 37.2322 & 60.000000 & 75/75\\
F006 & TuRBO-1 (authors) / 60 s & 36.5105 & 60.000000 & 75/75\\
F007 & NL-SHADE-LBC (budget-adapted) / 60 s & 38.4950 & 60.000000 & 75/75\\
F008 & SNO-PSO (authors simplified release) / 60 s & 38.5837 & 60.000000 & 75/75\\
F009 & SNO-DE (authors simplified release) / 60 s & 38.0996 & 60.000000 & 75/75\\
F010 & PSO / 300 s & 36.0129 & 300.000000 & 75/75\\
F011 & GP-BO (qLogNEI) / 300 s & 34.7766 & 300.000000 & 75/75\\
F012 & CMA-ES / 300 s & 35.2306 & 300.000000 & 75/75\\
F013 & TuRBO-1 (authors) / 300 s & 35.1541 & 300.000000 & 75/75\\
F014 & NL-SHADE-LBC (budget-adapted) / 300 s & 36.6910 & 300.000000 & 75/75\\
F015 & SNO-PSO (authors simplified release) / 300 s & 37.0257 & 300.000000 & 75/75\\
F016 & SNO-DE (authors simplified release) / 300 s & 36.2455 & 300.000000 & 75/75\\
F017 & PSO / 600 s & 35.5852 & 600.165056 & 75/75\\
F018 & GP-BO (qLogNEI) / 600 s & 34.4791 & 600.136418 & 75/75\\
F019 & CMA-ES / 600 s & 34.8870 & 600.161309 & 75/75\\
F020 & TuRBO-1 (authors) / 600 s & 34.8589 & 600.160747 & 75/75\\
F021 & NL-SHADE-LBC (budget-adapted) / 600 s & 35.9235 & 600.162828 & 75/75\\
F022 & SNO-PSO (authors simplified release) / 600 s & 36.6901 & 600.163324 & 75/75\\
F023 & SNO-DE (authors simplified release) / 600 s & 36.4580 & 600.163402 & 75/75\\
F024 & NPE-MAF / stored density96 / target / R4 mean & 43.9142 & 0.001131 & 75/75\\
F025 & NPE-NSF / stored density96 / target / R4 mean & 43.8472 & 0.003501 & 75/75\\
F026 & NPE-MDN / stored density96 / target / R4 mean & 45.5531 & 0.000579 & 75/75\\
F027 & NPE-MAF / stored density96 / source+target / R4 mean & 44.6590 & 0.001128 & 75/75\\
F028 & NPE-NSF / stored density96 / source+target / R4 mean & 43.1952 & 0.003499 & 75/75\\
F029 & NPE-MDN / stored density96 / source+target / R4 mean & 45.5850 & 0.000580 & 75/75\\
F030 & NPE-MAF / stored density96 / target / R4 individual & 45.8632 & 0.001130 & 75/75\\
F031 & NPE-NSF / stored density96 / target / R4 individual & 45.6595 & 0.003501 & 75/75\\
F032 & NPE-MDN / stored density96 / target / R4 individual & 44.4517 & 0.000580 & 75/75\\
F033 & NPE-MAF / stored density96 / source+target / R4 individual & 46.1874 & 0.001129 & 75/75\\
F034 & NPE-NSF / stored density96 / source+target / R4 individual & 44.5344 & 0.003481 & 75/75\\
F035 & NPE-MDN / stored density96 / source+target / R4 individual & 42.5352 & 0.000580 & 75/75\\
F036 & NPE-MAF / stored density116 / target / R4 mean & 41.7254 & 0.001210 & 75/75\\
F037 & NPE-NSF / stored density116 / target / R4 mean & 39.1073 & 0.003732 & 75/75\\
F038 & NPE-MDN / stored density116 / target / R4 mean & 43.6514 & 0.000596 & 75/75\\
F039 & NPE-MAF / stored density116 / source+target / R4 mean & 40.2810 & 0.001211 & 75/75\\
F040 & NPE-NSF / stored density116 / source+target / R4 mean & 39.9824 & 0.003728 & 75/75\\
F041 & NPE-MDN / stored density116 / source+target / R4 mean & 42.7472 & 0.000596 & 75/75\\
F042 & NPE-MAF / stored density116 / target / R4 individual & 40.9218 & 0.001213 & 75/75\\
F043 & NPE-NSF / stored density116 / target / R4 individual & 41.8852 & 0.003732 & 75/75\\
F044 & NPE-MDN / stored density116 / target / R4 individual & 42.5757 & 0.000595 & 75/75\\
F045 & NPE-MAF / stored density116 / source+target / R4 individual & 41.6309 & 0.001213 & 75/75\\
F046 & NPE-NSF / stored density116 / source+target / R4 individual & 41.3209 & 0.003708 & 75/75\\
F047 & NPE-MDN / stored density116 / source+target / R4 individual & 39.9053 & 0.000595 & 75/75\\
F048 & NPE-MAF / MCMC sample mean / target / R4 mean & 43.6434 & 12.623092 & 75/75\\
F049 & NPE-MAF / MCMC surrogate rank / target / R4 mean & 41.7304 & 12.627640 & 75/75\\
F050 & NPE-NSF / MCMC sample mean / target / R4 mean & 43.3954 & 53.982199 & 75/75\\
F051 & NPE-NSF / MCMC surrogate rank / target / R4 mean & 41.6785 & 53.987416 & 75/75\\
F052 & NPE-MDN / MCMC sample mean / target / R4 mean & 43.8160 & 13.828253 & 75/75\\
F053 & NPE-MDN / MCMC surrogate rank / target / R4 mean & 42.4627 & 13.832757 & 75/75\\
F054 & NPE-MAF / MCMC sample mean / source+target / R4 mean & 42.9965 & 12.710593 & 75/75\\
F055 & NPE-MAF / MCMC surrogate rank / source+target / R4 mean & 41.5159 & 12.716096 & 75/75\\
F056 & NPE-NSF / MCMC sample mean / source+target / R4 mean & 40.1570 & 54.805734 & 75/75\\
F057 & NPE-NSF / MCMC surrogate rank / source+target / R4 mean & 38.3470 & 54.812499 & 75/75\\
F058 & NPE-MDN / MCMC sample mean / source+target / R4 mean & 45.1040 & 9.980112 & 75/75\\
F059 & NPE-MDN / MCMC surrogate rank / source+target / R4 mean & 43.7110 & 9.985833 & 75/75\\
F060 & NPE-MAF / MCMC sample mean / target / R4 individual & 42.0899 & 12.736739 & 75/75\\
F061 & NPE-MAF / MCMC surrogate rank / target / R4 individual & 39.8608 & 12.742136 & 75/75\\
F062 & NPE-NSF / MCMC sample mean / target / R4 individual & 41.4720 & 57.077635 & 75/75\\
F063 & NPE-NSF / MCMC surrogate rank / target / R4 individual & 39.1237 & 57.083200 & 75/75\\
F064 & NPE-MDN / MCMC sample mean / target / R4 individual & --- & NA & 73/75\\
F065 & NPE-MDN / MCMC surrogate rank / target / R4 individual & --- & NA & 73/75\\
F066 & NPE-MAF / MCMC sample mean / source+target / R4 individual & 41.3300 & 12.549348 & 75/75\\
F067 & NPE-MAF / MCMC surrogate rank / source+target / R4 individual & 38.5559 & 12.555635 & 75/75\\
F068 & NPE-NSF / MCMC sample mean / source+target / R4 individual & 40.3083 & 56.948637 & 75/75\\
F069 & NPE-NSF / MCMC surrogate rank / source+target / R4 individual & 39.1536 & 56.956548 & 75/75\\
F070 & NPE-MDN / MCMC sample mean / source+target / R4 individual & 40.9871 & 10.950762 & 75/75\\
F071 & NPE-MDN / MCMC surrogate rank / source+target / R4 individual & 39.7285 & 10.956603 & 75/75\\
F072 & NPE-MAF / sample mean / target / R4 mean & --- & NA & 55/75\\
F073 & NPE-MAF / surrogate rank / target / R4 mean & --- & NA & 55/75\\
F074 & NPE-NSF / sample mean / target / R4 mean & --- & NA & 49/75\\
F075 & NPE-NSF / surrogate rank / target / R4 mean & --- & NA & 49/75\\
F076 & NPE-MDN / sample mean / target / R4 mean & --- & NA & 47/75\\
F077 & NPE-MDN / surrogate rank / target / R4 mean & --- & NA & 47/75\\
F078 & FMPE / sample mean / target / R4 mean & --- & NA & 47/75\\
F079 & FMPE / surrogate rank / target / R4 mean & --- & NA & 47/75\\
F080 & NPSE / sample mean / target / R4 mean & --- & NA & 52/75\\
F081 & NPSE / surrogate rank / target / R4 mean & --- & NA & 52/75\\
F082 & NPE-MAF / sample mean / source+target / R4 mean & --- & NA & 43/75\\
F083 & NPE-MAF / surrogate rank / source+target / R4 mean & --- & NA & 43/75\\
F084 & NPE-NSF / sample mean / source+target / R4 mean & --- & NA & 35/75\\
F085 & NPE-NSF / surrogate rank / source+target / R4 mean & --- & NA & 35/75\\
F086 & NPE-MDN / sample mean / source+target / R4 mean & --- & NA & 45/75\\
F087 & NPE-MDN / surrogate rank / source+target / R4 mean & --- & NA & 45/75\\
F088 & FMPE / sample mean / source+target / R4 mean & --- & NA & 47/75\\
F089 & FMPE / surrogate rank / source+target / R4 mean & --- & NA & 47/75\\
F090 & NPSE / sample mean / source+target / R4 mean & --- & NA & 53/75\\
F091 & NPSE / surrogate rank / source+target / R4 mean & --- & NA & 53/75\\
F092 & NPE-MAF / sample mean / target / R4 individual & --- & NA & 45/75\\
F093 & NPE-MAF / surrogate rank / target / R4 individual & --- & NA & 45/75\\
F094 & NPE-NSF / sample mean / target / R4 individual & --- & NA & 46/75\\
F095 & NPE-NSF / surrogate rank / target / R4 individual & --- & NA & 46/75\\
F096 & NPE-MDN / sample mean / target / R4 individual & --- & NA & 44/75\\
F097 & NPE-MDN / surrogate rank / target / R4 individual & --- & NA & 44/75\\
F098 & FMPE / sample mean / target / R4 individual & --- & NA & 61/75\\
F099 & FMPE / surrogate rank / target / R4 individual & --- & NA & 61/75\\
F100 & NPSE / sample mean / target / R4 individual & --- & NA & 43/75\\
F101 & NPSE / surrogate rank / target / R4 individual & --- & NA & 43/75\\
F102 & NPE-MAF / sample mean / source+target / R4 individual & --- & NA & 52/75\\
F103 & NPE-MAF / surrogate rank / source+target / R4 individual & --- & NA & 52/75\\
F104 & NPE-NSF / sample mean / source+target / R4 individual & --- & NA & 52/75\\
F105 & NPE-NSF / surrogate rank / source+target / R4 individual & --- & NA & 52/75\\
F106 & NPE-MDN / sample mean / source+target / R4 individual & --- & NA & 45/75\\
F107 & NPE-MDN / surrogate rank / source+target / R4 individual & --- & NA & 45/75\\
F108 & FMPE / sample mean / source+target / R4 individual & --- & NA & 70/75\\
F109 & FMPE / surrogate rank / source+target / R4 individual & --- & NA & 70/75\\
F110 & NPSE / sample mean / source+target / R4 individual & --- & NA & 51/75\\
F111 & NPSE / surrogate rank / source+target / R4 individual & --- & NA & 51/75\\
F112 & Joint surrogate optimized116 / target & 37.3166 & 0.007544 & 75/75\\
F113 & Joint surrogate uniform256 / target & 38.4908 & 0.009669 & 75/75\\
F114 & Joint surrogate optimized116 / source+target & 36.6085 & 0.009394 & 75/75\\
F115 & Joint surrogate uniform256 / source+target & 38.1289 & 0.017196 & 75/75\\
F116 & Risk compression9 / Direct & 35.4728 & 0.000122 & 75/75\\
F117 & Risk compression18 / Direct & 35.5553 & 0.000123 & 75/75\\
F118 & Response compression9 / Surrogate & 35.8276 & 0.000061 & 75/75\\
F119 & Response compression18 / Surrogate & 35.6014 & 0.000063 & 75/75\\
F120 & Response compression99.5variance / Surrogate & 35.6828 & 0.000065 & 75/75\\
F121 & Dense365 / Global fixed & 39.8001 & 0.000000 & 75/75\\
F122 & Dense365 / Asset fixed & 36.9663 & 0.000000 & 75/75\\
F123 & Uniform prior / Single random parameter & 45.4885 & 0.000001 & 75/75\\
F124 & Optimized116 / Single random candidate & 45.9224 & 0.000001 & 75/75\\
F125 & Compressed teacher9 / Direct9 & 35.6106 & 0.000055 & 75/75\\
\bottomrule
\end{longtable}

\subsection{Unlabelled-window controls}
At a fixed response bank, the training-window comparison retains the 1,159 source windows and varies the target-day contribution. The original target contribution has 95 windows, the dense version 365, and a non-overlapping extension 185. A repeated-95 control matches the dense row count while correcting weights so that each original window keeps its total contribution. The dense direct score was 34.01762 versus 34.20919 for the original data. The 185-window version had score 34.23277. Both uncertainty intervals leave the incremental quality benefit unresolved. Future-response prediction MSE is reported as an intermediate measure alongside native calibration energy score.

\section{Statistical analysis and research records}\label{app:materials}
Table entries use complete-cohort run means. For native loss $e_{ri}$ in run $r$ on target $i$, compute $\overline e_r=N^{-1}\sum_i e_{ri}$, then report the mean and sample SD of the $K$ run means. SD describes between-run variation. Bold identifies the lowest displayed mean, including displayed ties. Fixed controls have no between-fit SD and carry a dagger where indicated. Relative loss reductions use $100(1-L_M/L_{\rm ref})$ with a named reference.

The common cohorts contain 512 set-A targets in each low-dimensional task, 177 neural targets and 75 financial windows. Ours averages 2 bank means in predator--prey, 4 bank/fit configurations in traffic, 2 fits in neural circuits and 3 joint runs in finance. Each optimizer uses 3 seeds. Posterior rows use 4 bank/fit configurations in the low-dimensional tasks and 2 fits in neural circuits and finance. Financial component analyses use the separate 475-window cohort, including 3 response banks with 5 fits per bank. Supplement~\ref{app:evaluationcohorts} records selection and confirmation roles.

For low-dimensional intervals, average the fitted rules within each target and resample paired targets. Ratios use resampled mean errors. The first and second 512-target sets have independent observation and evaluation streams. All targets contribute to the primary comparison. The intervals condition on the supplied banks and fitted rules.

Neural intervals concern the complete-rhythm target population. The original analysis averages 2 fits before 9,999 target resamples. The 8-fit follow-up uses 19,999 crossed fit/target resamples with its bank and evaluation streams fixed. Financial intervals use 20,000 crossed date/stock resamples, averaging fitted rules within each input. Fixed blocks of 3 successive starts examine within-day dependence, and a sensitivity analysis resamples joint runs globally. The financial cohort comprises five stocks and five dates, four of them consecutive, that informed retrospective development. Its intervals describe that observed cohort. Intervals throughout are nominal and unadjusted for multiple comparisons. Crossed factors and overlapping windows retain their dependence through the stated resampling units.

The supplied data tables separate development, retrospective algorithm comparisons, set A, set B and the new neural cohort. They retain method identities, per-fit readouts and incomplete results. Broader exploratory records, including applications outside the present four-task scope, remain in the research repository. Application selection is documented in Supplement~\ref{app:selectionhistory}. The accompanying \path{supplementary_design_data.zip} contains the fixed candidate coordinates, group identifiers, normalization, simulated confirmation targets and financial initialization/configuration files, with a file manifest and README. These design data identify the instances used here. The supplied CSV files and result archives contain the numerical comparisons, and the manuscript source archive builds the main paper and supplement. Separate audit materials include selected experiment runners, implementation files and a dependency lock file. The new traffic confirmation package supplies both runners, protocols and final target, return and evaluation arrays. Additional project code is required to execute them, and the tandem comparison also uses the training response pool and fitted models retained in the project repository. Raw SZ000001 and Nasdaq TotalView-ITCH observations are also excluded. Repeating the native simulations requires those code dependencies and access to the corresponding market records. A complete external code release and an authorized raw-data access route are not yet available with this revision.

\begin{longtable}{@{}llrr@{}}
\caption{Memory comparisons on the 2 independent low-dimensional target sets. Expanded memories contain both direction-specific and joint-response records. Decision-data memories additionally receive the 512-query, 4-repeat decision-training records and use the corresponding refitted surrogate model. Every cell averages all 512 targets and 4 rules. Set-B decision-data controls were added after review without refitting. Separate sets are not pooled.}\label{tab:cachereplication}\\
\toprule Task & Procedure & Set A MSE & Set B MSE\\\midrule
\endfirsthead\toprule Task & Procedure & Set A MSE & Set B MSE\\\midrule\endhead
Predator--prey & Full local variant & 0.04193 & 0.05186\\
Predator--prey & Expanded empirical & 0.05900 & 0.08674\\
Predator--prey & Expanded surrogate & 0.16930 & 0.16838\\
Predator--prey & Decision-data empirical & 0.05815 & 0.08663\\
Predator--prey & Decision-data surrogate & 0.16186 & 0.17272\\
Traffic & Full local variant & 0.95217 & 0.97262\\
Traffic & Expanded empirical & 0.97676 & 0.99736\\
Traffic & Expanded surrogate & 0.95154 & 0.97255\\
Traffic & Decision-data empirical & 0.97674 & 0.99732\\
Traffic & Decision-data surrogate & 0.95545 & 0.97705\\
\bottomrule\end{longtable}

\subsection{Fit sensitivity and the error distribution}\label{app:distribution}
\begin{longtable}{@{}llrrrrr@{}}
\caption{Strong-memory error distributions on set B. Fits are averaged within each target before computing summaries. Wins count targets where the full local variant has lower error than the control in that row. All targets enter the primary means.}\label{tab:cachedistribution}\\
\toprule Task & Procedure & Mean & Median & 90th & 99th & Full local variant wins\\\midrule
\endfirsthead\toprule Task & Procedure & Mean & Median & 90th & 99th & Full local variant wins\\\midrule\endhead
Predator--prey & Full local variant & 0.05186 & 0.01151 & 0.08263 & 0.76870 & ---\\
Predator--prey & empirical memory & 0.08674 & 0.01513 & 0.10215 & 0.96796 & 472/512\\
Predator--prey & surrogate memory & 0.16838 & 0.02563 & 0.22687 & 3.28367 & 511/512\\
Traffic & Full local variant & 0.97262 & 0.94915 & 1.21557 & 1.38986 & ---\\
Traffic & empirical memory & 0.99736 & 0.97653 & 1.23914 & 1.42999 & 463/512\\
Traffic & surrogate memory & 0.97255 & 0.94925 & 1.20410 & 1.42106 & 245/512\\
\bottomrule\end{longtable}

On set B, dropping the 6 largest absolute paired gaps as a diagnostic leaves predator--prey improvements of 27.65\% against empirical memory and 67.39\% against surrogate memory. Dropping 26 leaves 24.21\% and 61.72\%. The primary estimates retain all targets. These trimmed subsets locate the source of the mean difference.

Table~\ref{tab:bankresults} shows the current local-memory comparisons separately for the 2 response banks. Target-bootstrap intervals use the conditional aggregation defined at the start of Supplement~\ref{app:materials}. For the neural full mixture, the 2 fit errors are 0.64227 and 0.63109, compared with empirical error 0.74887. Surrogate-only errors are 1.11553 and 0.93429. Both original fits show the same direction. Supplement~\ref{app:neuralfitstability} extends the initialization analysis.

\begin{longtable}{@{}lllrrr@{}}
\caption{Local-proposal quality by response bank on set B. Each bank averages its 2 fits within a target. Percentage changes are Full local variant/control minus one, with target-bootstrap 95\% intervals conditional on that bank. Two banks do not estimate population-level training variability.}\label{tab:bankresults}\\
\toprule
Task & Bank & Control & Full local variant & Control & Relative change (\%)\\\midrule
\endfirsthead
\toprule
Task & Bank & Control & Full local variant & Control & Relative change (\%)\\\midrule
\endhead
Predator--prey & 0 & Empirical & 0.03991 & 0.07633 & -47.71 [-60.40, -31.09]\\
Predator--prey & 0 & Surrogate & 0.03991 & 0.15183 & -73.71 [-79.33, -65.90]\\
Predator--prey & 1 & Empirical & 0.06381 & 0.09715 & -34.32 [-42.15, -24.77]\\
Predator--prey & 1 & Surrogate & 0.06381 & 0.18493 & -65.50 [-74.72, -56.79]\\
Traffic & 0 & Empirical & 0.97129 & 0.99750 & -2.63 [-2.90, -2.36]\\
Traffic & 0 & Surrogate & 0.97129 & 0.97310 & -0.19 [-0.31, -0.07]\\
Traffic & 1 & Empirical & 0.97394 & 0.99722 & -2.33 [-2.62, -2.06]\\
Traffic & 1 & Surrogate & 0.97394 & 0.97200 & +0.20 [+0.08, +0.33]\\
\bottomrule
\end{longtable}

\begin{longtable}{@{}lrrrrr@{}}
\caption{Distribution of predator--prey errors on set A. All 4 fits are averaged within each of the 512 targets before computing quantiles. The full local variant has lower error on 198 targets and higher error on 314. Its lower mean is concentrated in the upper tail, the median does not improve. This is the earlier multiple-proposal variant, not the primary rule.}\label{tab:postdistribution}\\
\toprule
Procedure & Mean & Median & 90th & 99th & Maximum\\\midrule
\endfirsthead
\toprule
Procedure & Mean & Median & 90th & 99th & Maximum\\\midrule
\endhead
Full local variant & 0.04193 & 0.01166 & 0.08538 & 0.53798 & 0.94221\\
MAF, 32-sample mean & 0.07256 & 0.01131 & 0.08684 & 0.94738 & 8.97695\\
\bottomrule
\end{longtable}

For predator--prey versus the R4-trained MAF 32-sample mean on set A, the two bank-level mean pairs are 0.03779 versus 0.06982 and 0.04606 versus 0.07530. A post-hoc sensitivity analysis removes the largest absolute paired target differences only to locate the source of the aggregate gain. Removing 6 of 512 targets changes the mean ratio from $-42.22\%$ to $-2.39\%$. Removing 26 changes it to $+5.41\%$. All 512 targets remain in the primary result. The post-hoc paired-gap analysis shows that the mean advantage is concentrated in a small set of large differences. The accompanying CSVs retain all per-bank and sensitivity rows.

The financial direct score-9, full-teacher model has mean energies 34.00169, 34.06498 and 34.06946 across the 3 response banks, averaging 5 fits within each bank. All use the same 475 windows and the date/stock analysis defined above.

\subsection{Additional local mechanism and financial timing analyses}\label{app:reviewmechanism}
The candidate-count-matched controls in Table~\ref{tab:mechanism} were added after independent manuscript review. They use all set-B observations, 4 existing fits and the same independent response streams. Simulating only previously unevaluated returned parameters required 228,800 predator--prey and 362,624 traffic responses, with no execution failures. Repeated parameters reuse their matching saved responses. Removing $c_B$ or $C_B$ leaves the original ridge and radius unchanged. The joint-response mean control removes both. The isotropic control replaces $C_B$ by $\operatorname{tr}(C_B)I/p$ and removes $c_B$. The other controls use 0.5 and 2 times the selected original mean-step radius. All replacements remain reported. The single-proposal timing omits both mean/conservative proposal calculation and surrogate ranking. All other input processing and physical-parameter conversion remain included. It is interleaved with the full local variant on 1 physical core, with 5 warm-ups and 31 repetitions per target/fit.

The covariance curvature sums over response-feature directions as $C_B=R^{-1}\sum_r(B_r-\overline B)(B_r-\overline B)^\top$. Although repeats are few, its rank is not generally bounded by $R-1$ when each $B_r$ has multiple feature columns. The matched controls leave the empirical effect of these finite-repeat covariance estimates unresolved.

For the financial primary full-teacher model versus compressed surrogate selection, the 475-window paired difference is $-0.0213\%$ with descriptive crossed date/stock/block interval $[-1.0680\%,0.7433\%]$. Per-query medians from 31 timing repetitions are paired within input, bank and fit. Over 7,125 such pairs, surrogate minus primary-direct time has mean 5.053 microseconds, median 4.727, interquartile range 4.096--5.679 and 5th--95th percentiles 3.396--7.713. The information-matched compressed-teacher direct saving has mean 4.983, median 4.647 and interquartile range 3.987--5.579. These distributions describe this implementation and profiling session.

The historical single-run financial table assembles quality and latency from their recorded experiments. In particular, the 0.122 ms measurement in F116 comes from CPU 200 with 28 interleaved direct/posterior paths. The compressed-surrogate time in F118 is from a later compression group on CPU 54. Those historical times belong to separate profiling groups. Current comparisons place both paths in the same randomized, interleaved profiling group on CPU 54. The uncompressed-direct comparison on all 475 windows has ES 34.03495 versus 34.04538 for score-9, with score-9 relative difference $+0.0306\%$ (descriptive interval $[-0.2591\%,0.3827\%]$). Its same-group return times are 0.06552 and 0.05872 ms. Its paired quality interval spans both gains and losses from compression.

\begin{table*}[t]
\centering\normalsize
\caption{Candidate-count-matched local mechanism controls. Each row replaces only the stochastic proposal in the full local variant, keeping the anchor, other two candidates, surrogate model and gate. Term deletions keep the original radius. The two radius controls change it by the stated factor. The mean-field and term-deletion controls use the same saved joint-response records as the stochastic field. Percentages are Full local variant/row minus one with paired target-bootstrap 95\% intervals, so negative values favour the full local variant. All 512 set-B targets and 4 fits are retained. These exploratory follow-up controls were specified after the earlier set-B results. No replacement was chosen as a new primary method.}\label{tab:mechanism}
\begin{tabular}{@{}lrrrr@{}}\toprule
Replacement & \multicolumn{2}{c}{Predator--prey} & \multicolumn{2}{c}{Traffic}\\
 & Loss & Change (\%) & Loss & Change (\%)\\\midrule
Joint-response mean field & 0.05166 & +0.40 [-0.44, +1.57] & 0.97281 & -0.02 [-0.04, +0.01]\\
Without $c_B$ & 0.05162 & +0.46 [-0.29, +1.59] & 0.97260 & +0.00 [-0.02, +0.02]\\
Without $C_B$ & 0.05190 & -0.07 [-0.20, +0.04] & 0.97287 & -0.03 [-0.04, -0.01]\\
Half-radius mean proposal & 0.05702 & -9.04 [-15.89, -2.42] & 0.97409 & -0.15 [-0.18, -0.12]\\
Double-radius mean proposal & 0.05915 & -12.32 [-23.88, +3.52] & 0.97371 & -0.11 [-0.14, -0.09]\\
Isotropic curvature penalty & 0.06001 & -13.57 [-35.42, +4.03] & 0.97237 & +0.03 [-0.00, +0.05]\\
\bottomrule\end{tabular}
\end{table*}

\subsection{Local-regression controls}\label{app:regression}
\begin{table*}[t]
\centering\normalsize
\caption{Local correction versus local regression in predator--prey calibration. Both regression controls receive the full shared response bank and choose neighbour count on 205 development inputs. Errors average 2 banks. The proposed rule also averages 2 fits per bank. Every target set contains 512 observations with 64 native evaluation responses per return. Timing interleaves all compared procedures on 1 CPU and uses set B. It is separate from Table~\ref{tab:maincomparison}. Both local rows are evaluated on both sets. Details and paired intervals are in Supplement~\ref{app:regression}.}\label{tab:localregression}
\begin{tabular}{@{}lrrr@{}}\toprule
Procedure & Set A MSE $\downarrow$ & Set B MSE $\downarrow$ & Time (ms) $\downarrow$\\\midrule
Regression-adjusted response samples & 0.09885 & 0.28831 & 2.0694\\
Local regression of response means & 0.16210 & 0.46767 & 0.3655\\
Full local variant & 0.04193 & 0.05186 & 0.1269\\
Ours & 0.04263 & 0.05371 & 0.0222\\
\bottomrule\end{tabular}
\end{table*}

The regression-adjusted point control fits unit parameters $u_j$ locally to standardized individual responses $z_j$, using all 642,240 shared augmented records per bank. A second version fits the same relation to the 90,315 candidate response means. For a query $x$, let $h$ be the distance to its $K$th nearest response, enlarged by a factor $1+10^{-8}$ and floored at $10^{-12}$. The row vector is $b_j=[1,(z_j-x)^\top/h]$, with Epanechnikov weight $w_j=\max\{1-\norm{z_j-x}^2/h^2,10^{-12}\}$. Weighted ridge regression solves for a parameter-vector intercept and slopes. The intercept is unpenalized. Each slope receives $10^{-3}$ times the average diagonal of the slope Gram matrix, floored at $10^{-12}$ before multiplication. The clipped intercept is returned. Equivalently, the unregularized form is the weighted mean of locally regression-adjusted parameter samples. This defines a regression-adjusted point-return control on the augmented bank.

Neighbour counts $K\in\{32,64,128,256,512,1024,2048\}$ are selected by mean native error on the same 205 development inputs, with 64 responses per proposed return. In banks 0 and 1, individual-response regression selected 2,048 and 128 neighbours. Mean-response regression selected 32 and 256. Neither target set participates in this selection. Both controls are deterministic given a bank. We average their 2 bank results and compare with the proposed rule averaged within each bank over its 2 fits. Paired target bootstraps condition on those banks. The set-A/B median errors are 0.01439/0.01404 for sample regression and 0.01701/0.01605 for mean regression. The proposed full local variant has medians 0.01166/0.01151. Its wins against mean regression are 356/512 and 360/512. All targets remain in mean errors, including large finite errors.

The review-requested regression experiment added 629,504 native responses across development and both test sets, with no failed return group. A separate timing pass interleaved both regressions, both full-rule fits and both primary fits on CPU 62. It used 5 warm-ups and 31 repetitions per query, including normalization, nearest-neighbour search or retrieval, regression or correction, and physical-parameter conversion. Individual-response regression took 3.3525 and 0.7862 ms in the 2 banks, reflecting their selected neighbourhoods. Mean regression took 0.2848 and 0.4462 ms. The full local variant averaged 0.1269 ms and the primary rule 0.0222 ms in this profiling group. These interleaved measurements supply Table~\ref{tab:localregression}.

\subsection{Current-cohort uninformed neural controls}\label{app:neuraltrivial}
The strongest historical constant is the repeated-response empirical-memory minimizer at the mean complete training response. Its exact stored conductance vector is reused without looking at the new target responses. The random control draws one independent vector from the simulator prior for each target, without screening. The midpoint sets every unit coordinate to 0.5. Each uses the same 16 return-evaluation random streams as the current methods. The earlier single-response-selected constant is retained as a weaker control (MSE 5.1546, completeness 64.08\%). All four controls required 9,856 additional native responses, with no execution failures. The full mixture improves all 154 targets against the weaker constant, midpoint and random control. It improves 147 against the strongest constant. Return-only timings include physical conductance conversion and, for the random row, a new random draw. They exclude preparation just as other query timings do. The strongest constant has 100\% completeness, so its quality gap concerns feature values.

\subsection{Neural validity-weight sensitivity}\label{app:weight}
The primary metric gives each unavailable feature weight 5 in the augmented response. Its squared contribution is $25/18$ per missing feature, in addition to the masked numeric-value discrepancy. This chosen penalty rewards reproduction of a complete observed rhythm. Table~\ref{tab:validityweight} rescales only the saved missing-feature contributions by $(w/5)^2$. It keeps all returned parameters and native responses unchanged. The ranking of those fixed weight-5 returns reverses between weights 2.5 and 5. This diagnostic evaluates fixed returns under different utilities. The reselection experiment below changes the selected parameters with the utility.
\begin{longtable}{@{}rrrrr@{}}
\caption{Sensitivity to the neural validity weight. These are evaluation-only rescoring results for the same full-budget returns, fitting and selection retain weight 5. Intervals are paired target-bootstrap intervals conditional on the 2 fitted models, without multiplicity adjustment. The values-only row still assigns 0 to invalid numeric features and is a diagnostic, not a preferred objective.}\label{tab:validityweight}\\
\toprule
Validity weight & Empirical MSE & Mixture MSE & Mixture / empirical $-1$ (\%) & 95\% interval (\%)\\\midrule\endfirsthead
\toprule Validity weight & Empirical MSE & Mixture MSE & Relative change (\%) & 95\% interval (\%)\\\midrule\endhead
0 & 0.48169 & 0.57411 & +19.19 & [+7.89, +67.69]\\
1 & 0.49237 & 0.57661 & +17.11 & [+6.99, +56.41]\\
2.5 & 0.54848 & 0.58975 & +7.52 & [+0.53, +25.11]\\
5 & 0.74887 & 0.63668 & -14.98 & [-38.17, -2.46]\\
10 & 1.55041 & 0.82438 & -46.83 & [-67.66, -26.94]\\
\bottomrule\end{longtable}

\subsection{Equal-budget spread memories and new coverage confirmation}\label{app:densecache}

Table~\ref{tab:densecache} compares concentrated and spread designs at the same retained budget of 642,240 responses on 512 targets in each of sets A, B and C. Set C was generated after the local rules and spread allocation were fixed. Means and SDs use 2 banks for empirical memories and Ours, or 4 bank/fit configurations for surrogate selections and the full local variant.

\begin{table}[H]
\centering\normalsize
\setlength{\tabcolsep}{8pt}
\caption{Candidate coverage in predator--prey calibration. Loss (mean $\pm$ SD). Bold marks the lowest mean in each cohort.}\label{tab:densecache}
\begin{tabularx}{180mm}{@{}L{48mm}CCC@{}}
\toprule
Method & Set A & Set B & Set C\\\midrule
Spread empirical memory & \makebox[13mm][r]{$\mathbf{0.0326}$}\,$\pm$\,\makebox[18mm][l]{$\mathbf{0.0014}$} & \makebox[13mm][r]{$\mathbf{0.0420}$}\,$\pm$\,\makebox[18mm][l]{$\mathbf{0.0007}$} & \makebox[13mm][r]{$\mathbf{0.0390}$}\,$\pm$\,\makebox[18mm][l]{$\mathbf{0.000015}$}\\
Spread surrogate selection & \makebox[13mm][r]{$0.1869$}\,$\pm$\,\makebox[18mm][l]{$0.0203$} & \makebox[13mm][r]{$0.1760$}\,$\pm$\,\makebox[18mm][l]{$0.0101$} & \makebox[13mm][r]{$0.1653$}\,$\pm$\,\makebox[18mm][l]{$0.0168$}\\
Empirical memory & \makebox[13mm][r]{$0.0590$}\,$\pm$\,\makebox[18mm][l]{$0.0065$} & \makebox[13mm][r]{$0.0867$}\,$\pm$\,\makebox[18mm][l]{$0.0147$} & \makebox[13mm][r]{$0.0755$}\,$\pm$\,\makebox[18mm][l]{$0.0151$}\\
Surrogate selection & \makebox[13mm][r]{$0.1693$}\,$\pm$\,\makebox[18mm][l]{$0.0418$} & \makebox[13mm][r]{$0.1684$}\,$\pm$\,\makebox[18mm][l]{$0.0232$} & \makebox[13mm][r]{$0.1679$}\,$\pm$\,\makebox[18mm][l]{$0.0103$}\\
Full local variant & \makebox[13mm][r]{$0.0419$}\,$\pm$\,\makebox[18mm][l]{$0.0048$} & \makebox[13mm][r]{$0.0519$}\,$\pm$\,\makebox[18mm][l]{$0.0138$} & \makebox[13mm][r]{$0.0482$}\,$\pm$\,\makebox[18mm][l]{$0.0060$}\\
Ours & \makebox[13mm][r]{$0.0426$}\,$\pm$\,\makebox[18mm][l]{$0.0052$} & \makebox[13mm][r]{$0.0537$}\,$\pm$\,\makebox[18mm][l]{$0.0183$} & \makebox[13mm][r]{$0.0489$}\,$\pm$\,\makebox[18mm][l]{$0.0079$}\\
\bottomrule
\end{tabularx}
\end{table}

The equal-budget follow-up draws independent parameter locations from the same clipped $0.5+0.1N(0,I)$ design as the original anchors. The independent location seeds are 20420261012 and 20420261013. Each of 2 banks retains exactly 642,240 responses at $160{,}560$, $80{,}280$, $40{,}140$ or $20{,}070$ locations for repeat budgets 4, 8, 16 or 32. Prefix sharing between the 4 designs reduces the actual generated union to 3,211,200 training responses across both banks. This union records the exploration cost across allocations. Surrogate models use the same 2-256-unit SiLU architecture, AdamW settings and development early stopping as the earlier surrogate selection, with seeds 7 and 17 (16 fits in total). Independent native development MSE on the same 205 inputs chooses the empirical and surrogate repeat budget separately. Empirical selection chooses 4 in both banks. Surrogate selection chooses 32 and 8. The complete grid below retains every result on A and B. Training counts exclude development and evaluation responses, as for the original memory. The follow-up added 1,326,016 such responses with no failed return groups.

A second follow-up fits local surrogate regressions using only the 4-repeat spread memory. At a retrieved parameter, a weighted affine response fit uses its $K\in\{32,64,128,256\}$ nearest parameters, with Epanechnikov distance weights and ridge $10^{-3}$ times the mean slope-Gram diagonal. The intercept is unpenalized. A prepared ridge inverse gives a correction capped at $\eta\in\{0.1,0.25,0.5,1\}$ times the distance to the 64th nearest parameter. Leaving the empirical return unchanged is an additional option. All 17 choices are assessed on the 205 development inputs with 64 independent native responses. Both banks select the unchanged empirical return. The selected dense-fit rule consequently preserves the empirical return for every query.

Set C is generated only after these decisions are saved, using an untruncated prior draw for 512 new targets (design seed 20520261012), with observation seed $205200000000+128i$ and return seed $205300000000+128i+r$ for zero-based target $i$ and replicate $r$. Each return receives 64 native responses. It compares the dense empirical and development-selected dense surrogate memories with both original local variants. The original concentrated empirical and surrogate memories were subsequently evaluated on all the same targets and streams, adding 181,824 native responses without failures. Their errors are 0.07547 and 0.16790. No rule was changed in this additional comparison. It also confirms that the selected dense local-fit option returns exactly the empirical parameters. The extra disabled-correction implementation took 0.02698 ms and does not improve the memory. This overhead can simply be omitted. The original primary and full-rule increases relative to dense empirical error are 25.33\% (conditional target interval, 8.87--43.44\%) and 23.32\% (7.36--40.62\%). The new cohort confirms the lower mean error from wider coverage. Paired intervals average banks within targets and use 9,999 target resamples conditional on these designs.

\begin{longtable}{@{}lrrrr@{}}
\caption{Complete spread-memory allocation grid. All rows retain 642,240 responses per bank, MSE averages the 2 banks, and surrogate rows also average 2 fits per bank. The selected surrogate row combines the development-selected allocation in each bank.}\label{tab:densegrid}\\
\toprule Memory & Repeats & Locations & Set A MSE & Set B MSE\\\midrule\endfirsthead
\toprule Memory & Repeats & Locations & Set A MSE & Set B MSE\\\midrule\endhead
Empirical & 4 & 160560 & 0.03261 & 0.04197\\
Empirical & 8 & 80280 & 0.03455 & 0.04767\\
Empirical & 16 & 40140 & 0.03885 & 0.05223\\
Empirical & 32 & 20070 & 0.04711 & 0.07648\\
Empirical & Selected & Varies & 0.03261 & 0.04197\\
Surrogate & 4 & 160560 & 0.16670 & 0.21594\\
Surrogate & 8 & 80280 & 0.17910 & 0.18430\\
Surrogate & 16 & 40140 & 0.17608 & 0.21713\\
Surrogate & 32 & 20070 & 0.18314 & 0.16344\\
Surrogate & Selected & Varies & 0.18695 & 0.17597\\
\bottomrule\end{longtable}

The dense local-fit follow-up used no new training responses, 512 new observation responses, and 850,176 responses for development and return evaluation. No execution failures occurred.

\subsection{Profiling groups}\label{app:timinggroups}
Table~\ref{tab:timinggroups} separates independent profiling sessions. Absolute microsecond timings vary with the contemporaneous set of interleaved procedures and memory state. In particular, the full predator--prey rule varies from about 0.080 to 0.127 ms across groups. Only within-row-group ratios or paired differences are interpreted. All current groups use 1 core of the AMD EPYC 9755 system, warmed batch-1 queries and randomized method order.

\begin{longtable}{@{}L{35mm}L{17mm}L{57mm}L{58mm}@{}}
\caption{Timing groups for current comparisons. Core numbers identify pinned logical CPU indices, not different processor models.}\label{tab:timinggroups}\\
\toprule Group & CPU & Compared procedures & Reported scope\\\midrule\endfirsthead
\toprule Group & CPU & Compared procedures & Reported scope\\\midrule\endhead
Memory/local, set A & 52 & Primary, full local, empirical and surrogate selections, separate spread-memory profile & Table~\ref{tab:maincomparison}, 5 warm-ups, 31 repetitions, saved primary returns checked against the evaluation archive\\
Local mechanism, B & 112 & Full and primary rules & Predator--prey 0.0798/0.0164 ms, traffic 0.1147/0.0493 ms\\
Local regression, B & 62 & Two regressions, both local variants & Table~\ref{tab:localregression}\\
Neural confirmation & 54 & All full-budget weight-5 smoothers, surrogate, empirical and constant controls & Earlier component analysis, main-table times use the new joint profile below\\
Neural joint profile & 54 & Main-table memory controls, Ours and MAF readouts & Tables~\ref{tab:overallcomparison} and~\ref{tab:maincomparison}, identical native-evaluated returns verified\\
Neural posterior & 56 & Actual MDN/MAF draws and full mixture, separately for each fit/cohort & Table~\ref{tab:neuralnpe}, 3 posterior and 31 direct repetitions\\
Spread memory, B & 52 & Selected spread memories, both local variants & Empirical 0.0200, one proposal 0.0168, full 0.0844 ms\\
Spread memory, C & 52 & Selected spread memories, both local variants & Table~\ref{tab:densecache}\\
Financial, current & 54 & Direct and compressed surrogate selections & Section~\mbox  {V-E}, 475/75-window paired comparisons\\
Historical financial & 200 / 54 & Earlier separately profiled procedures & Supplementary full grid, no cross-group ratio\\
\bottomrule\end{longtable}

\subsection{Neural smoothing, coverage and independent confirmation}\label{app:neuralconfirmation}
The added smoothing controls use the original candidate groups and validity counts. If feature $\ell$ is available $k_{j\ell}$ times among $n_j$ responses, Laplace smoothing is $(k_{j\ell}+1)/(n_j+2)$. Global-Beta smoothing is $(k_{j\ell}+4p_\ell)/(n_j+4)$, where $p_\ell$ is the response-count-weighted training validity rate. The neighbour variants replace $p_\ell$ with its count-weighted rate among the 16, 64 or 256 nearest distinct candidate locations in unit parameter space, excluding the candidate itself. All retain empirical numeric moments. The neural-validity control replaces only validity moments with their mixture estimate. All five weights and every smoother remain in Table~\ref{tab:neuralreselection}. None is selected from test results.

The original-cohort reselection added 16,016 responses for previously unevaluated returns. The separate spread prior bank generated 69,367 responses at independent uniform unit coordinates (parameter seed 20820261012. Native seed $2080000000+i$). It has 699 complete responses. The confirmation observations use parameter seed 20920261012 and native seed $2090000000+i$, yielding 177 complete observations among 16,384 attempts. Every eligible observation is retained, without filtering on the return quality. Return evaluation uses seed $2091000000+64i+r$, where $i$ is the original prior-batch index and $r=0,\ldots,15$. The models, primary weight, entire utility grid and smoother definitions were fixed before this batch was generated. The confirmation used those fixed models and normalization. The later fixed-configuration training-seed extension is reported separately in Supplement~\ref{app:neuralfitstability}.

Evaluating the prior-memory control on the original cohort required 2,480 additional responses. The complete confirmation grid added 37,664 responses, and the fixed-parameter/surrogate-only follow-up added 7,856. No execution failed. Incomplete rhythms remain valid simulator outcomes and enter the discrepancy. At primary weight 5, the 2 mixture fits have confirmation errors 0.93089 and 0.85049, versus empirical error 1.04968. Uncertainty intervals average those fits within each target before 9,999 target resamples and condition on the fitted models. Initialization variability is assessed separately in Supplement~\ref{app:neuralfitstability}.

The empirical weight plateau was checked from returned coordinates and stored validity moments. The weight-5, weight-10 and weight-20 empirical selections are identical for all 177 confirmation targets. All selected candidates have empirical feature-validity probabilities equal to one. 2 selections have 4 stored responses, 3 have 9, 151 have 16 and 21 have 21. Thus the plateau is not confined to once-observed candidates. Independent native responses reduce their complete-response rate to 97.07\%. The probability-estimation controls test how additional information can distinguish reliability despite those all-success empirical records.

The complete timing pass interleaves all weight-5 controls on CPU 54, including physical decoding, with 5 warm-ups and 31 measured repetitions. It records returned parameters as well as timings. The exact index has different pruning costs for different score geometries. The 0.0101 ms mixture time and 0.0149 ms 16-neighbour time reflect their geometry-dependent pruning. All preparation is excluded from these query times.

\begin{longtable}{@{}lp{48mm}rrrr@{}}
\caption{Neural calibration after reselection at each validity weight. Each weight is used both to choose a parameter and to score its independent native responses. Values denotes masked numeric MSE, and Complete denotes complete returned responses. No weight or smoother is selected using these results.}\label{tab:neuralreselection}\\
\toprule Cohort & Rule & Weight & Loss & Values & Complete (\%)\\\midrule\endfirsthead
\toprule Cohort & Rule & Weight & Loss & Values & Complete (\%)\\\midrule\endhead
Original & Empirical & 1 & 0.5331 & 0.5124 & 95.82\\
Original & Empirical & 2.5 & 0.6143 & 0.5164 & 96.96\\
Original & Empirical & 5 & 0.7489 & 0.4817 & 97.65\\
Original & Empirical & 10 & 1.5691 & 0.4913 & 97.65\\
Original & Empirical & 20 & 4.8023 & 0.4913 & 97.65\\
Original & Full neural mix & 1 & 0.4760 & 0.4624 & 97.46\\
Original & Full neural mix & 2.5 & 0.5610 & 0.5297 & 98.90\\
Original & Full neural mix & 5 & 0.6367 & 0.5741 & 99.45\\
Original & Full neural mix & 10 & 0.8323 & 0.7150 & 99.74\\
Original & Full neural mix & 20 & 0.9802 & 0.9081 & 99.96\\
Original & Neural validity only & 1 & 0.5289 & 0.5162 & 97.63\\
Original & Neural validity only & 2.5 & 0.6005 & 0.5358 & 98.15\\
Original & Neural validity only & 5 & 0.6511 & 0.5209 & 99.01\\
Original & Neural validity only & 10 & 0.7832 & 0.6840 & 99.78\\
Original & Neural validity only & 20 & 1.0638 & 0.8834 & 99.90\\
Original & Laplace validity & 1 & 0.5317 & 0.5146 & 96.63\\
Original & Laplace validity & 2.5 & 0.6316 & 0.5360 & 97.04\\
Original & Laplace validity & 5 & 0.7647 & 0.5381 & 98.01\\
Original & Laplace validity & 10 & 1.4852 & 0.5743 & 98.01\\
Original & Laplace validity & 20 & 4.0227 & 0.5956 & 98.09\\
Original & Global Beta validity & 1 & 0.5338 & 0.5159 & 96.47\\
Original & Global Beta validity & 2.5 & 0.6317 & 0.5450 & 97.36\\
Original & Global Beta validity & 5 & 0.9090 & 0.5883 & 97.48\\
Original & Global Beta validity & 10 & 1.4418 & 0.5850 & 98.09\\
Original & Global Beta validity & 20 & 4.0612 & 0.6340 & 98.09\\
Original & Neighbours 16 validity & 1 & 0.5376 & 0.5151 & 95.74\\
Original & Neighbours 16 validity & 2.5 & 0.6074 & 0.5365 & 97.93\\
Original & Neighbours 16 validity & 5 & 0.7802 & 0.6072 & 98.82\\
Original & Neighbours 16 validity & 10 & 0.8494 & 0.7096 & 99.72\\
Original & Neighbours 16 validity & 20 & 1.1066 & 0.8180 & 99.84\\
Original & Neighbours 64 validity & 1 & 0.5502 & 0.5267 & 95.66\\
Original & Neighbours 64 validity & 2.5 & 0.6106 & 0.5307 & 97.61\\
Original & Neighbours 64 validity & 5 & 0.8620 & 0.6078 & 98.09\\
Original & Neighbours 64 validity & 10 & 1.1338 & 0.7595 & 99.19\\
Original & Neighbours 64 validity & 20 & 1.1474 & 0.9309 & 99.88\\
Original & Neighbours 256 validity & 1 & 0.5526 & 0.5263 & 95.01\\
Original & Neighbours 256 validity & 2.5 & 0.6169 & 0.5291 & 97.32\\
Original & Neighbours 256 validity & 5 & 0.8509 & 0.5899 & 98.13\\
Original & Neighbours 256 validity & 10 & 1.0707 & 0.7866 & 99.39\\
Original & Neighbours 256 validity & 20 & 1.2213 & 1.0229 & 99.92\\
Original & Spread prior empirical & 1 & 0.8418 & 0.7080 & 72.36\\
Original & Spread prior empirical & 2.5 & 1.5318 & 0.7022 & 72.40\\
Original & Spread prior empirical & 5 & 4.0205 & 0.7022 & 72.40\\
Original & Spread prior empirical & 10 & 13.9756 & 0.7022 & 72.40\\
Original & Spread prior empirical & 20 & 53.7956 & 0.7022 & 72.40\\
Confirmation & Empirical & 1 & 0.7309 & 0.7033 & 94.46\\
Confirmation & Empirical & 2.5 & 0.9063 & 0.7803 & 96.01\\
Confirmation & Empirical & 5 & 1.0497 & 0.7054 & 97.07\\
Confirmation & Empirical & 10 & 2.0825 & 0.7054 & 97.07\\
Confirmation & Empirical & 20 & 6.2139 & 0.7054 & 97.07\\
Confirmation & Full neural mix & 1 & 0.7306 & 0.7151 & 96.77\\
Confirmation & Full neural mix & 2.5 & 0.7800 & 0.7343 & 98.45\\
Confirmation & Full neural mix & 5 & 0.8907 & 0.7857 & 99.06\\
Confirmation & Full neural mix & 10 & 1.1111 & 0.9227 & 99.58\\
Confirmation & Full neural mix & 20 & 1.3444 & 1.0933 & 99.86\\
Confirmation & Neural validity only & 1 & 0.7282 & 0.7027 & 95.04\\
Confirmation & Neural validity only & 2.5 & 0.8100 & 0.7201 & 97.37\\
Confirmation & Neural validity only & 5 & 0.8714 & 0.7576 & 99.01\\
Confirmation & Neural validity only & 10 & 1.0715 & 0.8989 & 99.61\\
Confirmation & Neural validity only & 20 & 1.3114 & 1.0917 & 99.88\\
Confirmation & Laplace validity & 1 & 0.7237 & 0.6992 & 95.20\\
Confirmation & Laplace validity & 2.5 & 0.8362 & 0.7224 & 96.43\\
Confirmation & Laplace validity & 5 & 1.0548 & 0.7596 & 97.53\\
Confirmation & Laplace validity & 10 & 1.9520 & 0.7789 & 97.53\\
Confirmation & Laplace validity & 20 & 5.4713 & 0.7789 & 97.53\\
Confirmation & Global Beta validity & 1 & 0.7242 & 0.7000 & 95.27\\
Confirmation & Global Beta validity & 2.5 & 0.8466 & 0.7289 & 96.29\\
Confirmation & Global Beta validity & 5 & 1.0666 & 0.7675 & 97.49\\
Confirmation & Global Beta validity & 10 & 1.9520 & 0.7789 & 97.53\\
Confirmation & Global Beta validity & 20 & 5.4414 & 0.8117 & 97.56\\
Confirmation & Neighbours 16 validity & 1 & 0.7339 & 0.7057 & 94.46\\
Confirmation & Neighbours 16 validity & 2.5 & 0.8240 & 0.7320 & 97.39\\
Confirmation & Neighbours 16 validity & 5 & 1.0013 & 0.7934 & 98.59\\
Confirmation & Neighbours 16 validity & 10 & 1.0725 & 0.9156 & 99.65\\
Confirmation & Neighbours 16 validity & 20 & 1.3468 & 1.0330 & 99.82\\
Confirmation & Neighbours 64 validity & 1 & 0.7290 & 0.7027 & 94.92\\
Confirmation & Neighbours 64 validity & 2.5 & 0.8310 & 0.7246 & 96.72\\
Confirmation & Neighbours 64 validity & 5 & 1.0813 & 0.7841 & 97.67\\
Confirmation & Neighbours 64 validity & 10 & 1.2921 & 1.0096 & 99.36\\
Confirmation & Neighbours 64 validity & 20 & 1.2165 & 1.1538 & 99.96\\
Confirmation & Neighbours 256 validity & 1 & 0.7281 & 0.6984 & 94.17\\
Confirmation & Neighbours 256 validity & 2.5 & 0.8272 & 0.7198 & 96.68\\
Confirmation & Neighbours 256 validity & 5 & 1.1571 & 0.8422 & 97.67\\
Confirmation & Neighbours 256 validity & 10 & 1.2474 & 1.0355 & 99.54\\
Confirmation & Neighbours 256 validity & 20 & 1.3125 & 1.2497 & 99.96\\
Confirmation & Spread prior empirical & 1 & 1.1433 & 0.9757 & 65.36\\
Confirmation & Spread prior empirical & 2.5 & 2.0235 & 0.9757 & 65.36\\
Confirmation & Spread prior empirical & 5 & 5.1669 & 0.9757 & 65.36\\
Confirmation & Spread prior empirical & 10 & 17.7404 & 0.9757 & 65.36\\
Confirmation & Spread prior empirical & 20 & 68.0347 & 0.9757 & 65.36\\
\bottomrule\end{longtable}

\subsection{Full local-rule deletion analysis}
Table~\ref{tab:ablation} explains the choice of the primary rule. On set B, the earlier full local variant reduces predator--prey MSE from 0.07694 for fixed shrinkage to 0.05186. Removing its joint-stream proposal raises MSE to 0.05379, a 3.59\% relative full-rule reduction (unadjusted target interval, conditional on the fitted rules, 0.66--6.99\%). This comparison supports the additional candidate. The covariance controls are reported separately below.

\begin{table*}[t]
\centering\normalsize
\caption{Ablation of the earlier full local variant. All variants use the same fitted models, observations and native response streams on confirmation set B. Each pair of columns reports MSE and the percentage change of the full local variant relative to that row, with a paired target-bootstrap 95\% interval. Negative changes favour the full local variant. The primary method is the last row. Removing a proposal leaves the other proposals and scoring rule fixed. Predator--prey retrieval is empirical. Traffic bank 1 has identical mean/shrinkage/anchor candidates, so some deletions leave its return unchanged by construction. The always-select control keeps both alternatives and removes their acceptance test against the fixed-shrinkage proposal.}\label{tab:ablation}
\begin{tabular}{@{}lrrrr@{}}
\toprule
Query rule & \multicolumn{2}{c}{Predator--prey} & \multicolumn{2}{c}{Traffic}\\ & MSE $\downarrow$ & Full / row $-1$ (\%) & MSE $\downarrow$ & Full / row $-1$ (\%)\\\midrule
Full rule & 0.05186 & --- & 0.97262 & ---\\
Ordinary surrogate ranking & 0.05715 & -9.25 [-14.53, -4.89] & 0.97284 & -0.02 [-0.04, -0.01]\\
Remove mean proposal & 0.05334 & -2.78 [-7.46, +2.62] & 0.97246 & +0.02 [-0.00, +0.03]\\
Remove joint-stream proposal & 0.05379 & -3.59 [-6.99, -0.66] & 0.97405 & -0.15 [-0.18, -0.12]\\
Always select an alternative & 0.05212 & -0.50 [-2.28, +0.49] & 0.97254 & +0.01 [+0.00, +0.02]\\
Fixed-shrinkage proposal only & 0.07694 & -32.60 [-42.01, -22.74] & 0.97460 & -0.20 [-0.24, -0.17]\\
Mean proposal only & 0.05426 & -4.42 [-8.21, -1.23] & 0.97395 & -0.14 [-0.17, -0.10]\\
1 proposal (primary) & 0.05371 & -3.45 [-8.20, +2.21] & 0.97232 & +0.03 [+0.00, +0.06]\\
\bottomrule
\end{tabular}
\end{table*}

Most of the gain of the full local variant over ordinary surrogate ranking is reproduced by suppressing the conservative candidate and always selecting an alternative. The fitted acceptance gate then changes only four of 2,048 returns on set B, corresponding to two targets with opposite effects. Equal-candidate controls that remove covariance curvature, the noise cross-term or both have unresolved differences from the full local variant. The simpler primary return omits the extra candidate comparison (Supplement~\ref{app:reviewmechanism}).

Returning only the joint-stream correction gives MSE 0.04263 on set A and 0.05371 on set B, which are 1.69\% and 3.57\% above the full local variant, with conditional intervals $[-0.37\%,4.09\%]$ and $[-2.12\%,8.96\%]$. The simpler return avoids the extra comparison, with paired mechanism-group times of 0.0164 versus 0.0798 ms in predator--prey and 0.0493 versus 0.1147 ms in traffic. Both variants lose to the spread empirical memory on predator--prey. The simplified rule accelerates use of the original bank, while the spread memory improves the allocation of responses.

Reducing acceptance supervision for the full local variant from 4,096 queries with 16 responses to 512 with 4 preserves all predator--prey returns on set A and changes traffic mean error by about $-0.002\%$. This 32-fold reduction concerns nominal decision labels only. Shared response records and field development remain the main offline investment.

\subsection{Earlier default neural posterior-draw grid}\label{app:neuralnpe}
MDN and MAF use the \texttt{sbi} 0.26.0 NPE defaults on all 69,367 normalized parameter/encoded-response rows. The support prior is uniform on the 31-dimensional unit cube, but the training rows retain the original augmented design. The resulting point-return workflows are evaluated under that empirical training design. Training uses batch size 200, learning rate 0.0005, the default 10\% row validation split and stopping patience of 20 epochs. Internal validation splits rows and can contain repeated parameter groups, while external target cohorts remain separate. Both families and seeds complete training within the 3,600-second cap. The recorded GPU fit times are 276.1 and 212.3 seconds for MDN and 399.4 and 390.1 seconds for MAF, in seed-7/seed-17 order, on NVIDIA RTX PRO 5000 GPUs. These durations measure fitting after simulation. Architectures and stopping parameters are not tuned.

For each target, 32 and 256 posterior draws supply two outputs, their physical conductance mean and the sample with minimum expected response error from the existing surrogate model. The synaptic log coordinates are exponentiated before physical averaging and transformed back afterwards. Default rejection against the support prior is retained. A readout that fails its 120-second limit is recorded. No out-of-support sample is substituted and no failed target is removed from a full-cohort mean. Query repeats use 3 actual posterior draws for timing, interleaved with 31 measurements of the direct rule after warm-up on CPU 56. Their median time is reported, while the first returned parameter is used for native quality evaluation. Training and model loading are excluded. Both cohorts use 16 independent native responses at the same target-specific seeds as the memory controls. All incomplete native rhythms enter the discrepancy.

Failed timed calls occurred for MDN seed 7 on the confirmation cohort (6 calls on 1 target), for MDN seed 17 on the original cohort (6 calls on 1 target) and for MDN seed 17 on the confirmation cohort (18 calls on 3 targets).
The individual coverage rows below determine which families can be pooled under this original sampling protocol. Supplement~\ref{app:comparisonprotocol} reports extended sampling attempts, the bounded MDN mode in the extended inventory, and the current bounded-coordinate mean readouts. The record files retain every timeout, training state and return. The family table aggregates only complete procedures.

\begin{longtable}{@{}llrlrrr@{}}
\caption{Every neural NPE fit and readout. Coverage counts targets with a returned parameter, finite native evaluation and a finite timing. Full-cohort means are NA if any target is missing. Successful-target means do not replace these missing means.}\label{tab:neuralnpefits}\\
\toprule Cohort & Family & Seed & Readout & Coverage & Loss & ms\\\midrule\endfirsthead
\toprule Cohort & Family & Seed & Readout & Coverage & Loss & ms\\\midrule\endhead
Original & MDN & 7 & 32 mean & 154/154 & 4.5037 & 1.036\\
Original & MDN & 7 & 32 rank & 154/154 & 2.8109 & 1.225\\
Original & MDN & 7 & 256 mean & 154/154 & 4.6348 & 2.043\\
Original & MDN & 7 & 256 rank & 154/154 & 2.1619 & 2.937\\
Original & MDN & 17 & 32 mean & 153/154 & --- & NA\\
Original & MDN & 17 & 32 rank & 153/154 & --- & NA\\
Original & MDN & 17 & 256 mean & 153/154 & --- & NA\\
Original & MDN & 17 & 256 rank & 153/154 & --- & NA\\
Original & MAF & 7 & 32 mean & 154/154 & 5.8406 & 65.309\\
Original & MAF & 7 & 32 rank & 154/154 & 4.2034 & 65.492\\
Original & MAF & 7 & 256 mean & 154/154 & 5.3621 & 220.722\\
Original & MAF & 7 & 256 rank & 154/154 & 4.1151 & 221.605\\
Original & MAF & 17 & 32 mean & 154/154 & 4.7696 & 59.968\\
Original & MAF & 17 & 32 rank & 154/154 & 3.8904 & 60.150\\
Original & MAF & 17 & 256 mean & 154/154 & 4.8895 & 300.857\\
Original & MAF & 17 & 256 rank & 154/154 & 2.9333 & 301.742\\
Confirmation & MDN & 7 & 32 mean & 176/177 & --- & NA\\
Confirmation & MDN & 7 & 32 rank & 176/177 & --- & NA\\
Confirmation & MDN & 7 & 256 mean & 176/177 & --- & NA\\
Confirmation & MDN & 7 & 256 rank & 176/177 & --- & NA\\
Confirmation & MDN & 17 & 32 mean & 174/177 & --- & NA\\
Confirmation & MDN & 17 & 32 rank & 174/177 & --- & NA\\
Confirmation & MDN & 17 & 256 mean & 174/177 & --- & NA\\
Confirmation & MDN & 17 & 256 rank & 174/177 & --- & NA\\
Confirmation & MAF & 7 & 32 mean & 177/177 & 6.6624 & 106.078\\
Confirmation & MAF & 7 & 32 rank & 177/177 & 5.0392 & 106.264\\
Confirmation & MAF & 7 & 256 mean & 177/177 & 6.9055 & 285.713\\
Confirmation & MAF & 7 & 256 rank & 177/177 & 4.4905 & 286.603\\
Confirmation & MAF & 17 & 32 mean & 177/177 & 5.9117 & 139.751\\
Confirmation & MAF & 17 & 32 rank & 177/177 & 4.1535 & 139.936\\
Confirmation & MAF & 17 & 256 mean & 177/177 & 5.8208 & 841.729\\
Confirmation & MAF & 17 & 256 rank & 177/177 & 3.1492 & 842.618\\
\bottomrule\end{longtable}

\begin{longtable}{@{}llrrrr@{}}
\caption{Error and latency distributions for complete neural posterior families. Median error averages the 2 fits within each target before taking the median. Relative error is mixture/NPE minus one, in percent, with a conditional 95\% interval from 9,999 target resamples, negative favours the mixture. The timing columns summarize target/fit medians of 3 actual posterior readouts. These intervals condition on the 2 fitted models and are not multiplicity adjusted.}\label{tab:neuralnpedistribution}\\
\toprule Cohort & Readout & Median error & Relative error [95\% CI] & Median ms & 95th ms\\\midrule\endfirsthead
\toprule Cohort & Readout & Median error & Relative error [95\% CI] & Median ms & 95th ms\\\midrule\endhead
Original & MAF, 32 mean & 1.3516 & -88.00 [-94.62, -79.62] & 36.63 & 101.14\\
Original & MAF, 32 rank & 1.7343 & -84.27 [-93.00, -73.13] & 36.81 & 101.31\\
Original & MAF, 256 mean & 0.7744 & -87.58 [-94.35, -79.05] & 109.82 & 543.15\\
Original & MAF, 256 rank & 0.9528 & -81.93 [-91.97, -69.10] & 110.69 & 544.04\\
Confirmation & MAF, 32 mean & 1.8008 & -85.83 [-93.70, -77.31] & 36.65 & 219.31\\
Confirmation & MAF, 32 rank & 1.1935 & -80.62 [-91.11, -69.92] & 36.83 & 219.51\\
Confirmation & MAF, 256 mean & 2.1117 & -86.00 [-93.76, -77.55] & 112.24 & 940.40\\
Confirmation & MAF, 256 rank & 1.1044 & -76.68 [-88.82, -64.93] & 113.12 & 941.33\\
\bottomrule\end{longtable}

\begin{table*}[t]
\centering\normalsize
\caption{Default posterior point readouts in neural calibration. Models share 69,367 training responses. Mean averages sampled physical parameters. Rank selects a sample using the shared surrogate score. Each number specifies actual posterior draws. Complete families average 2 fits. Incomplete families have no pooled mean. Quality uses 16 new responses at completeness weight 5. Median query times include sampling and decoding. Ratios use the mixture measured in the same profiling group. All fitting and readout failures are recorded in Supplement~\ref{app:neuralnpe}.}\label{tab:neuralnpe}
\begin{tabular}{@{}lrrrrr@{}}\toprule
Procedure & Original MSE & \multicolumn{4}{c}{Confirmation cohort}\\
 & (154 targets) & Loss & Complete (\%) & Median (ms) & Median/Ours\\\midrule
NPE--MDN (incomplete family) & --- & --- & --- & --- & ---\\
NPE--MAF, 32 mean & 5.3051 & 6.2871 & 66.54 & 36.65 & 4499$\times$\\
NPE--MAF, 32 rank & 4.0469 & 4.5964 & 73.76 & 36.83 & 4521$\times$\\
NPE--MAF, 256 mean & 5.1258 & 6.3631 & 66.86 & 112.24 & 13777$\times$\\
NPE--MAF, 256 rank & 3.5242 & 3.8199 & 77.75 & 113.12 & 13885$\times$\\
Ours, full mixture & 0.6367 & 0.8907 & 99.06 & 0.00815 & 1$\times$\\
\bottomrule\end{tabular}
\end{table*}

\subsection{Neural target distributions and individual fits}\label{app:neuraldistribution}
The follow-up analysis uses the saved native responses, with no new model or simulator runs. The complete weight grid is selected and evaluated at matching weights. Its numeric and validity contributions are reconstructed as numeric error plus the recorded weight-5 penalty times $(\omega/5)^2$. The same 9,999 target-bootstrap draws as the confirmation analysis (seed 208888) provide paired intervals. At weight 5, the mean gain of the mixture over empirical selection occurs despite 133 worsened targets and a higher median. Reductions in the high-error part of the distribution dominate the average. Against the 16-neighbour smoother, wins and losses are nearly balanced, with unresolved differences for both fitted surrogates. These are the original 2 fitted rules. The later 8-fit follow-up below estimates initialization variability separately.

\begin{longtable}{@{}rllrrr@{}}
\caption{Target-error distributions on neural confirmation. Errors use the declared weight in each row for selection and evaluation. Fits are averaged within the 177 targets before taking medians or counting wins/ties/losses (W/T/L). Relative error is mixture/control minus one in percent, with a conditional 95\% target interval. Ties use absolute tolerance $10^{-12}$.}\label{tab:neuraldistributions}\\
\toprule Weight & Control & Mixture median & Control median & W/T/L & Relative error [95\% CI]\\\midrule\endfirsthead
\toprule Weight & Control & Mixture median & Control median & W/T/L & Relative error [95\% CI]\\\midrule\endhead
1 & Empirical & 0.0686 & 0.0548 & 31/55/91 & -0.04 [-2.81, +3.89]\\
1 & 16 neighbours & 0.0686 & 0.0577 & 41/62/74 & -0.45 [-3.70, +2.40]\\
2.5 & Empirical & 0.0876 & 0.0577 & 37/28/112 & -13.94 [-39.54, -0.10]\\
2.5 & 16 neighbours & 0.0876 & 0.0853 & 73/41/63 & -5.34 [-17.12, +0.78]\\
5 & Empirical & 0.1330 & 0.0577 & 37/7/133 & -15.15 [-37.48, -3.18]\\
5 & 16 neighbours & 0.1330 & 0.1468 & 85/9/83 & -11.05 [-34.36, +5.25]\\
10 & Empirical & 0.2440 & 0.0577 & 36/3/138 & -46.65 [-68.07, -26.90]\\
10 & 16 neighbours & 0.2440 & 0.2200 & 67/2/108 & +3.59 [-7.38, +22.08]\\
20 & Empirical & 0.3597 & 0.0577 & 38/1/138 & -78.36 [-88.10, -65.24]\\
20 & 16 neighbours & 0.3597 & 0.2687 & 27/1/149 & -0.18 [-22.51, +30.48]\\
\bottomrule\end{longtable}

\begin{longtable}{@{}rlrrrr@{}}
\caption{Individual surrogate fits on neural confirmation at weight 5. Relative error is mixture/control minus one in percent. Intervals resample targets within each fixed fit, not a population of training seeds. W/T/L counts improved, tied and worsened targets.}\label{tab:neuralperfitcontrols}\\
\toprule Seed & Control & Mixture MSE & Control MSE & Relative error [95\% CI] & W/T/L\\\midrule\endfirsthead
\toprule Seed & Control & Mixture MSE & Control MSE & Relative error [95\% CI] & W/T/L\\\midrule\endhead
7 & Empirical & 0.9309 & 1.0497 & -11.32 [-33.19, +1.95] & 33/8/136\\
17 & Empirical & 0.8505 & 1.0497 & -18.98 [-43.27, -6.87] & 37/24/116\\
7 & 16 neighbours & 0.9309 & 1.0013 & -7.04 [-31.00, +13.33] & 75/16/86\\
17 & 16 neighbours & 0.8505 & 1.0013 & -15.07 [-38.85, +0.43] & 96/23/58\\
\bottomrule\end{longtable}

The two recorded full-surrogate fits took 26.2293 and 37.8925 seconds, excluding generation of the shared responses. Neighbour smoothing fits no neural model but still performs offline neighbour searches and prepares its exact query index. No paired complete preparation-time comparison was run. The historical construction cost of the enriched candidate design is not priced by the 69,367-retained-response prior-spreading control. The reported coverage contrast compares the supplied enriched design with uniform-prior sampling.

In the earlier neural NPE timing group, the direct per-query median is 0.008147 ms after pooling the target/fit medians. The corresponding MAF medians are 36.6501, 36.8333, 112.2406 and 113.1243 ms for 32-mean, 32-rank, 256-mean and 256-rank. Ratios of these medians are 4,499, 4,521, 13,777 and 13,885, whereas the medians of the individual paired ratios are 4,756, 4,779, 13,861 and 13,992. These summaries differ from ratios of mean times because of the sampling tail.

\subsection{Minimum standalone response-budget control}\label{app:minimalcache}
The primary joint field consumes 40 responses at each of 10,035 anchors, comprising 4 retrieval, 4 joint-center and 32 paired perturbation responses. To compare with this 401,400-response minimum, the control takes the first 100,350 independently spread locations of each existing bank and the first 4 responses per location. The prefix and repeat count are fixed arithmetically, without selection on target errors. All 512 existing set-C observations and their original 64-response return streams are used. The experiment adds 23,360 return-evaluation responses, reusing 659 exactly matching parameter/target response groups. It adds no training responses and has no execution failures. This follow-up uses the already examined set-C cohort.

The two prefix banks have MSE 0.04224 and 0.04160. Their mean 0.04192 is below the primary MSE 0.04894. The correction is 16.75\% higher, with a 95\% interval of 2.72--30.32\% from 19,999 paired target resamples (seed 21320261012), averaging banks/fits within targets. Correction improves 289 targets and worsens 223 despite its higher mean. Timings interleave both prefix banks and all 4 fixed local fits on CPU 55, using 5 warm-ups and 31 repeats. Mean times are 0.01350 versus 0.01397 ms. This tests retained response consumption. Inherited full-rule development, model preparation and historical design work are still additional costs.

\subsection{Training-initialization stability on the neural confirmation cohort}\label{app:neuralfitstability}
The added seeds are 27, 37, 47, 57, 67 and 77, specified together before fitting. They use exactly the original full-budget training and early-stopping split, normalization and 3-layer network, AdamW learning rate $5\times10^{-4}$ and weight decay $10^{-4}$, batches of 512, gradient clipping at 5, and at most 12,000 steps. Validation occurs every 200 steps with patience 10 checks. All six complete within the 1,200-second per-fit limit. Their recorded 1-CPU fit times range from 34.0 to 42.7 seconds after the shared responses exist. No confirmation observation participates in fitting or stopping. Full mixing, validity-only mixing and surrogate-only scoring are all evaluated at the existing $\tau=4$ and primary completeness weight 5. All eight models, including seeds 7 and 17, are evaluated as individual returns.

All 177 existing confirmation targets remain. New returns receive 16 native responses at the same $2091000000+64i+r$ streams, with $i$ the original observation-pool index. Exactly matching prior evaluations are reused. The follow-up adds 7,136 native responses and has no execution failures. It estimates initialization variability on the fixed bank and existing target cohort.

Intervals use 19,999 resamples (seed 21220261012). Seed-only intervals resample 8 model indices and hold all target outcomes fixed. Target-only intervals average fits and resample 177 target indices. Crossed intervals independently resample fit and target indices and then compute the ratio of average errors. The untrained empirical and neighbour controls do not vary with model seed. The intervals condition on the training bank, evaluation streams and complete-rhythm population. The crossed interval against the neighbour smoother spans zero, retaining uncertainty across both fits and targets.

\begin{longtable}{@{}llrrrr@{}}
\caption{Neural initialization stability across 8 fits. Relative mean-error changes and 95\% intervals are percentages, negative favours the row procedure. Seed-only intervals condition on the targets, target-only intervals condition on the fitted models, and crossed intervals resample both. Empirical and 16-neighbour control means are 1.04968 and 1.00135.}\label{tab:neuralseedstability}\\
\toprule Procedure & Control & Change (\%) & Seed-only CI & Target-only CI & Crossed CI\\\midrule\endfirsthead
\toprule Procedure & Control & Change (\%) & Seed-only CI & Target-only CI & Crossed CI\\\midrule\endhead
Full mix & Empirical & -16.20 & [-17.89, -14.37] & [-38.67, -4.52] & [-38.74, -4.28]\\
Full mix & 16-neighbour & -12.16 & [-13.93, -10.24] & [-34.70, +2.78] & [-34.75, +3.31]\\
Validity only & Empirical & -16.22 & [-18.09, -14.35] & [-38.00, -5.28] & [-38.08, -5.07]\\
Validity only & 16-neighbour & -12.17 & [-14.14, -10.22] & [-34.58, +2.76] & [-34.85, +3.49]\\
Surrogate only & Empirical & +18.95 & [+8.74, +29.71] & [+2.45, +44.79] & [-1.33, +50.82]\\
Surrogate only & 16-neighbour & +24.69 & [+13.99, +35.97] & [-3.46, +79.70] & [-5.29, +82.74]\\
\bottomrule\end{longtable}

\begin{longtable}{@{}llrll@{}}
\caption{Every fitted neural rule in the initialization follow-up. Error changes are row/control minus one, in percent, with 95\% target-bootstrap intervals conditional on that individual model. No seed is discarded, these are not ensemble predictions.}\label{tab:neuralallfits}\\
\toprule Procedure & Seed & Loss & Change vs empirical [CI] & Change vs 16-neighbour [CI]\\\midrule\endfirsthead
\toprule Procedure & Seed & Loss & Change vs empirical [CI] & Change vs 16-neighbour [CI]\\\midrule\endhead
Full mix & 7 & 0.93089 & -11.32 [-33.18, +1.98] & -7.04 [-30.64, +13.13]\\
Full mix & 17 & 0.85049 & -18.98 [-43.41, -6.78] & -15.07 [-38.67, +0.45]\\
Full mix & 27 & 0.84779 & -19.23 [-43.90, -6.80] & -15.34 [-40.42, +1.13]\\
Full mix & 37 & 0.88491 & -15.70 [-37.92, -4.08] & -11.63 [-34.29, +5.26]\\
Full mix & 47 & 0.85835 & -18.23 [-42.58, -5.81] & -14.28 [-37.88, +0.95]\\
Full mix & 57 & 0.89005 & -15.21 [-37.54, -2.75] & -11.11 [-33.18, +4.84]\\
Full mix & 67 & 0.90530 & -13.75 [-35.48, -0.85] & -9.59 [-31.39, +8.03]\\
Full mix & 77 & 0.86886 & -17.23 [-40.96, -4.83] & -13.23 [-36.08, +2.64]\\
Validity only & 7 & 0.90598 & -13.69 [-34.85, -1.32] & -9.52 [-32.51, +9.07]\\
Validity only & 17 & 0.83683 & -20.28 [-45.63, -8.04] & -16.43 [-40.44, -1.50]\\
Validity only & 27 & 0.89431 & -14.80 [-36.11, -3.94] & -10.69 [-35.80, +11.25]\\
Validity only & 37 & 0.92031 & -12.32 [-31.68, -2.28] & -8.09 [-31.10, +14.57]\\
Validity only & 47 & 0.85265 & -18.77 [-43.47, -6.48] & -14.85 [-38.55, +0.29]\\
Validity only & 57 & 0.87474 & -16.67 [-40.17, -4.21] & -12.64 [-34.91, +2.14]\\
Validity only & 67 & 0.90068 & -14.19 [-35.90, -1.94] & -10.05 [-32.00, +7.11]\\
Validity only & 77 & 0.85016 & -19.01 [-43.94, -6.54] & -15.10 [-38.65, -0.13]\\
Surrogate only & 7 & 1.22373 & +16.58 [-9.02, +37.34] & +22.21 [-11.21, +58.26]\\
Surrogate only & 17 & 1.18337 & +12.74 [-8.31, +50.03] & +18.18 [-10.55, +77.76]\\
Surrogate only & 27 & 1.17805 & +12.23 [-9.58, +42.93] & +17.65 [-11.16, +69.19]\\
Surrogate only & 37 & 1.42352 & +35.62 [+8.93, +91.55] & +42.16 [+4.75, +137.58]\\
Surrogate only & 47 & 1.02375 & -2.47 [-21.13, +12.20] & +2.24 [-22.75, +31.73]\\
Surrogate only & 57 & 1.06819 & +1.76 [-15.09, +17.27] & +6.67 [-17.49, +40.69]\\
Surrogate only & 67 & 1.39754 & +33.14 [+4.18, +71.65] & +39.57 [+0.24, +108.53]\\
Surrogate only & 77 & 1.49084 & +42.03 [+9.86, +115.69] & +48.88 [+7.78, +163.07]\\
\bottomrule\end{longtable}

\section{Current-cohort comparison protocol}\label{app:comparisonprotocol}
Table~\ref{tab:fourtaskcomplete} reports the full four-task comparison, with loss above query time in each cell. Table~\ref{tab:maincomparison} retains the extended task-specific inventory. Loss is always evaluated using new native simulator responses at the returned parameter. The metric is standardized MSE for predator--prey and traffic, completeness-weighted standardized MSE for neural circuits, and the energy loss of Eq.~(19) for financial responses. The same target, transformation, parameter bounds and evaluation stream are used for all methods within a task. The supplied \path{comparison_summary.csv} is the numeric source for these tables and Table~\ref{tab:maincomparison}. \path{comparison_run_means.csv} retains their statistical units. Loss means in the primary comparison have four decimal places. Only SD precision is increased when four decimals would hide nonzero variation. The main comparison reports all computation times in milliseconds. Historical tables use their explicitly labelled units.

\paragraph{Common iterative optimizers}
The native comparison uses PSO, CMA-ES, GP-BO and TuRBO-1 on all 512 predator--prey, 512 traffic and 177 neural targets. Each method has optimizer seeds 7, 17 and 27, giving 14,412 search queries. The shared unit-box centre is evaluated first and counts toward the limit of 128 candidate evaluations. Every search objective averages 4 native responses. A 600-second deadline limits the complete query, including optimizer construction and input-to-parameter computation. Code loading and independent return evaluation are outside this timer. The comparison evaluates the best return available at these caps. Financial runs retain the existing, larger 512-evaluation/600-second budget on 75 windows and the same 3 seeds.

The scikit-opt PSO implementation uses population 10, inertia 0.8 and cognitive/social weights 0.5. The pycma CMA-ES implementation uses population 10 and initial unit-coordinate step 0.30. BoTorch GP-BO uses five initial random points and noisy log expected improvement. The original-author TuRBO-1 code uses $2p$ initialization points, where $p$ is the parameter dimension. There is no selection of settings or seeds from test loss. The main tables report the best valid incumbent at the stopping budget. A numerical simulator failure is logged and rejected during search. A rejection value is never reported as an evaluation loss. The per-query archive retains candidate histories, stopping reasons, returned parameters, elapsed times and independent response losses.

Objective streams are shared across candidates and methods within a query/run. Their seed is $2151000000+10000000t+10000s+8i$, with task index $t\in\{0,1,2\}$, optimizer seed $s\in\{7,17,27\}$ and target index $i$. Independent evaluation uses the original set-A stream $192200000000+10000000t'+256i$ with the existing task codes $t'=1,2$ for predator--prey and traffic, respectively. For neural circuits it uses $2091000000+64j+r$, where $j$ is the original prior-pool index and $r=0,\ldots,15$. Evaluation randomness never enters optimization. All finite evaluations remain in the complete-cohort mean, regardless of whether they improve on Ours.

\paragraph{Traffic spread memory}
Each of 2 new independent banks contains 120,420 uniform unit-box parameter draws and 4 responses per draw. Seeds 21520261015 and 21520261016 generate the locations. This uses 481,680 retained responses per bank, matching the expanded traffic bank without local-perturbation clusters. The $R=4$ allocation and empirical score are fixed before evaluation. No additional grid is selected on set A. Training-only transforms are inherited unchanged. All 512 set-A targets receive 64 independent responses at each returned parameter using the same streams as the other methods. The new training and evaluation totals are 963,360 and 65,536 responses. Both runs complete with loss values of 0.97079045 and 0.97166183. Exact retrieval, normalization and physical conversion take mean 2.2305 ms across the banks. The larger candidate set increases this index workload.

\paragraph{Unified neural timing}
A single profile on CPU 54 interleaves empirical, surrogate, spread and 16-neighbour selection, Ours, and the original unit-coordinate MAF mean/ranked readouts on all 177 targets. Prepared methods use 5 warm-ups and 31 measurements. Each posterior readout uses 32 actual samples, with 3 repeated timed queries. The shared draw cost is charged in full to each readout. Every fast return and the first posterior return is checked against its saved native-evaluation parameter, giving 2,478 verified parameter checks. The first posterior draw determines the evaluated parameter. Timing repetitions do not select a different return. Ours has mean time $0.0099465\pm0.0008547$ ms across the 2 fit profiles. Its unchanged mean loss is $0.8906908\pm0.0568546$. The full numeric timing data accompany the manuscript.

\paragraph{Posterior fitting and point decisions}
The low-dimensional \texttt{sbi} comparison retains the finite model checkpoint with the lowest internal validation loss, including when training stops at its 3,600-second cap or a numerical error. Each bank provides 361,260 predator--prey or 280,980 traffic training rows, capping retained responses at 4 per parameter and excluding repeated development-selected groups. CMPE and sCM use these same rows. The 10 unsuccessful R4 configurations from the earlier grid are rerun with their original bank, seed, architecture and optimizer defaults, comprising 3 predator--prey MDN fits, 4 predator--prey NSF fits, 1 predator--prey NPSE fit and 2 traffic NSF fits. Checkpoints are selected using development loss only. Every available model is evaluated on all 512 targets with 32 actual posterior draws per query, the physical sample mean, and independent $R=64$ responses. Earlier successful fits already retain their best validation checkpoint and are reused. The source records contain stopping reasons, epochs, validation losses and full target-level responses.

For the original neural MDN in unit coordinates, extending rejection sampling from 120 to 600 seconds still leaves 176/177 and 174/177 complete targets for the 2 models. A separate vectorized slice-MCMC readout uses the library defaults of 20 chains, 200 warmup steps, thinning 1 and resampling initialization. The 32-draw physical mean is evaluated on every target for both original MDN models. Each model completes 176/177 queries. The same remaining target exceeds 600 seconds. Chain initialization and warmup enter the query timer. These unsuccessful sampling procedures have no full-cohort loss estimate.

MDN (MAP) instead optimizes the learned unit-coordinate conditional density within $[0,1]^{31}$. The label denotes a mode in unit-box parameter coordinates. A query evaluates the unit-box centre and 64 scrambled Sobol points, selects the 8 highest-density starts, and runs bounded L-BFGS-B for at most 200 iterations per start. The fixed settings are function tolerance $10^{-10}$, gradient tolerance $10^{-6}$ and at most 20 line-search steps. Stored network weights are cast to double precision for density and gradient calculations. The best finite density candidate is returned and converted to physical conductances. These settings are fixed before any MAP target evaluation. All targets use this same procedure. The 600-second cap includes input encoding, initialization, every optimization step and physical conversion. Timings use the median of 3 identical deterministic returns. Quality uses 16 independent native responses at the first return. The MDN models are unchanged.

\paragraph{Bounded-coordinate completion and consistency models}
All five \texttt{sbi} families are fitted on the full neural bank and on the financial source-plus-public simulation bank using seeds 7 and 17. The neural inputs are the original 36 encoded feature/validity coordinates, with 69,367 response rows. Financial fitting uses 16,848 parameter/condition pairs and the mean of 4 responses per pair. Its 48 conditioning coordinates contain the 40-coordinate current book and 8 simulated response residuals during training. At query time, the latter are replaced by the prediction from the common training-only future-response model. The response predictor is the fixed seed-7 model from the original financial benchmark. The two new fitting seeds vary the posterior model only. The three financial optimizer runs vary prediction and search seeds jointly. No subsequent evaluation response is a conditioning input.

The financial posterior bank combines 256 SZ000001 source conditions from 2 January--2 September 2019 with 95 conditions from the five Nasdaq stocks on 8 December 2025. Each condition has 48 independently sampled uniform unit-box parameters and 4 responses per parameter, giving $(256+95)\times48=16,848$ pairs and 67,392 simulated responses. All seven posterior families use this bank. The response teacher for Ours and surrogate selection instead uses the fixed 116-candidate dictionary on 64 source and 20 target conditions, giving 9,744 pairs and 38,976 responses. These 84 conditions are a subset of the posterior bank's 351 conditions, but none of the 116 dictionary vectors equals a posterior training vector. The banks therefore share conditions rather than candidate/condition simulation records. The direct-score and future-response models use 1,159 source real windows from the same 2019 period plus 365 target real windows from 8 December 2025. These 1,524 measured windows supply outcome labels, separately from either simulation bank. The comparison shares query information and response scales while retaining these different offline training designs.

For each parameter coordinate, fitting transforms $u$ to $z=\log\{u/(1-u)\}$ after clipping training values to $[10^{-6},1-10^{-6}]$. Sampling applies the inverse sigmoid before physical conversion. The transformed prior is the logit image of the original uniform unit-box prior. The change prevents rejection caused solely by samples outside the parameter box. These bounded-coordinate fits are recorded separately from the original untransformed fits. Architectures and optimizer settings remain the \texttt{sbi} defaults. A 3,600-second fit cap and the lowest finite internal validation loss determine the checkpoint, without evaluation-target selection. The original box-coordinate MDN and MAF procedures remain separately identified in Table~\ref{tab:maincomparison}.

CMPE~\cite{schmitt2024consistency} and the conditional sCM implementation~\cite{lu2025consistency} use BayesFlow 2.0.14, Keras 3.15.1 and the PyTorch backend. They use the default TimeMLP networks and default noise schedules. Parameters are logit transformed in all tasks. Parameters and conditioning inputs are standardized on a random 90\% training split, leaving 10\% for monitoring. Each fit uses 10,000 Adam updates at learning rate $10^{-3}$ with batch size 256 and a 3,600-second cap. The final finite checkpoint is used, so no evaluation score or changing discretization loss selects a training step. CMPE receives \path{total_steps}=10,000 for its progressive consistency schedule. Sampling uses the implementation defaults of 11 steps for CMPE and 15 for sCM. The low-dimensional fits use both original banks and seeds 7 and 17. Neural and financial fits use seeds 7 and 17.

Each query draws 32 samples and averages physical parameters, including exponentiation before averaging log-coordinate synaptic conductances and financial scale parameters. All transformations and sampling steps enter the single-thread CPU query timer. The first draw set determines the evaluated return. 3 timed repetitions provide a median latency without selecting a different parameter. The complete original cohorts and independent native streams are retained, covering 512 targets with 64 responses for each low-dimensional task, 177 targets with 16 responses for neural circuits, and 75 windows with 16 responses for finance. The new experiment records include every returned parameter, fit-level and target-level loss, query timing, model configuration and checkpoint identifier.

The bounded-coordinate completion comprises 20 fitted models, and the consistency comparison adds 24. All 44 provide evaluable returns for their full cohorts. Their 11,720 returned parameters receive 580,736 independent native responses, with no additional training simulations. The accompanying data include all training durations and stopping reasons, alongside the existing low-dimensional fits.

\paragraph{Execution outcomes}
Table~\ref{tab:comparisonfailures} reports execution outcomes separately from the metric tables. A dash in a loss or Time cell indicates that at least one required fit/query did not complete under the stated procedure. Complete-cohort loss is reserved for procedures returning every required parameter under their specified implementations and caps. Financial stored-density and MCMC procedures that completed the full cohort retain their numerical loss and Time in Table~\ref{tab:finfull} under their explicit readout names.

\begin{longtable}{@{}L{25mm}L{24mm}L{22mm}L{29mm}L{68mm}@{}}
\caption{Execution outcomes for posterior procedures, separate from loss and Time. A completed fit requires an evaluable return for every target. Historical financial counts refer to 256-draw queries from 1 fitted model, the new main-table procedures use 32 draws. Original unsuccessful training attempts remain in the earlier grid and source ledger.}\label{tab:comparisonfailures}\\
\toprule Task & Procedure & Complete / planned & Statistical unit & Recorded outcome\\\midrule\endfirsthead
\toprule Task & Procedure & Complete / planned & Statistical unit & Recorded outcome\\\midrule\endhead
Predator--prey & CMPE, sCM & 4 / 4 each & 512-target fits & New consistency fits, all targets evaluated\\
Traffic & CMPE, sCM & 4 / 4 each & 512-target fits & New consistency fits, all targets evaluated\\
Neural circuit & All 7 main SBI & 2 / 2 each & 177-target fits & Bounded coordinates, all targets evaluated\\
Financial & All 7 main SBI & 2 / 2 each & 75-window fits & Bounded coordinates, all windows evaluated\\

Predator--prey & MDN & 4 / 4 & 512-target fits & Best finite validation checkpoint, original architecture, seed and 3,600 s cap\\
Predator--prey & NSF & 4 / 4 & 512-target fits & Best finite validation checkpoint, original architecture, seed and 3,600 s cap\\
Predator--prey & NPSE & 4 / 4 & 512-target fits & Best finite validation checkpoint, original architecture, seed and 3,600 s cap\\
Traffic & NSF & 4 / 4 & 512-target fits & Best finite validation checkpoint, original architecture, seed and 3,600 s cap\\
Neural circuit & MDN (box) & 0 / 2 & 177-target fits & Rejection sampling completes 176/177 and 174/177 targets complete, remaining queries exceed 600 s\\
Neural circuit & MDN (MCMC) & 0 / 2 & 177-target fits & Default slice MCMC completes 176/177 targets per fit, 1 query per fit exceeds 600 s\\
Neural circuit & MDN (MAP) & 2 / 2 & 177-target fits & Same bounded mode readout on all 177 targets per fit\\
Financial & MDN (original) & 45 / 75 & 256-draw queries & 30 query timeouts, 600 s cap\\
Financial & MAF (original) & 52 / 75 & 256-draw queries & 23 query timeouts, 600 s cap\\
Financial & NSF (original) & 52 / 75 & 256-draw queries & 14 timeouts and 9 spline inverse assertions, 600 s cap\\
Financial & FMPE (original) & 70 / 75 & 256-draw queries & 5 query timeouts, 600 s cap\\
Financial & NPSE (original) & 51 / 75 & 256-draw queries & 24 query timeouts, 600 s cap\\
\bottomrule\end{longtable}

\paragraph{Reading a financial loss difference}
The energy loss uses Euclidean distances after scaling each of the 8 response summaries. The scale for feature $k$ is $\sigma_k=\max\{\operatorname{SD}_{\mathrm{train}}(Y_k-b_k(c)),10^{-6}\}$, with sample SD computed on the 1,159 original source-training windows. The current-book baseline $b(c)$ is 0 for the first four features and uses the current spread, log ask depth, log bid depth and depth imbalance for the remaining features. These fixed scales are inherited by every later model and optimizer. Neither new evaluation windows nor individual methods refit them. Table~\ref{tab:financialscales} gives the values. Price and quantity conventions are specified in Supplement~\ref{app:finance}.

\begin{longtable}{@{}L{70mm}L{52mm}r@{}}
\caption{Financial feature scales used by every method. Price units retain a 0.01 tick after the common price shift, depth uses canonical quantities. Full precision is supplied in \texttt{financial\_metric\_scales.csv}.}\label{tab:financialscales}\\
\toprule Response feature & Unit before scaling & Scale\\\midrule\endfirsthead
\toprule Response feature & Unit before scaling & Scale\\\midrule\endhead
Final mid-price change & Price units & 0.03138252\\
Mean mid-price change & Price units & 0.01897684\\
SD of successive mid-price changes & Price units & 0.00181542\\
Fraction of nonzero mid-price changes & Dimensionless & 0.11683434\\
Mean spread & Price units & 0.00305999\\
Mean log ask depth & Dimensionless log depth & 0.21983054\\
Mean log bid depth & Dimensionless log depth & 0.22314479\\
Mean depth imbalance & Dimensionless & 0.14160298\\
\bottomrule\end{longtable}

For a deterministic response with only its final mid-price change displaced by 1 price tick (0.01), financial loss is
\[\frac{2(0.01)}{0.03138252}\approx0.6373.\]
This example isolates one coordinate. With stochastic responses or discrepancies in several features, the Euclidean norm and the response-dispersion term jointly determine loss. A correctly matched stochastic response distribution can still have positive expected energy loss because the observations themselves vary. Standardization combines the physical features into a dimensionless response criterion.

In the single-run 75-window comparison, changing the stock-fixed loss of 36.9663 to the direct loss of 35.4728 is a 4.04\% reduction. In the separate three-run comparison, changing the direct mean of 35.6001 to the GP-BO mean of 34.5429 is a 2.97\% reduction relative to Ours. Conversely, Ours has 3.06\% higher loss than GP-BO because that comparison uses GP-BO in the denominator. These percentages describe relative changes in the response criterion. For low-dimensional MSE, the square root gives discrepancy in training-standard-deviation units, so loss 0.04 corresponds to 0.2 such units per feature. The losses measure their respective response objectives and are compared within tasks.

\begingroup
\renewcommand{\arraystretch}{1.00}
\begin{longtable}{@{}L{65mm}rr@{}}
\caption{Full current-cohort results, including timing variability and task-specific controls. Loss matches the main matrix. Time is in milliseconds, SD is across the stated run or profiling means. Fixed rules have no fitting SD. Dashes denote incomplete historical procedures, detailed in Table~\ref{tab:comparisonfailures}.}\label{tab:maincomparison}\\
\toprule Method & Loss $\downarrow$ & Time (ms) $\downarrow$\\\midrule\endfirsthead
\multicolumn{3}{l}{Table \thetable\ continued}\\\toprule Method & Loss $\downarrow$ & Time (ms) $\downarrow$\\\midrule\endhead
\bottomrule\endfoot
\multicolumn{3}{l}{\textit{Predator--prey, 512 targets}}\\
PSO & $0.7783\pm0.0060$ & \rev{$650\,\pm\,300$}\\
CMA-ES & $0.5895\pm0.1062$ & \rev{$366\,\pm\,69$}\\
GP-BO & $0.7042\pm0.0286$ & \rev{$28\,800\,\pm\,1000$}\\
TuRBO-1 & $0.5361\pm0.0137$ & \rev{$8350\,\pm\,230$}\\
MDN & $0.1598\pm0.0314$ & \rev{$0.440\,\pm\,0.037$}\\
MAF & $0.0726\pm0.0175$ & \rev{$1.7937\,\pm\,0.0062$}\\
NSF & $0.0594\pm0.0111$ & \rev{$2.680\,\pm\,0.021$}\\
FMPE & $0.1032\pm0.0173$ & \rev{$54.1\,\pm\,1.2$}\\
NPSE & $0.0520\pm0.0087$ & \rev{$157.3\,\pm\,1.7$}\\
CMPE & $0.2745\pm0.0705$ & \rev{$38.82\,\pm\,0.30$}\\
sCM & $0.1717\pm0.0506$ & \rev{$51.76\,\pm\,0.25$}\\
Ours & $0.0426\pm0.0052$ & \rev{$0.016848\,\pm\,0.000074$}\\
Empirical memory & $0.0590\pm0.0065$ & \rev{$0.01219\,\pm\,0.00014$}\\
Surrogate selection & $0.1693\pm0.0418$ & \rev{$0.0238\,\pm\,0.0028$}\\
Full local variant & $0.0419\pm0.0048$ & \rev{$0.08452\,\pm\,0.00021$}\\
Spread empirical memory & $0.0326\pm0.0014$ & \rev{$0.010961\,\pm\,0.000093$}\\
MAF (ranked) & $0.0957\pm0.0251$ & \rev{$1.9246\,\pm\,0.0065$}\\
\addlinespace[4pt]
\multicolumn{3}{l}{\textit{Traffic, 512 targets}}\\
PSO & $0.9643\pm0.0044$ & \rev{$2257.8\,\pm\,5.1$}\\
CMA-ES & $0.9566\pm0.0015$ & \rev{$2278\,\pm\,17$}\\
GP-BO & $0.9582\pm0.0008$ & \rev{$38\,800\,\pm\,2200$}\\
TuRBO-1 & $0.9563\pm0.0018$ & \rev{$10\,290\,\pm\,190$}\\
MDN & $0.9603\pm0.0002$ & \rev{$0.5223\,\pm\,0.0060$}\\
MAF & $0.9599\pm0.0001$ & \rev{$2.470\,\pm\,0.067$}\\
NSF & $0.9598\pm0.0001$ & \rev{$3.45\,\pm\,0.16$}\\
FMPE & $0.9605\pm0.0005$ & \rev{$76.4\,\pm\,3.1$}\\
NPSE & $0.9602\pm0.0001$ & \rev{$223\,\pm\,23$}\\
CMPE & $0.9635\pm0.0023$ & \rev{$38.71\,\pm\,0.11$}\\
sCM & $0.9607\pm0.0005$ & \rev{$51.74\,\pm\,0.24$}\\
Ours & $0.9518\pm0.0019$ & \rev{$0.0689\,\pm\,0.0071$}\\
Empirical memory & $0.9768\pm0.0001$ & \rev{$0.716\,\pm\,0.024$}\\
Surrogate selection & $0.9515\pm0.0007$ & \rev{$0.1377\,\pm\,0.0037$}\\
Full local variant & $0.9522\pm0.0021$ & \rev{$0.151\,\pm\,0.011$}\\
MAF (ranked) & $0.9573\pm0.0003$ & \rev{$2.612\,\pm\,0.066$}\\
Spread empirical memory & $0.9712\pm0.0006$ & \rev{$2.23\,\pm\,0.12$}\\
\addlinespace[4pt]
\multicolumn{3}{l}{\textit{Neural circuit, 177 targets}}\\
PSO & $8.1145\pm3.9187$ & \rev{$396\,400\,\pm\,2000$}\\
CMA-ES & $4.4041\pm1.8814$ & \rev{$403\,450\,\pm\,670$}\\
GP-BO & $4.9766\pm2.7743$ & \rev{$568\,700\,\pm\,1200$}\\
TuRBO-1 & $3.1006\pm1.5835$ & \rev{$455\,700\,\pm\,3900$}\\
MDN & $9.0585\pm0.1588$ & \rev{$0.5081\,\pm\,0.0011$}\\
MAF & $7.4332\pm0.1554$ & \rev{$13.092\,\pm\,0.090$}\\
NSF & $6.1619\pm0.0828$ & \rev{$3.696\,\pm\,0.021$}\\
FMPE & $9.2273\pm0.3269$ & \rev{$26.8\,\pm\,1.5$}\\
NPSE & $6.7794\pm0.3803$ & \rev{$164.6\,\pm\,1.5$}\\
CMPE & $9.4304\pm0.5793$ & \rev{$39.81\,\pm\,0.24$}\\
sCM & $8.1974\pm0.3825$ & \rev{$52.78\,\pm\,0.36$}\\
Ours & $0.8907\pm0.0569$ & \rev{$0.00995\,\pm\,0.00085$}\\
Empirical memory & $1.0497$ & \rev{$0.01589\,\pm\,0.00021$}\\
Surrogate selection & $1.2035\pm0.0285$ & \rev{$0.0080\,\pm\,0.0018$}\\
Spread empirical memory & $5.1669$ & \rev{$0.013166\,\pm\,0.000051$}\\
16-neighbour validity & $1.0013$ & \rev{$0.014720\,\pm\,0.000010$}\\
MDN (box) & --- & ---\\
MAF (box) & $6.2871\pm0.5308$ & \rev{$124\,\pm\,25$}\\
MAF (box, ranked) & $4.5964\pm0.6263$ & \rev{$124\,\pm\,25$}\\
MDN (MCMC) & --- & ---\\
MDN (MAP) & $9.0141\pm1.7306$ & \rev{$683.44\,\pm\,0.90$}\\
\addlinespace[4pt]
\multicolumn{3}{l}{\textit{Financial, 75 windows}}\\
PSO & $36.1573\pm0.8401$ & \rev{$600\,229\,\pm\,55$}\\
CMA-ES & $34.6887\pm0.1722$ & \rev{$600\,227\,\pm\,57$}\\
GP-BO & $34.5429\pm0.0556$ & \rev{$600\,128\,\pm\,11$}\\
TuRBO-1 & $34.8454\pm0.0823$ & \rev{$600\,158.0\,\pm\,3.7$}\\
MDN & $41.0954\pm0.1059$ & \rev{$0.538\,\pm\,0.012$}\\
MAF & $38.9568\pm0.3051$ & \rev{$3.0345\,\pm\,0.0077$}\\
NSF & $39.6606\pm0.3297$ & \rev{$2.9655\,\pm\,0.0085$}\\
FMPE & $36.9147\pm0.1949$ & \rev{$18.95\,\pm\,0.23$}\\
NPSE & $38.1050\pm1.0360$ & \rev{$158.00\,\pm\,0.68$}\\
CMPE & $37.7118\pm0.3587$ & \rev{$39.04\,\pm\,0.34$}\\
sCM & $38.9950\pm1.3210$ & \rev{$53.59\,\pm\,0.50$}\\
Ours & $35.6001\pm0.1335$ & \rev{$0.05624\,\pm\,0.00052$}\\
Surrogate selection & $35.8609\pm0.0810$ & \rev{$0.06102\,\pm\,0.00034$}\\
Stock-fixed parameter & $36.9663$ & \rev{$0.000463$}\\
NL-SHADE-LBC & $35.9345\pm0.1253$ & \rev{$600\,227\,\pm\,56$}\\
SNO-PSO & $36.4983\pm0.1691$ & \rev{$600\,227\,\pm\,55$}\\
SNO-DE & $36.3039\pm0.1400$ & \rev{$600\,228\,\pm\,56$}\\
\end{longtable}
\endgroup

\subsection{Financial search trajectories and one-shot attainment}\label{app:financialcurves}
Figure~3 uses all 75 financial windows. Search curves include PSO, CMA-ES, GP-BO, TuRBO-1, NL-SHADE-LBC, SNO-PSO and SNO-DE, each with joint model/search seeds 7, 17 and 27. For each window/run, every recorded improvement of the search objective already has an independent 16-response evaluation against the real subsequent response. First attainment is the earliest recorded incumbent with evaluation loss no greater than the corresponding direct fit, available within 600 seconds. The subsequent real response and the direct evaluation loss never guide search or stopping. All 225 pairs remain in each search denominator. Current-incumbent matches can fall below first-attainment counts because the search objective and independent evaluation loss differ.

Checkpoints are reconstructed every 60 seconds from 60 through 600 seconds. At an intermediate checkpoint, the selected candidate is the last recorded objective incumbent available by that time. A repeated incumbent retains its measured loss. The final checkpoint uses the originally returned parameter and recorded complete query time, including final-call and termination overhead, consistently with Table~\ref{tab:overallcomparison}. This adds 7 observed checkpoints per algorithm without rerunning or interpolating a search. The recovered 60-, 300- and 600-second values agree exactly with the previous comparison. There are 15,750 window/run/checkpoint values. Pointwise bars are sample SDs of the 3 complete-window run means.

The posterior curves use MDN, MAF, NSF, FMPE, NPSE, CMPE and sCM, with 2 complete fits per method and 75 windows per fit. Both posterior fits condition on the fixed seed-7 history predictor and are compared with the corresponding seed-7 direct fit on each window. This pairing holds the history-prediction configuration fixed. Search instead varies its predictor and optimizer jointly across 3 runs. The parameter is the physical mean of the first 32-draw query, and latency is the median of 3 warmed complete queries. A successful one-shot return contributes to the matching fraction at that measured latency and remains unchanged afterwards. An unsuccessful return never contributes. Each curve records one posterior decision per window and fit.

Table~\ref{tab:financialattainment} reports first matches and final-return matches. Ours defines its paired quality reference and therefore reaches 100\% when its measured query completes. This follows from the paired-reference definition. The accompanying \path{financial_minute_window_values.csv}, \path{financial_incumbent_trajectories.csv} and \path{financial_all_attainment_records.csv} retain the checkpoints, objective incumbents, paired thresholds and individual matching times. No new model fitting, native simulation or search was needed for this reconstruction.

\begin{table}[H]
\centering\normalsize
\caption{Financial matching counts for Figure~3(a). First matches occur by 600 seconds. Median times use matching pairs only. Endpoint counts use the final return. Posterior methods make one return.}\label{tab:financialattainment}
\begin{tabularx}{180mm}{@{}Yrrrr@{}}\toprule
Method & Window/run pairs & First matches & Final matches & Median match time (s)\\\midrule
PSO & 225 & 132 & 108 & 102.10\\
CMA-ES & 225 & 186 & 153 & 62.15\\
GP-BO & 225 & 183 & 160 & 42.93\\
TuRBO-1 & 225 & 172 & 146 & 54.12\\
NL-SHADE-LBC & 225 & 130 & 96 & 56.63\\
SNO-PSO & 225 & 100 & 82 & 82.16\\
SNO-DE & 225 & 119 & 97 & 45.51\\
MDN & 150 & 31 & 31 & 0.000527\\
MAF & 150 & 40 & 40 & 0.00301\\
NSF & 150 & 45 & 45 & 0.00295\\
FMPE & 150 & 46 & 46 & 0.0185\\
NPSE & 150 & 48 & 48 & 0.158\\
CMPE & 150 & 46 & 46 & 0.039\\
sCM & 150 & 41 & 41 & 0.0535\\
Ours & 225 & 225 & 225 & $5.61\times10^{-5}$\\
\bottomrule\end{tabularx}
\end{table}

\section{Extended component and posterior comparisons}\label{app:extendedcomparisons}
The main comparison fixes the same four optimization algorithms and seven posterior-estimation methods across tasks. The complete task-specific inventory remains in Table~\ref{tab:maincomparison}, including posterior point readouts, memory controls and the additional financial optimizers. Historical procedures retain their original full-cohort values and recorded failures under explicit readout or coordinate labels.

Posterior methods in the main matrix share the 32-draw physical-mean decision described in Section~\mbox  {IV-D}. Supplement~\ref{app:comparisonprotocol} specifies the new bounded-coordinate fits and CMPE/sCM implementation. The extended table also retains the original unit-box neural models, the surrogate-ranked MAF readout and the bounded MDN mode readout under explicit names. Their sampling and numerical failures remain visible in Table~\ref{tab:comparisonfailures}. Earlier sample-count and training-budget grids retain their original settings in Supplement~\ref{app:neuralnpe}. Their common endpoint is native response quality of one returned parameter under the supplied simulation designs.

\subsection{Neural components, variability and response budget}
Table~\ref{tab:neuralextended} retains the complete component decomposition, the 8-fit initialization study, the separate 154-target response-budget cohort and target-level trade-offs. The main-text diagnostic summarizes the eight-fit follow-up. The original 177-target component comparison remains here. Validity-only mixing can match or improve the observed full-mixture mean. The spread-prior and reduced-repeat controls also bound the contribution of the supplied training pool and simulation budget.

\begin{longtable}{@{}lrrrr@{}}
\caption{Neural reliability and response-budget ablations. Panels A--C report Loss and its numeric/validity components as mean $\pm$ sample SD of run means at weight 5, using 16 new responses per return. Complete is the percentage with all 18 features. Bold marks each best displayed metric, including ties, $\dagger$ denotes a fixed rule. A uses 2 fits and B 8 fits on the same 177 targets. C uses a separate 154-target cohort. D reports target medians and wins/ties/losses after averaging fits within each target (tie tolerance $10^{-12}$).}\label{tab:neuralextended}\\
\toprule
Method & Loss & Numeric loss & Validity penalty & Complete (\%)\\\midrule
\midrule\multicolumn{5}{l}{\textit{A. Component controls}}\\
Empirical memory & $1.0497^{\dagger}$ & $\mathbf{0.7054^{\dagger}}$ & $0.3443^{\dagger}$ & $97.07^{\dagger}$\\
\textbf{Validity-only mixture} & $\mathbf{0.8714\,\pm\,0.0489}$ & $0.7576\,\pm\,0.0212$ & $0.1138\,\pm\,0.0277$ & $99.01\,\pm\,0.25$\\
Laplace validity & $1.0548^{\dagger}$ & $0.7596^{\dagger}$ & $0.2952^{\dagger}$ & $97.53^{\dagger}$\\
Global Beta validity & $1.0666^{\dagger}$ & $0.7675^{\dagger}$ & $0.2992^{\dagger}$ & $97.49^{\dagger}$\\
16-neighbour validity & $1.0013^{\dagger}$ & $0.7934^{\dagger}$ & $0.2079^{\dagger}$ & $98.59^{\dagger}$\\
Spread empirical memory & $5.1669^{\dagger}$ & $0.9757^{\dagger}$ & $4.1912^{\dagger}$ & $65.36^{\dagger}$\\
Surrogate selection & $1.2035\,\pm\,0.0285$ & $1.0584\,\pm\,0.0896$ & $0.1452\,\pm\,0.0610$ & $98.80\,\pm\,0.40$\\
Fixed parameter & $1.3603^{\dagger}$ & $1.3603^{\dagger}$ & $\mathbf{0.0000^{\dagger}}$ & $\mathbf{100.00^{\dagger}}$\\
Ours & $0.8907\,\pm\,0.0569$ & $0.7857\,\pm\,0.0333$ & $0.1050\,\pm\,0.0236$ & $99.06\,\pm\,0.22$\\
\midrule\multicolumn{5}{l}{\textit{B. Training variability}}\\
\textbf{Validity-only mixture} & $\mathbf{0.8795\,\pm\,0.0304}$ & $\mathbf{0.7529\,\pm\,0.0092}$ & $0.1265\,\pm\,0.0259$ & $98.92\,\pm\,0.21$\\
Surrogate selection & $1.2486\,\pm\,0.1709$ & $1.0172\,\pm\,0.0515$ & $0.2314\,\pm\,0.1660$ & $98.01\,\pm\,1.50$\\
Ours & $0.8796\,\pm\,0.0289$ & $0.7707\,\pm\,0.0174$ & $\mathbf{0.1089\,\pm\,0.0188}$ & $\mathbf{99.04\,\pm\,0.17}$\\
\midrule\multicolumn{5}{l}{\textit{C. Response budget}}\\
Empirical memory (full budget) & $0.7489^{\dagger}$ & $\mathbf{0.4817^{\dagger}}$ & $0.2672^{\dagger}$ & $97.65^{\dagger}$\\
Ours ($R=4$) & $0.8760\,\pm\,0.0147$ & $0.7633\,\pm\,0.0211$ & $0.1127\,\pm\,0.0064$ & $98.99\,\pm\,0.06$\\
\textbf{Ours (full budget)} & $\mathbf{0.6367\,\pm\,0.0079}$ & $0.5741\,\pm\,0.0326$ & $\mathbf{0.0626\,\pm\,0.0247}$ & $\mathbf{99.45\,\pm\,0.20}$\\
\midrule\multicolumn{5}{l}{\textit{D. Target-level trade-offs}}\\
Reference & Fits & Ours median & Ref. median & Wins / ties / losses\\
Empirical memory & 2 & 0.1330 & 0.0577 & 37/7/133\\
16-neighbour validity & 2 & 0.1330 & 0.1468 & 85/9/83\\
Empirical memory & 8 & 0.1312 & 0.0577 & 37/4/136\\
16-neighbour validity & 8 & 0.1312 & 0.1468 & 80/2/95\\
\bottomrule

\end{longtable}

\subsection{Financial teacher, compression and real-data controls}
Table~\ref{tab:financeextended} retains all teacher, output-compression and real-training-data comparisons. The main text shows teacher matching and training-data controls. Output compression is reported here with the same cohort and query-time definitions.

\begin{longtable}{@{}lrr@{}}
\caption{Financial component and real-data ablations. Loss is the energy loss in Eq.~(19), reported as mean $\pm$ sample SD of run means. Time is mean query latency, with full run statistics in the accompanying CSV. Bold marks the best displayed value in each metric. Panel A uses 475 windows and 3 banks with 5 fits per bank, pooled descriptively. Panel B uses the 75-window, 3-run cohort. Panel C uses 475 evaluation windows and 5 fits on a fixed response bank, retaining 1,159 source windows and varying target-day records. Repeated rows control row count without adding observations. Query times include record summarization, comparisons are within each panel.}\label{tab:financeextended}\\
\toprule
Method & Loss $\downarrow$ & Time (ms) $\downarrow$\\\midrule
\midrule\multicolumn{3}{l}{\textit{A. Teacher matching}}\\
Ours (compressed teacher) & $34.0602\,\pm\,0.0580$ & $\mathbf{5.54\times10^{-2}}$\\
Surrogate selection & $34.0526\,\pm\,0.0550$ & $6.04\times10^{-2}$\\
\textbf{Ours} & $\mathbf{34.0454\,\pm\,0.0726}$ & $\mathbf{5.54\times10^{-2}}$\\
\midrule\multicolumn{3}{l}{\textit{B. Output compression}}\\
\textbf{Ours (116 scores)} & $\mathbf{35.5509\,\pm\,0.1174}$ & $6.35\times10^{-2}$\\
Ours (18 scores) & $35.6567\,\pm\,0.0884$ & $5.66\times10^{-2}$\\
Surrogate selection (928 outputs) & $35.8566\,\pm\,0.2312$ & $7.67\times10^{-2}$\\
Surrogate selection (9 coordinates) & $35.8609\,\pm\,0.0810$ & $6.10\times10^{-2}$\\
Surrogate selection (18 coordinates) & $35.8012\,\pm\,0.1876$ & $6.27\times10^{-2}$\\
Ours & $35.6001\,\pm\,0.1335$ & $\mathbf{5.62\times10^{-2}}$\\
\midrule\multicolumn{3}{l}{\textit{C. Real training data}}\\
Ours (95 windows) & $34.2092\,\pm\,0.0221$ & $\mathbf{6.06\times10^{-2}}$\\
Ours (95 windows repeated to 365 rows) & $34.2038\,\pm\,0.0539$ & $6.13\times10^{-2}$\\
Ours (185 non-overlapping windows) & $34.2328\,\pm\,0.0189$ & $6.17\times10^{-2}$\\
\textbf{Ours (365 overlapping windows)} & $\mathbf{34.0176\,\pm\,0.0554}$ & $6.26\times10^{-2}$\\
\bottomrule

\end{longtable}

\section{Date separation and historical evaluation roles}\label{app:tcssdates}
The reported Nasdaq models use 8 December 2025 for training and 9--12 December 2025 plus 12 June 2026 for evaluation. Each evaluation date contains 95 non-overlapping windows, comprising 19 windows for each of five stocks. The search subset contains three fixed windows per stock/date. Evaluation outcomes were not fitting targets for the reported models, but the dates were consulted during dictionary and readout development. The TCSS analyses reuse all of these returns without further fitting or selection.

A separate historical 40-date experiment used SZ000001 in 2019. Its chronological division comprised 164 training dates through 2 September, 40 development dates from 3 September through 5 November and 40 later evaluation dates. That experiment evaluated an earlier dual-head distilled neural model, a forest policy and native 48-candidate search. Mean energy scores were 4.71634, 4.15815 and 4.08952, respectively. The distilled model failed the prespecified 5\% quality margin. Its records explicitly disclose prior research exposure of the underlying data. This historical diagnostic records the earlier dual-head model and its chronological evaluation protocol. The current rank-nine forest head uses the Nasdaq protocol described above.

\section{Financial behaviour and incremental-cost diagnostics}\label{app:tcssdiagnostics}
The additional analysis reuses 21,375 recorded parameter returns, representing nine bank/representation combinations, five fits and 475 windows. Exact coincident returns share 1,432 stored response files. Each file contains the same sixteen evaluation streams for its fixed query and parameter. The analysis uses the stored evaluations with fixed models and retains every observable, date, stock, bank and fit.

\begin{figure}[H]
\centering\includegraphics[width=0.8\textwidth]{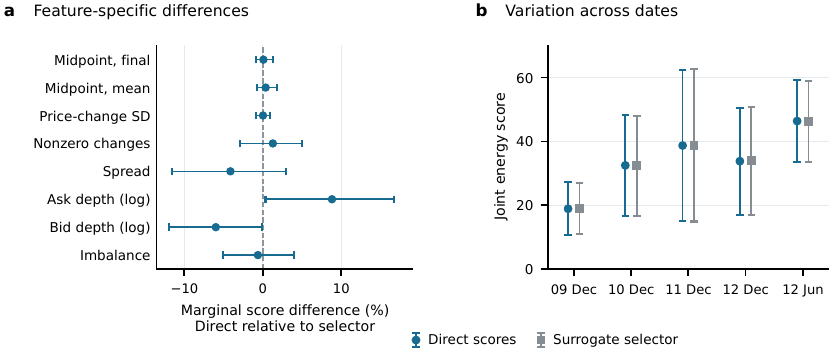}
\caption{Financial response diagnostics on 475 windows. (a) Marginal energy-score differences for the matched direct model relative to conditional surrogate selection. Negative values favour direct scoring. Points average banks and fits within windows, and bars are exploratory 95\% crossed-date/stock bootstrap intervals without multiplicity adjustment. Depth is log transformed. (b) Joint error averaged over 19 windows per stock/date and the same fitted rules. Error bars show SD across five stocks. December dates are in 2025 and the June date is in 2026.}\label{fig:financialdiagnostics}
\end{figure}

For standardized scalar observable $k$, its marginal score is
\begin{equation}
e_{ik}=\frac{2}{R}\sum_{r=1}^R|z_{irk}-Y_{ik}|
-\frac{1}{R(R-1)}\sum_{r\ne q}|z_{irk}-z_{iqk}|,\quad R=16.
\end{equation}
The eight existing calibration observables supply the marginal response diagnostics. Each score uses scalar absolute distance, while the joint energy score uses Euclidean distance across features. The marginal scores are therefore reported individually. Technical bank/fit replicates are averaged within each window. Stock/date cells average nineteen windows. The paired bootstrap independently samples five date indices and five stock indices with replacement for each of 20,000 draws (seed 20261028), using identical draws across methods. Percentile intervals are exploratory and unadjusted for the eight observable comparisons. Figure~\ref{fig:financialdiagnostics}b displays SD across the five stocks.
\begin{table}[tb]
\centering\small
\caption{Matched-teacher financial behaviour diagnostics on all 475 windows. Direct and selector columns give marginal energy scores. Differences and exploratory percentile intervals are percentages relative to the selector. Depth observables are log transformed.}\label{tab:behaviordetail}
\begin{tabular}{@{}lrrrr@{}}\toprule
Observable & Direct & Selector & Difference (\%) & 95\% interval\\\midrule
Terminal midpoint change & 17.0297 & 17.0161 & +0.08 & [-0.86, +1.36]\\
Mean midpoint change & 16.8896 & 16.8283 & +0.36 & [-0.73, +1.77]\\
Price-change volatility & 18.5432 & 18.5385 & +0.03 & [-0.87, +0.95]\\
Price-movement frequency & 1.5647 & 1.5450 & +1.28 & [-2.95, +5.04]\\
Mean spread & 4.6720 & 4.8730 & -4.12 & [-11.52, +2.98]\\
Ask-side depth & 2.6988 & 2.4802 & +8.81 & [+0.34, +16.77]\\
Bid-side depth & 2.4321 & 2.5868 & -5.98 & [-11.92, -0.15]\\
Book imbalance & 2.2123 & 2.2261 & -0.62 & [-5.03, +3.97]\\
\bottomrule\end{tabular}
\end{table}

\begin{table}[tb]
\centering\small
\caption{Financial joint energy score by evaluation date, averaging all banks and fits within windows. Direct uses the matched teacher.}\label{tab:datedetail}
\begin{tabular}{@{}lrrrr@{}}\toprule
Date & Windows & Direct & Selector & Difference (\%)\\\midrule
2025-12-09 & 95 & 18.9006 & 19.0048 & -0.548\\
2025-12-10 & 95 & 32.4908 & 32.3441 & +0.454\\
2025-12-11 & 95 & 38.7204 & 38.7596 & -0.101\\
2025-12-12 & 95 & 33.8093 & 33.9113 & -0.301\\
2026-06-12 & 95 & 46.3797 & 46.2434 & +0.295\\
\bottomrule\end{tabular}
\end{table}

For incremental cost, the recorded total time to prepare the matched score head averages 2,896.44 ms across fifteen bank/fit combinations (range 2,810.79--2,968.04 ms). Mean matched-direct and selector query times are 0.0554267 and 0.0604096 ms, giving $Q^*\simeq581{,}276$. The calculation assumes that the common response teacher already exists and that its observation regime remains useful. It includes measured head preparation and model serialization. It excludes shared simulation/teacher preparation, query-library compilation, loading and later refresh. Assigning zero preparation cost to the selector's separate future model makes this a conservative head-offset accounting comparison. The incremental estimate combines historical preparation timings with the interleaved query profile. Compressed model-file sizes average 16,637,568 and 13,716,619 bytes for direct and selector paths.

\section{Neural extension and full four-task evidence}\label{app:neuralextension}
The neural-circuit experiment extends the calibration method to outputs with missing features. The network predicts feature values and their availability, which are combined in the following scoring equations.
\paragraph{Neural response and validity network}
The amortized surrogate predicts numeric means, log variances and availability probabilities. The numeric loss $L_{\mathrm{num}}$, binary cross-entropy loss $L_{\mathrm{valid}}$ and combined objective $L_N$ are defined in Eqs.~(5)--(7). For feature $k$, logit $\gamma_k$ gives $p_k=(1+e^{-\gamma_k})^{-1}$. The numeric term uses available values, while the validity term uses every indicator. Early stopping and calibration-score selection use separate development groups.

For the neural task, encode a response as $(a\odot v,\omega a)$ and use $\nu=d$, where $\omega$ expresses the penalty for a missing feature. For a complete target $x$, write $e^{\mathrm v}$ for the expected error when a feature is valid and $e^{\mathrm m}$ for its loss when missing, with

\begin{align}\label{eq:validityparts}
e^{\mathrm v}_{j\ell}(x)&=(\mu_{j\ell}-x_\ell)^2+\sigma^2_{j\ell},\\
e^{\mathrm m}_\ell(x)&=x_\ell^2+\omega^2.\nonumber
\end{align}

Weighting these two cases by the predicted validity probability gives

\begin{equation}\label{eq:validityrisk}
\widehat r_j^{\mathrm{sur}}(x)=\frac1d\sum_{\ell=1}^{d}
\left[p_{j\ell}e^{\mathrm v}_{j\ell}(x)+(1-p_{j\ell})e^{\mathrm m}_\ell(x)\right].
\end{equation}

Compute empirical scores from the same encoding and apply Eq.~(10). The primary setting is $\omega=5$. Sensitivity experiments rebuild scores and reselect parameters at each weight.

\subsection{Complete four-task comparison}\label{app:fourtask}
The complete comparison retains the original predator--prey, traffic, neural-circuit and financial values, including fixed posterior-mean readouts and iterative optimizers. Task-specific simple controls remain in the component sections of this supplement and the main Results.
\begin{table}[t]
\centering\normalsize
\setlength{\tabcolsep}{4pt}
\caption{Calibration error and query time across four tasks. Each cell gives mean error above time (ms), both as mean $\pm$ sample SD across complete-cohort runs. Lower is better. Bold marks the lowest mean per task and metric. Cohorts contain 512 predator--prey and 512 traffic development targets, 177 neural confirmation targets and 75 retrospective financial windows. Posterior methods use 32-sample physical means. Run counts are 3 for search, 4/4/2/2 for posterior methods and 2/4/2/3 for ours, in column order. Preparation is reported separately in Section~\mbox  {V-G}.}\label{tab:fourtaskcomplete}\label{tab:overallcomparison}
\begin{tabularx}{\textwidth}{@{}L{23mm}CCCC@{}}
\toprule
Method & Predator--prey & Traffic & Neural circuit & Financial\\
\midrule
PSO & \resultcell{$0.7783\,\pm\,0.0060$}{$650\,\pm\,300$} & \resultcell{$0.9643\,\pm\,0.0044$}{$2257.8\,\pm\,5.1$} & \resultcell{$8.1145\,\pm\,3.9187$}{$396\,400\,\pm\,2000$} & \resultcell{$36.1573\,\pm\,0.8401$}{$600\,229\,\pm\,55$}\\[3pt]
CMA-ES & \resultcell{$0.5895\,\pm\,0.1062$}{$366\,\pm\,69$} & \resultcell{$0.9566\,\pm\,0.0015$}{$2278\,\pm\,17$} & \resultcell{$4.4041\,\pm\,1.8814$}{$403\,450\,\pm\,670$} & \resultcell{$34.6887\,\pm\,0.1722$}{$600\,227\,\pm\,57$}\\[3pt]
GP-BO & \resultcell{$0.7042\,\pm\,0.0286$}{$28\,800\,\pm\,1000$} & \resultcell{$0.9582\,\pm\,0.0008$}{$38\,800\,\pm\,2200$} & \resultcell{$4.9766\,\pm\,2.7743$}{$568\,700\,\pm\,1200$} & \resultcell{$\mathbf{34.5429\,\pm\,0.0556}$}{$600\,128\,\pm\,11$}\\[3pt]
TuRBO-1 & \resultcell{$0.5361\,\pm\,0.0137$}{$8350\,\pm\,230$} & \resultcell{$0.9563\,\pm\,0.0018$}{$10\,290\,\pm\,190$} & \resultcell{$3.1006\,\pm\,1.5835$}{$455\,700\,\pm\,3900$} & \resultcell{$34.8454\,\pm\,0.0823$}{$600\,158.0\,\pm\,3.7$}\\[3pt]
\midrule
MDN & \resultcell{$0.1598\,\pm\,0.0314$}{$0.440\,\pm\,0.037$} & \resultcell{$0.9603\,\pm\,0.0002$}{$0.5223\,\pm\,0.0060$} & \resultcell{$9.0585\,\pm\,0.1588$}{$0.5081\,\pm\,0.0011$} & \resultcell{$41.0954\,\pm\,0.1059$}{$0.538\,\pm\,0.012$}\\[3pt]
MAF & \resultcell{$0.0726\,\pm\,0.0175$}{$1.7937\,\pm\,0.0062$} & \resultcell{$0.9599\,\pm\,0.0001$}{$2.470\,\pm\,0.067$} & \resultcell{$7.4332\,\pm\,0.1554$}{$13.092\,\pm\,0.090$} & \resultcell{$38.9568\,\pm\,0.3051$}{$3.0345\,\pm\,0.0077$}\\[3pt]
NSF & \resultcell{$0.0594\,\pm\,0.0111$}{$2.680\,\pm\,0.021$} & \resultcell{$0.9598\,\pm\,0.0001$}{$3.45\,\pm\,0.16$} & \resultcell{$6.1619\,\pm\,0.0828$}{$3.696\,\pm\,0.021$} & \resultcell{$39.6606\,\pm\,0.3297$}{$2.9655\,\pm\,0.0085$}\\[3pt]
FMPE & \resultcell{$0.1032\,\pm\,0.0173$}{$54.1\,\pm\,1.2$} & \resultcell{$0.9605\,\pm\,0.0005$}{$76.4\,\pm\,3.1$} & \resultcell{$9.2273\,\pm\,0.3269$}{$26.8\,\pm\,1.5$} & \resultcell{$36.9147\,\pm\,0.1949$}{$18.95\,\pm\,0.23$}\\[3pt]
NPSE & \resultcell{$0.0520\,\pm\,0.0087$}{$157.3\,\pm\,1.7$} & \resultcell{$0.9602\,\pm\,0.0001$}{$223\,\pm\,23$} & \resultcell{$6.7794\,\pm\,0.3803$}{$164.6\,\pm\,1.5$} & \resultcell{$38.1050\,\pm\,1.0360$}{$158.00\,\pm\,0.68$}\\[3pt]
CMPE & \resultcell{$0.2745\,\pm\,0.0705$}{$38.82\,\pm\,0.30$} & \resultcell{$0.9635\,\pm\,0.0023$}{$38.71\,\pm\,0.11$} & \resultcell{$9.4304\,\pm\,0.5793$}{$39.81\,\pm\,0.24$} & \resultcell{$37.7118\,\pm\,0.3587$}{$39.04\,\pm\,0.34$}\\[3pt]
sCM & \resultcell{$0.1717\,\pm\,0.0506$}{$51.76\,\pm\,0.25$} & \resultcell{$0.9607\,\pm\,0.0005$}{$51.74\,\pm\,0.24$} & \resultcell{$8.1974\,\pm\,0.3825$}{$52.78\,\pm\,0.36$} & \resultcell{$38.9950\,\pm\,1.3210$}{$53.59\,\pm\,0.50$}\\[3pt]
\midrule
Ours & \resultcell{$\mathbf{0.0426\,\pm\,0.0052}$}{$\mathbf{0.016848\,\pm\,0.000074}$} & \resultcell{$\mathbf{0.9518\,\pm\,0.0019}$}{$\mathbf{0.0689\,\pm\,0.0071}$} & \resultcell{$\mathbf{0.8907\,\pm\,0.0569}$}{$\mathbf{0.00995\,\pm\,0.00085}$} & \resultcell{$35.6001\,\pm\,0.1335$}{$\mathbf{0.05624\,\pm\,0.00052}$}\\[3pt]
\bottomrule
\end{tabularx}
\end{table}

\subsection{Fixed comparator settings}\label{app:tcsssettings}
PSO uses a population of 10, inertia 0.8 and cognitive/social weights 0.5.
CMA-ES uses a population of 10 and initial unit-coordinate step 0.30.
GP-BO uses 5 initial random points and noisy log expected improvement, and TuRBO-1 uses $2p$ initialization points for $p$ parameter dimensions.
The search seeds are 7, 17 and 27.
These configurations were fixed without selection on evaluation loss.
The posterior models implemented in \texttt{sbi} retain their recorded default architectures and training settings, with a $3.6\times10^6$ ms fit cap and the lowest finite internal-validation-loss checkpoint.
CMPE and sCM use the BayesFlow TimeMLP defaults, 10,000 Adam updates at learning rate $10^{-3}$, batch size 256 and a 90/10 training/monitoring split under the same fit cap.
Their query samplers use 11 and 15 steps, respectively, and all posterior readouts average 32 physical-parameter draws.
The full model-specific settings and parameter transformations remain in Supplement~\ref{app:comparisonprotocol}.

\section{Confirmation data and preparation accounting}\label{app:revisionevidence}

\subsection{Independent targets and subsequent component analyses}
The predator--prey set-C observations and the 177 complete-rhythm neural observations were generated after fixing their original comparison rules. The paired intervals for those comparisons resample targets after averaging the original bank or fit results within targets. They therefore describe target variability conditional on the prepared rules. The minimum-response-budget comparison on set C and the eight-fit neural extension subsequently reused these observations. They assess budget allocation and fitting variability, respectively. Table~\ref{tab:confirmationroles} records these roles alongside the financial data.

\begin{longtable}{@{}p{34mm}p{28mm}p{104mm}@{}}
\caption{Evaluation roles of the recorded cohorts. New simulator draws used to evaluate a return are distinct from targets generated after fixing a method.}\label{tab:confirmationroles}\\
\toprule Cohort & Size & Role and configuration boundary\\\midrule\endfirsthead
\toprule Cohort & Size & Role and configuration boundary\\\midrule\endhead
Predator--prey A/B & 512 each & Development comparisons used to select the primary local rule.\\
Predator--prey C & 512 & Generated after fixing the local rules and wider-memory allocation. The later 401,400-response prefix control uses this same cohort.\\
Traffic A/B & 512 each & Development comparisons used for component and primary-rule choices.\\
Traffic confirmation & 64 & Generated after freezing the small-bank local, expanded empirical and expanded surrogate rules.\\
Traffic inverse comparison & 64 & A separate new cohort generated after freezing those three rules and both tandem-style inverse fits.\\
Neural confirmation & 177 & All complete observations from a new 16,384-draw prior batch. Original models, scoring weight and smoother definitions were fixed before generation. Six further fitting seeds were later assessed on these same targets.\\
Financial evaluation & 475 windows & Outcomes excluded from fitting but consulted during dictionary and readout development. The 75-window search comparison is a subset.\\
\bottomrule
\end{longtable}

The original neural two-fit comparison lowers mean error by 15.15\% relative to empirical selection, with a paired 95\% interval of 3.18--37.48\%. Median error rises from 0.05768 to 0.13297, while the 90th percentile falls from 1.76895 to 0.88824. Against validity-only mixing, the full mixture has 2.21\% higher mean error, with an interval of $[-1.41\%,7.56\%]$. Against the 16-neighbour validity estimate, it has 11.05\% lower mean error, with an interval of $[-5.25\%,34.36\%]$ for the reduction. These intervals use 9,999 paired target resamples and the original fixed fits. Supplement~\ref{app:neuralfitstability} reports the separate fit-and-target analysis for eight fits.

\subsection{Small-budget traffic confirmation}\label{app:trafficpilot}

The small traffic experiment uses one bank with 512 anchors and the simulator, response features and physical ranges specified in Section~IV. All settings, code dependencies and random seeds were fixed before new targets were generated. The response scale and the original 1,843/205 normalization/development split were reconstructed from 2,048 groups with four responses each. The bank contains four retrieval responses and 28 joint-field responses per anchor. Thus, preparation uses 8,192 normalization/development responses and 16,384 bank responses. The local field uses common random streams at each centre and its six coordinate perturbations.

The primary rule retains the original anchor-only response fit, $\tau=64$, and a correction bounded by half the distance to the 64th other anchor. The two simple controls use all 3,584 centre and perturbation locations. Empirical selection averages squared response discrepancy over the stored repeats, pooling coincident centres. Surrogate selection uses a separate response model fitted to all pooled locations. Each response model uses seed 7, the architecture and optimization settings in Section~\mbox  {IV-C}, and the same 205 development groups for early stopping. There is one fit per model. All three rules receive the same prepared response pool.

For this experiment, all three rules use an exact \texttt{cKDTree} index on lifted response moments. For candidate mean $\mu_j$ and scalar second moment $s_j=\E\norm{z_j}^2$, the indexed point is $(\mu_j,\sqrt{s_j-\norm{\mu_j}^2})$ and the query is $(x,0)$. Their squared distance equals the estimated expected squared response discrepancy. Exact searches use zero approximation tolerance and are checked against exhaustive scores before target generation. The local rule additionally applies its prepared affine correction and physical projection. This common implementation supplies the timings in this small-bank comparison.

All fitted query artifacts were saved and hashed before drawing 64 new uniform-prior targets. Target parameters use seed 202611040164. Observation and return-evaluation streams start at 421100000000 and 421200000000, respectively, with a stride of 128 per target. All 192 parameter returns were frozen before independent evaluation, using 64 responses per return and shared evaluation streams across methods. No target was filtered or replaced. The experiment used 36,928 responses in total, of which 12,352 generated the new observations and evaluated returns. All returned parameters were feasible and all simulations completed.

Table~\ref{tab:trafficpilot} reports all three rules. The three prespecified paired comparisons use 20,000 target-bootstrap resamples with seed 202611040264. Intervals condition on this bank and the two fitted models and have no multiplicity adjustment. The preparation budget and query implementation differ from the full-bank comparison in Table~\ref{tab:overallcomparison}, which remains a separate experiment.

\begin{table}[H]
\centering\small
\caption{Fixed-rule traffic confirmation on 64 new targets. Latency averages per-target medians of 31 interleaved complete queries after five warm-ups. The paired intervals describe relative error reduction by the row method against its stated comparator.}\label{tab:trafficpilot}
\begin{tabular}{@{}lrrrr@{}}
\toprule Method & Mean error & Median error & 90th percentile & Query time (ms)\\\midrule
Local correction & 0.96926 & 0.93810 & 1.21675 & 0.02631\\
Expanded empirical & 0.99096 & 0.94571 & 1.23704 & 0.06183\\
Expanded surrogate & 0.98249 & 0.94266 & 1.23632 & 0.05634\\
\midrule
\multicolumn{2}{@{}l}{Paired comparison} & \multicolumn{2}{c}{Reduction and 95\% interval (\%)} & Wins / 64\\
\multicolumn{2}{@{}l}{Local against empirical} & \multicolumn{2}{c}{$2.19\;[0.68,3.57]$} & 50\\
\multicolumn{2}{@{}l}{Local against surrogate} & \multicolumn{2}{c}{$1.35\;[-0.06,2.64]$} & 43\\
\multicolumn{2}{@{}l}{Surrogate against empirical} & \multicolumn{2}{c}{$0.85\;[-0.29,1.92]$} & 39\\
\bottomrule
\end{tabular}
\end{table}

\subsection{Direct inverse prediction on a second set of traffic targets}\label{app:traffictandem}

This comparison reuses the response pool, normalization, expanded response surrogate and three prepared rules from Supplement~\ref{app:trafficpilot}. It adds two fixed inverse fits and 64 new evaluation targets. Following tandem training~\cite{liu2018tandem}, the inverse is optimized through a separately fitted, frozen forward model. We adapt the response loss to stochastic calibration by including both the predicted squared mean discrepancy and response variance. Gradients pass through the teacher with respect to its inputs while its parameters remain fixed. Parameter labels are used to fit the forward model; the inverse itself is trained from response error.

The inverse has 36 inputs, two 256-unit SiLU hidden layers and three sigmoid outputs in the unit parameter box. Its loss averages $(\mu_T(g(x))-x)^2+\sigma_T^2(g(x))$ across response coordinates and training examples. The teacher is the existing expanded surrogate. Training samples are drawn uniformly from all 16,384 individual bank responses, giving eight responses at each anchor and four at each perturbation location. Early stopping uses the existing 205 development groups, with all four responses per group and equal group weights. These groups also served in fitting the teacher.

Both initialization seeds, 7 and 19, were specified before fitting. AdamW uses learning rate $10^{-3}$, weight decay $10^{-4}$, batches of 256 and a gradient-norm cap of 5. Fitting runs for at most 4,000 updates, evaluates development error every 100 updates and stops after eight checks without improvement. The retained checkpoints occur at updates 1,800 and 2,100 after 2,600 and 2,900 updates, respectively. Both fits enter the results. Learned weights are exported for single-thread NumPy evaluation, including input normalization and physical parameter conversion. Exported outputs were checked against both double- and single-precision network evaluations before target generation.

All five query rules were saved and hashed before target generation. New parameters use seed 202611040664. Observation and evaluation streams start at 431100000000 and 431200000000, respectively, with stride 128 per target. All 320 returns were frozen before evaluation with 64 fresh responses per return, sharing evaluation streams across methods. All 64 targets were retained and all returns were feasible. The comparison adds 20,544 simulator responses, comprising 64 observations and 20,480 return evaluations. It adds no training simulations or teacher refits.

\begin{table}[H]
\centering\small
\caption{Second traffic confirmation on 64 new targets. The five rules use the same response pool. Latency averages per-target medians of 31 interleaved complete queries after five warm-ups.}\label{tab:traffictandem}
\begin{tabular}{@{}lrrrr@{}}
\toprule Method & Mean error & Median error & 90th percentile & Query time (ms)\\\midrule
Local correction & 0.99295 & 0.98200 & 1.16131 & 0.02676\\
Expanded empirical & 1.02282 & 1.02442 & 1.19878 & 0.06346\\
Expanded surrogate & 0.99944 & 0.98553 & 1.19044 & 0.05938\\
Tandem inverse, seed 7 & 0.99752 & 0.98836 & 1.15645 & 0.01832\\
Tandem inverse, seed 19 & 0.99929 & 0.99837 & 1.16085 & 0.01831\\
\bottomrule
\end{tabular}
\end{table}

The primary contrast compares local error with the mean of the two inverse-fit errors within each target. This averages losses for statistical comparison; each inverse is a separate deployed rule. The estimated reduction is 0.546\%, with a paired 95\% interval of $[-0.297\%,1.429\%]$. All ten pairwise comparisons and three comparisons with the two-fit mean were specified in advance and retained. Intervals use 20,000 paired target-bootstrap resamples with seed 202611040464, conditional on the bank, frozen teacher and two inverse fits, without multiplicity adjustment.

\begin{table}[H]
\centering\small
\caption{All prespecified paired contrasts on the second traffic cohort. Differences are $100(L_{\rm row}/L_{\rm comparator}-1)$, so negative values favour the row method.}\label{tab:traffictandempairs}
\begin{tabular}{@{}llrr@{}}
\toprule Row method & Comparator & Difference (\%) & 95\% interval (\%)\\\midrule
Local & Empirical & $-2.920$ & $[-4.240,-1.671]$\\
Local & Surrogate & $-0.649$ & $[-1.512,0.210]$\\
Local & Inverse seed 7 & $-0.458$ & $[-1.317,0.379]$\\
Local & Inverse seed 19 & $-0.634$ & $[-1.586,0.266]$\\
Empirical & Surrogate & $2.339$ & $[1.275,3.532]$\\
Empirical & Inverse seed 7 & $2.535$ & $[1.458,3.733]$\\
Empirical & Inverse seed 19 & $2.354$ & $[1.349,3.461]$\\
Surrogate & Inverse seed 7 & $0.192$ & $[-0.549,0.930]$\\
Surrogate & Inverse seed 19 & $0.015$ & $[-0.789,0.776]$\\
Inverse seed 7 & Inverse seed 19 & $-0.177$ & $[-0.627,0.262]$\\\midrule
Local & Two-fit mean & $-0.546$ & $[-1.429,0.297]$\\
Empirical & Two-fit mean & $2.445$ & $[1.431,3.578]$\\
Surrogate & Two-fit mean & $0.103$ & $[-0.629,0.821]$\\
\bottomrule
\end{tabular}
\end{table}

The teacher's predicted mean-mismatch and variance terms are 0.4950 and 0.3482 for seed 7, and 0.4964 and 0.3502 for seed 19. Their sums are below the independently evaluated simulator errors of 0.9975 and 0.9993. Both inverses place at least one unit coordinate within 0.001 of a boundary on 6 of the 64 targets. These recorded diagnostics describe the fixed returns and were not used to select a fit.

\subsection{Traffic preparation and repeated-query cost}\label{app:trafficcost}

The traffic experiment records preparation from simulation through saved query artifacts. Simulation uses eight single-thread workers on an AMD EPYC 9755. Fitting, field and index construction, and queries use one core. Table~\ref{tab:trafficprepcost} sums each rule's required recorded stages, charging the full shared response pool to every rule. Processor time sums parent and joined-worker process times, including worker initialization and communication. Target generation, independent return evaluation and repeated timing probes are research operations and are accounted separately from preparation.

Preparation stages include design generation, normalization/development simulation, scale fitting, bank simulation, moment construction, model fitting where used, prediction of candidate moments, construction of local maps where used, exact indexing, verification and serialization. Both model checkpoint and query-artifact writes are timed. The common simulation stages take 11.3478 s of elapsed time and 89.9128 processor seconds. The software uses installed SciPy binaries and requires no custom query-library compilation.

Each method's saved artifact is loaded in three fresh instances of the experimental process. Median process-start times include Python and library imports and artifact loading. Object deserialization is also measured separately, taking 0.418, 0.758 and 0.767 ms for the local, empirical and surrogate rules, respectively. These are fresh process and object measurements, with the operating-system file cache unchanged. The process loads the full experimental runtime; it is not a minimal deployment loader.

\begin{table}[H]
\centering\small
\caption{Required traffic preparation stages and query-process startup. Each preparation was measured once. Processor seconds are the sum over active processes. The response pool contains 24,576 preparation responses, including scale/development groups.}\label{tab:trafficprepcost}
\begin{tabular}{@{}lrrr@{}}
\toprule Cost & Local correction & Expanded empirical & Expanded surrogate\\\midrule
Preparation wall time (s) & 14.8803 & 11.7497 & 13.9050\\
Preparation processor time (s) & 93.3236 & 90.3147 & 92.4688\\
Fresh query-process start (s) & 1.3255 & 1.3304 & 1.3366\\
Fresh query-process processor time (s) & 1.3226 & 1.3271 & 1.3339\\
Preparation with responses in memory (s) & 3.5325 & 0.4019 & 2.5572\\
\bottomrule
\end{tabular}
\end{table}

For cumulative elapsed cost, Table~\ref{tab:trafficcumulative} adds required preparation, one fresh query-process start including loading, and $Q$ times the measured mean query latency. It is a sum of recorded stages, not an uninterrupted deployment measurement. The same-response-in-memory calculation removes only the 11.3478 s spent regenerating the shared responses. It retains normalization, moment construction, fitting, maps, indices, saving and fresh query-process loading. The measured whole research job lasted 35.7398 s; 0.0901 s of manifest, hashing and output overhead was not allocated to individual rules.

\begin{table}[H]
\centering\small
\caption{Cumulative elapsed cost of the small traffic calibration workflows in seconds, including one query-process start. The query counts were fixed before inspecting results. All rows rebuild the response pool once per refresh cycle.}\label{tab:trafficcumulative}
\begin{tabular}{@{}rrrr@{}}
\toprule Queries per refresh & Local correction & Expanded empirical & Expanded surrogate\\\midrule
1 & 16.2058 & 13.0802 & 15.2417\\
100 & 16.2084 & 13.0863 & 15.2473\\
1,000 & 16.2321 & 13.1420 & 15.2980\\
10,000 & 16.4689 & 13.6985 & 15.8050\\
100,000 & 18.8365 & 19.2633 & 20.8754\\
1,000,000 & 42.5126 & 74.9116 & 71.5795\\
\bottomrule
\end{tabular}
\end{table}

The cumulative calculation uses the observed quality of each fixed rule; it does not assume equal calibration error. Rebuilding after fewer queries repeatedly incurs preparation, whereas retaining a suitable model spreads this cost across more parameter updates. The small-bank timings quantify this trade-off under the stated worker schedule and query implementation.

\subsection{Inverse fitting and complete workflow costs}\label{app:tandemcost}

The five-rule comparison uses a common minimal deployment module based on NumPy and SciPy, without importing the simulator or the fitting library. Each rule is loaded in three fresh processes, with median process elapsed and processor times recorded. Process elapsed times are 0.18490, 0.18517 and 0.18438 s for the local, empirical and surrogate rules, and 0.14324 and 0.14161 s for the two inverse fits. These measurements include object loading and retain the operating-system file cache.

Required preparation stages for the three existing rules are the recorded stages in Table~\ref{tab:trafficprepcost}. An inverse uses the shared response-generation, normalization and expanded-teacher fitting stages, totalling 13.88735 s. Loading the existing training data and teacher, fitting, writing the checkpoint, exporting and verifying one inverse adds 8.34322 s for seed 7 or 9.16300 s for seed 19. The fitting stages themselves take 8.20045 and 9.01989 s. Each deployment is charged one teacher fit and one inverse fit. Index construction for surrogate selection is excluded from the inverse workflow because the inverse does not use it.

\begin{table}[H]
\centering\small
\caption{Complete traffic workflow costs with the common minimal query runtime. $P$ includes required preparation and one fresh process start with loading. Cumulative elapsed costs sum those recorded stages and $Q$ times the warm complete-query latency.}\label{tab:tandemcost}
\begin{tabular}{@{}lrrrr@{}}
\toprule Method & $P$ (wall s) & $P$ (processor s) & $C(10^5)$ (s) & $C(10^6)$ (s)\\\midrule
Local correction & 15.06516 & 93.50698 & 17.74114 & 41.82493\\
Expanded empirical & 11.93491 & 90.49834 & 18.28095 & 75.39538\\
Expanded surrogate & 14.08942 & 92.65165 & 20.02717 & 73.46695\\
Tandem inverse, seed 7 & 22.37381 & 100.93509 & 24.20613 & 40.69703\\
Tandem inverse, seed 19 & 23.19196 & 101.75232 & 25.02301 & 41.50254\\
\bottomrule
\end{tabular}
\end{table}

The totals combine previously recorded shared preparation with the new inverse-fitting and deployment measurements. They estimate repeated use from measured stages and warm query rates. A first cold query and an uninterrupted end-to-end rebuild were not timed. With the raw response pool already available in memory, subtracting the 11.34780 s of response generation from every workflow retains normalization, moment construction, fitting, serialization and startup. These cost calculations use the fixed rules at the errors reported in Table~\ref{tab:traffictandem}.

Local correction offsets its additional elapsed preparation after approximately 85,291 queries against empirical selection and 29,915 against surrogate selection. The two inverse fits offset their additional preparation relative to local correction after approximately 866,308 and 961,843 queries. Processor-time crossings occur later for the inverses, at approximately 1.21 and 1.47 million queries. These workload calculations use each separately measured fit and query implementation; they do not extrapolate the small bank to the original full-bank benchmark.

\subsection{Financial preparation records and refresh workload}\label{app:financialrefresh}
The retained financial bank contains 38,976 responses. Training-only candidate generation adds 8,080 calls, giving a recorded lower bound of 47,056 calls before inherited-dictionary construction and rejected-candidate screening. The surviving records do not provide a complete elapsed preparation time for the primary full-response-teacher model. The measured 2.896 s score-head job belongs to the matched compressed-teacher control. Its timer covers loading the teacher and training data, predicting candidate responses, forming relative labels, computing the low-rank representation, fitting the forest and saving the model. Shared simulation and teacher fitting, deployment export, compilation and loading are additional operations.

Table~\ref{tab:refreshcost} applies the recorded mean head-preparation time of 2.896439 s and query times of 0.0554267 and 0.0604096 ms. It represents the extra score-head investment after the shared response model is available. The conditional selector's separate future-response-model preparation is assigned zero. The historical preparation and query measurements came from different sessions on the same processor family. If the head is replaced every $Q$ queries, this additional investment recurs at every refresh.

\begin{table}[H]
\centering\small
\caption{Incremental financial cost for a fixed number of queries before a score-head refresh. Times are seconds. Direct scoring includes the recorded extra-head preparation; shared preparation and deployment costs are outside this comparison.}\label{tab:refreshcost}
\begin{tabular}{@{}rrrr@{}}
\toprule Queries per refresh & Extra head plus direct queries & Conditional queries & Direct minus conditional\\\midrule
95 & 2.901705 & 0.005739 & 2.895966\\
475 & 2.922767 & 0.028695 & 2.894073\\
10,000 & 3.450707 & 0.604096 & 2.846610\\
100,000 & 8.439113 & 6.040963 & 2.398150\\
1,000,000 & 58.323174 & 60.409629 & $-2.086455$\\
\bottomrule
\end{tabular}
\end{table}

At 95 queries per refresh, the matched direct head saves 0.000473 s of query computation before the next fit, compared with its 2.896 s preparation. The same measured query saving offsets the extra head after approximately 581,276 queries. This calculation concerns the choice between two prepared scoring rules. A comparison with iterative search also requires the shared preparation and deployment times.

\section{Additional fitting settings and response diagnostics}\label{app:submissiondetails}

\subsection{Simulator mechanics and observations}\label{app:submissionmechanics}

Predator--prey calibration estimates four Lotka--Volterra rates from 18 noisy population features~\cite{lueckmann2021benchmarking}. Traffic calibration estimates desired speed $v_{0} \in \lbrack 12,32\rbrack$ m/s, time headway $T \in \lbrack 0.4,2.4\rbrack$ s and standstill spacing $s_{0} \in \lbrack 1,9\rbrack$ m in the stochastic intelligent-driver model~\cite{treiber2000congested}. The traffic simulator follows 24 vehicles on a 260 m ring for 180 steps of 0.2 s. It includes a periodic bottleneck, acceleration noise and measurement noise. Eight vehicle probes produce 36 gap and speed summaries.

An order-driven simulator represents 125 liquidity providers and 125 liquidity demanders interacting through a price--time priority book~\cite{belcak2022fast}. Seven calibrated parameters govern two quote scales, provider and demander buy probabilities, arrival and cancellation rates, and trading intensity. Each query combines eight summaries of the preceding five minutes with forty coordinates of the current ten-level book. These 48 inputs determine parameters for the next 300 s. We compare simulated and observed outcomes using midpoint changes, price-change variability and activity, spread, log depth on each side, and imbalance. Supplement~\ref{app:finance} defines the summaries, physical ranges and initialization.

\subsection{Model fitting and selection}\label{app:submissionfitting}

For traffic and predator--prey, the surrogates have two 256-unit SiLU layers~\cite{elfwing2017silu} with separate mean and log-variance heads. Their loss $L_{F}$ uses $\lambda_{\rm var} = 0.01$, variance floor $10^{- 5}$ and log-variance bounds $\lbrack - 12,8\rbrack$. AdamW~\cite{loshchilov2019adamw} uses learning rate $10^{- 3}$, weight decay $10^{- 4}$, batch size 256 and gradient-norm cap 5. Training allows at most 4,000 updates, with development loss checked every 100 updates. It stops after eight checks without improvement, and we retain the lowest-loss checkpoint. \roundedit{The original retrieval surrogate is fitted on 10,035 four-response anchor groups per bank; inherited feature scales use 1,843 separate groups, and 205 held-out groups provide development responses.} Retrieval weights and local radii were fixed before sets A and B. For incomplete neural responses, the surrogate minimizes $L_{N}$ from Eq.~(7), with settings given in Supplementary Table~\ref{tab:hyperparameters}.

The financial response, future-response and score models each use 100 bootstrap trees, with all features available at every split. They use minimum leaf size 1 and no depth limit. The response forest maps 40 book coordinates to $116 \times 8 = 928$ candidate means. The score forest instead maps 48 inputs to nine coordinates, which decode into 116 scores. Score construction uses $\nu = 8$, candidate index 8 as the reference and column-scale floor 0.001. Rank 9 was retained during retrospective development. Supplementary Tables~\ref{tab:modelstructures} and~\ref{tab:hyperparameters} list the architectures and distinguish fixed settings from selected ones.

\subsection{Fixed comparator settings}\label{app:submissioncomparators}

PSO uses population 10, inertia 0.8 and cognitive/social weights 0.5. CMA-ES uses population 10 and initial unit-coordinate step 0.30. GP-BO uses five initial random points and noisy log expected improvement~\cite{ament2023logei}, while TuRBO-1 uses $2p$ initialization points. The sbi posterior models use the recorded default architectures. Under a one-hour fit cap, we retain their lowest finite internal-validation-loss checkpoint. CMPE and sCM use BayesFlow TimeMLP defaults, 10,000 Adam updates at learning rate $10^{- 3}$, batch size 256 and a 90/10 training/monitoring split under the same fit cap. Their final finite checkpoints use 11 and 15 sampling steps, respectively. These settings were fixed without selection on evaluation error.

\subsection{Comparison workflows and budgets}\label{app:submissioncomparison}

The four tasks test repeated calibration across different response dimensions, parameter spaces and observation sources (Table~I). Every method returns feasible parameters and is evaluated with independent simulator responses. The main comparison measures calibration error and complete query time. Component studies then examine coverage, local correction, incomplete outputs and learning from historical observations.

We compare the prepared optimizer with PSO, CMA-ES, GP-BO and TuRBO-1 on all four tasks. Each search starts at the unit-box centre, and this evaluation counts toward its budget. Searches use seeds 7, 17 and 27. Financial searches allow 512 candidate evaluations with four responses each. The other tasks allow 128 evaluations with four responses each. All searches have a time limit of 10 minutes.

In finance, all optimizers receive the same future-response prediction from the 48 available inputs. They minimize standardized squared error between this prediction and a four-response simulation mean. The subsequent market outcome is reserved for evaluation. We record improvements in the search objective to form incumbent histories, then reconstruct returns at one-minute intervals. These checkpoints identify when the cohort mean first falls below the direct reference. This comparison also includes budget-adapted NL-SHADE-LBC~\cite{stanovov2022nlshade}, SNO-PSO and SNO-DE~\cite{tsai2026sno}. Supplement~\ref{app:financialsearch} specifies search and evaluation streams.

Empirical selection uses stored responses, while surrogate selection uses predicted responses. Wider memories distribute the same retained-response budget across more parameter locations. Local and neural variants change correction and validity estimation. In finance, conditional surrogate selection predicts the future response before comparing candidate means. It uses the same observations and response surrogate as the matched direct model. Static controls select one parameter globally or per stock by average surrogate-estimated training error, then reuse it across windows.

\subsection{Finite-sample evaluation metrics}\label{app:submissionmetrics}

We evaluate returned parameters using the expected calibration error $\R$ in Eq.~(2), which averages the response loss $\ell$. For target $i$, independent simulations give an estimate $\ell_{i}$, and the reported mean error is $N^{- 1}\sum_{i = 1}^{N}\ell_{i}$. Lower values indicate better agreement within a task. The training objectives $L_{F}$ and $L_{N}$ fit the surrogate and are distinct from this evaluation. For traffic and predator--prey, we use

\begin{equation}\label{eq:evaluationmse}
\ell_i=\frac{1}{R_{\rm eval}d}\sum_{r=1}^{R_{\rm eval}}\sum_{k=1}^{d}
\left(\frac{z_{irk}-x_{ik}}{\sigma_k}\right)^2,
\end{equation}

where training responses determine $\sigma_{k}$, $R_{\rm eval} = 64$, and $d = 36$ or 18. For neural evaluation, we average the squared error of the $(a \odot v,5a)$ encoding over 16 responses and divide by $d = 18$. Each missing feature contributes $\frac{25}{d}$ through its validity coordinate. We also report numeric error and complete-response rate to distinguish fit quality from feature availability.

Financial simulated summaries $z_{ir}$ and outcome $Y_{i}$ use the same training scales and current-book baseline. With $R_{\rm eval} = 16$, define

\begin{equation}\label{eq:energydistances}
\begin{aligned}
D_{{\rm obs},i}&=\frac{1}{R_{\rm eval}}\sum_{r=1}^{R_{\rm eval}}\norm{z_{ir}-Y_i},\\
D_{{\rm sim},i}&=\frac{1}{R_{\rm eval}(R_{\rm eval}-1)}
\sum_{r\ne q}\norm{z_{ir}-z_{iq}}.
\end{aligned}
\end{equation}

These distances give the per-window calibration error

\begin{equation}\label{eq:suppenergy}
\ell_i=\widehat{\mathrm{ES}}_i=2D_{{\rm obs},i}-D_{{\rm sim},i}.
\end{equation}

This criterion is twice the conventional energy score, a proper scoring rule for multivariate distributions~\cite{gneiting2007strictly}.

\subsection{Timing and aggregation}\label{app:submissiontiming}

Alongside calibration error, we report computation times in milliseconds, with search budgets and checkpoint intervals expressed in minutes. Query latency is measured on one AMD EPYC 9755 CPU core at batch size one. Timing includes normalization, selection or inference, physical conversion and financial record summarization. Prepared methods use five warm-ups and 31 interleaved repetitions per target. Posterior methods use three complete sampled queries, while search is timed once per target/seed. Financial forests use exported C tree traversal. Model loading, disk access and preparation precede query timing.

Tables summarize variability as mean ± sample SD across complete-cohort run means. For prepared timings, we first take each target's median. Search and financial direct comparisons have three runs. Financial and neural posterior comparisons have two fits, while low-dimensional posterior comparisons have four bank/fit configurations. The proposed optimizer uses two bank means for predator--prey, four bank/fit configurations for traffic and two fits for neural circuits. Financial component studies use three banks with five fits each, or five fits at one bank for training-window controls. For paired financial intervals, we average fitted rules within windows before taking 20,000 crossed date/stock bootstrap draws. These nominal intervals describe the five-stock, five-date cohort. Supplement~\ref{app:materials} specifies the remaining aggregation rules.

\subsection{Matched-quality attainment}\label{app:submissionattainment}

The matched-quality time in Eq.~(20) is
\[
T_{\rm match}=\min\{t\in\mathcal T\mid\ell_i(\theta_t^{\rm search})\leq\ell_i(\widehat\theta^{\rm direct})\},
\]
where $\mathcal{T}$ contains the recorded incumbent times and both returns are evaluated independently. Non-matches remain in the denominator of the matching fraction. Median times and paired ratios use successful matches. We also compare quality at the cohort level, recording the first one-minute checkpoint whose mean error falls below the direct mean. Its elapsed search time is reported in minutes.

\subsection{Financial response diagnostics}\label{app:submissiondiagnostics}

\clearpage
\bibliographystyle{IEEEtran}
\bibliography{references}